\documentclass[aps,prx,twocolumn,nofootinbib,nopreprintnumbers,showpacs,floatfix,longbibliography,superscriptaddress]{revtex4-1}
\usepackage{color}
\usepackage{calc}
\usepackage{mathtools,graphicx}
\usepackage{pifont}
\usepackage{bm}
\usepackage{microtype}
\usepackage{booktabs}
\usepackage{times}
\usepackage[varg]{txfonts}
\usepackage[colorlinks, pdfborder={0 0 0}]{hyperref}
\usepackage[utf8]{inputenc}
\usepackage[caption=false]{subfig}
\usepackage{paralist}
\usepackage[normalem]{ulem}
\usepackage{multirow}
\usepackage{etoolbox}
\definecolor{LinkColor}{rgb}{0.75, 0, 0}
\definecolor{CiteColor}{rgb}{0, 0.5, 0.5}
\definecolor{UrlColor}{rgb}{0, 0, 0.75}
\hypersetup{linkcolor=LinkColor}
\hypersetup{citecolor=CiteColor}
\hypersetup{urlcolor=UrlColor}
\allowdisplaybreaks
\DeclareFontFamily{OT1}{pzc}{}
\DeclareFontShape{OT1}{pzc}{m}{it}{<-> s * [1.10] pzcmi7t}{}
\DeclareMathAlphabet{\mathpzc}{OT1}{pzc}{m}{it}

\AtBeginDocument{%
    \newwrite\bibnotes
    \def\bibnotesext{Notes.bib}
    \immediate\openout\bibnotes=\jobname\bibnotesext
    \immediate\write\bibnotes{@CONTROL{REVTEX41Control}}
    \immediate\write\bibnotes{@CONTROL{%
    apsrev41Control,author="08",editor="1",pages="0",title="0",year="1"}}
     \if@filesw
     \immediate\write\@auxout{\string\citation{apsrev41Control}}%
    \fi
}%

\newtoggle{fullauthorlist}
\toggletrue{fullauthorlist}

\newtoggle{endauthorlist}
\toggletrue{endauthorlist}

\newcommand{\macro}[1]{\textcolor{black}{#1}}

\newcommand{\DISTANCECOMPACTOneFiveZeroNineOneFourCat}{\macro{\ensuremath{440_{-170}^{+150}}}} 
\newcommand{\MTOTSCOMPACTOneFiveZeroNineOneFourCat}{\macro{\ensuremath{66.1_{-3.3}^{+3.8}}}} 
\newcommand{\MFINALSavgCOMPACTOneFiveZeroNineOneFourCat}{\macro{\ensuremath{63.1_{-3.0}^{+3.4}}}} 
\newcommand{\SPINFINALCOMPACTOneFiveZeroNineOneFourCat}{\macro{\ensuremath{0.69_{-0.04}^{+0.05}}}} 
\newcommand{\PEMATCHSNRCOMPACTOneFiveZeroNineOneFourCat}{\macro{\ensuremath{25.3_{-0.2}^{+0.1}}}} 

\newcommand{\DISTANCECOMPACTOneFiveOneZeroOneTwoCat}{\macro{\ensuremath{1080_{-490}^{+550}}}} 
\newcommand{\MTOTSCOMPACTOneFiveOneZeroOneTwoCat}{\macro{\ensuremath{37.2_{-3.9}^{+10.6}}}} 
\newcommand{\MFINALSavgCOMPACTOneFiveOneZeroOneTwoCat}{\macro{\ensuremath{35.6_{-3.8}^{+10.8}}}} 
\newcommand{\SPINFINALCOMPACTOneFiveOneZeroOneTwoCat}{\macro{\ensuremath{0.67_{-0.11}^{+0.13}}}} 
\newcommand{\PEMATCHSNRCOMPACTOneFiveOneZeroOneTwoCat}{\macro{\ensuremath{9.2_{-0.4}^{+0.3}}}} 

\newcommand{\DISTANCECOMPACTOneFiveOneTwoTwoSixCat}{\macro{\ensuremath{450_{-190}^{+180}}}} 
\newcommand{\MTOTSCOMPACTOneFiveOneTwoTwoSixCat}{\macro{\ensuremath{21.5_{-1.5}^{+6.2}}}} 
\newcommand{\MFINALSavgCOMPACTOneFiveOneTwoTwoSixCat}{\macro{\ensuremath{20.5_{-1.5}^{+6.4}}}} 
\newcommand{\SPINFINALCOMPACTOneFiveOneTwoTwoSixCat}{\macro{\ensuremath{0.74_{-0.05}^{+0.07}}}} 
\newcommand{\PEMATCHSNRCOMPACTOneFiveOneTwoTwoSixCat}{\macro{\ensuremath{12.4_{-0.3}^{+0.2}}}} 
\newcommand{\SPINMAXMINOneFiveOneTwoTwoSixCat}{\macro{\ensuremath{0.28}}} 

\newcommand{\DISTANCECOMPACTOneSevenZeroOneZeroFourCat}{\macro{\ensuremath{990_{-430}^{+440}}}} 
\newcommand{\MTOTSCOMPACTOneSevenZeroOneZeroFourCat}{\macro{\ensuremath{51.0_{-4.1}^{+5.3}}}} 
\newcommand{\MFINALSavgCOMPACTOneSevenZeroOneZeroFourCat}{\macro{\ensuremath{48.9_{-4.0}^{+5.1}}}} 
\newcommand{\SPINFINALCOMPACTOneSevenZeroOneZeroFourCat}{\macro{\ensuremath{0.66_{-0.11}^{+0.08}}}} 
\newcommand{\PEMATCHSNRCOMPACTOneSevenZeroOneZeroFourCat}{\macro{\ensuremath{14.0_{-0.3}^{+0.2}}}} 

\newcommand{\DISTANCECOMPACTOneSevenZeroSixZeroEightCat}{\macro{\ensuremath{320_{-110}^{+120}}}} 
\newcommand{\MTOTSCOMPACTOneSevenZeroSixZeroEightCat}{\macro{\ensuremath{18.6_{-0.7}^{+3.2}}}} 
\newcommand{\MFINALSavgCOMPACTOneSevenZeroSixZeroEightCat}{\macro{\ensuremath{17.8_{-0.7}^{+3.4}}}} 
\newcommand{\SPINFINALCOMPACTOneSevenZeroSixZeroEightCat}{\macro{\ensuremath{0.69_{-0.04}^{+0.04}}}} 
\newcommand{\PEMATCHSNRCOMPACTOneSevenZeroSixZeroEightCat}{\macro{\ensuremath{15.6_{-0.3}^{+0.2}}}} 

\newcommand{\DISTANCECOMPACTOneSevenZeroSevenTwoNineCat}{\macro{\ensuremath{2840_{-1360}^{+1400}}}} 
\newcommand{\MTOTSCOMPACTOneSevenZeroSevenTwoNineCat}{\macro{\ensuremath{84.4_{-11.1}^{+15.8}}}} 
\newcommand{\MFINALSavgCOMPACTOneSevenZeroSevenTwoNineCat}{\macro{\ensuremath{79.5_{-10.2}^{+14.7}}}} 
\newcommand{\SPINFINALCOMPACTOneSevenZeroSevenTwoNineCat}{\macro{\ensuremath{0.81_{-0.13}^{+0.07}}}} 
\newcommand{\PEMATCHSNRCOMPACTOneSevenZeroSevenTwoNineCat}{\macro{\ensuremath{10.8_{-0.5}^{+0.4}}}} 
\newcommand{\SPINMAXMINOneSevenZeroSevenTwoNineCat}{\macro{\ensuremath{0.27}}} 

\newcommand{\DISTANCECOMPACTOneSevenZeroEightZeroNineCat}{\macro{\ensuremath{1030_{-390}^{+320}}}} 
\newcommand{\MTOTSCOMPACTOneSevenZeroEightZeroNineCat}{\macro{\ensuremath{59.0_{-4.1}^{+5.4}}}} 
\newcommand{\MFINALSavgCOMPACTOneSevenZeroEightZeroNineCat}{\macro{\ensuremath{56.3_{-3.8}^{+5.2}}}} 
\newcommand{\SPINFINALCOMPACTOneSevenZeroEightZeroNineCat}{\macro{\ensuremath{0.70_{-0.09}^{+0.08}}}} 
\newcommand{\PEMATCHSNRCOMPACTOneSevenZeroEightZeroNineCat}{\macro{\ensuremath{12.7_{-0.3}^{+0.2}}}} 

\newcommand{\DISTANCECOMPACTOneSevenZeroEightOneFourCat}{\macro{\ensuremath{600_{-220}^{+150}}}} 
\newcommand{\MTOTSCOMPACTOneSevenZeroEightOneFourCat}{\macro{\ensuremath{55.9_{-2.6}^{+3.4}}}} 
\newcommand{\MFINALSavgCOMPACTOneSevenZeroEightOneFourCat}{\macro{\ensuremath{53.2_{-2.4}^{+3.2}}}} 
\newcommand{\SPINFINALCOMPACTOneSevenZeroEightOneFourCat}{\macro{\ensuremath{0.72_{-0.05}^{+0.07}}}} 
\newcommand{\PEMATCHSNRCOMPACTOneSevenZeroEightOneFourCatLI}{\macro{\ensuremath{17.8_{-0.3}^{+0.3}}}} 

\newcommand{\DISTANCECOMPACTOneSevenZeroEightOneEightCat}{\macro{\ensuremath{1060_{-380}^{+420}}}} 
\newcommand{\MTOTSCOMPACTOneSevenZeroEightOneEightCat}{\macro{\ensuremath{62.2_{-4.1}^{+5.2}}}} 
\newcommand{\MFINALSavgCOMPACTOneSevenZeroEightOneEightCat}{\macro{\ensuremath{59.4_{-3.8}^{+4.9}}}} 
\newcommand{\SPINFINALCOMPACTOneSevenZeroEightOneEightCat}{\macro{\ensuremath{0.67_{-0.08}^{+0.07}}}} 
\newcommand{\PEMATCHSNRCOMPACTOneSevenZeroEightOneEightCat}{\macro{\ensuremath{11.9_{-0.4}^{+0.3}}}} 

\newcommand{\DISTANCECOMPACTOneSevenZeroEightTwoThreeCat}{\macro{\ensuremath{1940_{-900}^{+970}}}} 
\newcommand{\MTOTSCOMPACTOneSevenZeroEightTwoThreeCat}{\macro{\ensuremath{68.7_{-8.1}^{+10.8}}}} 
\newcommand{\MFINALSavgCOMPACTOneSevenZeroEightTwoThreeCat}{\macro{\ensuremath{65.4_{-7.4}^{+10.1}}}} 
\newcommand{\SPINFINALCOMPACTOneSevenZeroEightTwoThreeCat}{\macro{\ensuremath{0.72_{-0.12}^{+0.09}}}} 
\newcommand{\PEMATCHSNRCOMPACTOneSevenZeroEightTwoThreeCat}{\macro{\ensuremath{12.0_{-0.3}^{+0.2}}}} 

\newcommand{\FARperYearPyCBCGWOneFiveZeroNineOneFour}{\macro{\ensuremath{<1.5 \times 10^{-5} }}} 
\newcommand{\FARperYearGstLALGWOneFiveZeroNineOneFour}{\macro{\ensuremath{<1.0 \times 10^{-7} }}} 
\newcommand{\FARperYearcWBGWOneFiveZeroNineOneFour}{\macro{\ensuremath{<1.6 \times 10^{-4} }}} 

\newcommand{\FARperYearPyCBCGWOneFiveOneZeroOneTwo}{\macro{\ensuremath{0.17}}} 
\newcommand{\FARperYearGstLALGWOneFiveOneZeroOneTwo}{\macro{\ensuremath{7.9 \times 10^{-3} }}} 
\newcommand{\FARperYearcWBGWOneFiveOneZeroOneTwo}{\macro{--}} 

\newcommand{\FARperYearPyCBCGWOneFiveOneTwoTwoSix}{\macro{\ensuremath{<1.7 \times 10^{-5} }}} 
\newcommand{\FARperYearGstLALGWOneFiveOneTwoTwoSix}{\macro{\ensuremath{<1.0 \times 10^{-7} }}} 
\newcommand{\FARperYearcWBGWOneFiveOneTwoTwoSix}{\macro{\ensuremath{0.02}}} 

\newcommand{\FARperYearPyCBCGWOneSevenZeroOneZeroFour}{\macro{\ensuremath{<1.4 \times 10^{-5} }}} 
\newcommand{\FARperYearGstLALGWOneSevenZeroOneZeroFour}{\macro{\ensuremath{<1.0 \times 10^{-7} }}} 
\newcommand{\FARperYearcWBGWOneSevenZeroOneZeroFour}{\macro{\ensuremath{2.9 \times 10^{-4} }}} 

\newcommand{\FARperYearPyCBCGWOneSevenZeroSixZeroEight}{\macro{\ensuremath{<3.1 \times 10^{-4} }}} 
\newcommand{\FARperYearGstLALGWOneSevenZeroSixZeroEight}{\macro{\ensuremath{<1.0 \times 10^{-7} }}} 
\newcommand{\FARperYearcWBGWOneSevenZeroSixZeroEight}{\macro{\ensuremath{1.4 \times 10^{-4} }}} 

\newcommand{\FARperYearPyCBCGWOneSevenZeroSevenTwoNine}{\macro{\ensuremath{1.4}}} 
\newcommand{\FARperYearGstLALGWOneSevenZeroSevenTwoNine}{\macro{\ensuremath{0.18}}} 
\newcommand{\FARperYearcWBGWOneSevenZeroSevenTwoNine}{\macro{\ensuremath{0.02}}} 

\newcommand{\FARperYearPyCBCGWOneSevenZeroEightZeroNine}{\macro{\ensuremath{1.4 \times 10^{-4} }}} 
\newcommand{\FARperYearGstLALGWOneSevenZeroEightZeroNine}{\macro{\ensuremath{<1.0 \times 10^{-7} }}} 
\newcommand{\FARperYearcWBGWOneSevenZeroEightZeroNine}{\macro{--}} 

\newcommand{\FARperYearPyCBCGWOneSevenZeroEightOneFour}{\macro{\ensuremath{<1.2 \times 10^{-5} }}} 
\newcommand{\FARperYearGstLALGWOneSevenZeroEightOneFour}{\macro{\ensuremath{<1.0 \times 10^{-7} }}} 
\newcommand{\FARperYearcWBGWOneSevenZeroEightOneFour}{\macro{\ensuremath{<2.1 \times 10^{-4} }}} 

\newcommand{\FARperYearPyCBCGWOneSevenZeroEightOneEight}{\macro{--}} 
\newcommand{\FARperYearGstLALGWOneSevenZeroEightOneEight}{\macro{\ensuremath{4.2 \times 10^{-5} }}} 
\newcommand{\FARperYearcWBGWOneSevenZeroEightOneEight}{\macro{--}} 

\newcommand{\FARperYearPyCBCGWOneSevenZeroEightTwoThree}{\macro{\ensuremath{<3.3 \times 10^{-5} }}} 
\newcommand{\FARperYearGstLALGWOneSevenZeroEightTwoThree}{\macro{\ensuremath{<1.0 \times 10^{-7} }}} 
\newcommand{\FARperYearcWBGWOneSevenZeroEightTwoThree}{\macro{\ensuremath{2.1 \times 10^{-3} }}} 

\newcommand{\MinimumImprovementFromPreviousBoundsOTwoTGR}{\macro{\ensuremath{1.1}}}
\newcommand{\MaximumImprovementFromPreviousBoundsOTwoTGR}{\macro{\ensuremath{2.5}}}

\newcommand{\NinetyPercentCombinedGravitonMassBoundScaledOTwoTGR}{4.7}
\newcommand{\NinetyPercentCombinedGravitonMassBoundOTwoTGRImprovementFromGWOneSevenZeroOneZeroFourPaper}{1.6}

\newcommand{\IMRP}{\textsc{IMRPhenomPv2}}
\newcommand{\SEOB}{\textsc{SEOBNRv4}}
\newcommand{\SEOBROM}{\textsc{SEOBNRv4\_ROM}}

\newcommand{\cmark}{\ding{51}}

\newcommand{\linf}{\textsc{LALInference}}

\newcommand{\bw}{\textsc{BayesWave}}
\newcommand{\gstlal}{\textsc{GstLAL}}
\newcommand{\pycbc}{\textsc{PyCBC}}
\newcommand{\cwb}{\textsc{cWB}}

\newcommand{\twoc}[1]{\multicolumn{2}{c}{#1}}

\newcommand{\threec}[1]{\multicolumn{3}{c}{#1}}

\newcommand{\sixc}[1]{\multicolumn{6}{c}{#1}}
\DeclareMathOperator{\sign}{sign}

\newcommand{\detections}{\cite{GW150914_paper, TheLIGOScientific:2016qqj, GW151226, O1:BBH, GW170104, GW170608, GW170814paper}}

\begin{document}

\title{Tests of General Relativity with the Binary Black Hole Signals from the LIGO-Virgo Catalog GWTC-1}

\iftoggle{endauthorlist}{
 %
 %
 \let\mymaketitle\maketitle
 \let\myauthor\author
 \let\myaffiliation\affiliation
 \author{The LIGO Scientific Collaboration and the Virgo Collaboration}
}{
 %
 %
 \iftoggle{fullauthorlist}{

%

\author{B.~P.~Abbott}
\affiliation{LIGO, California Institute of Technology, Pasadena, CA 91125, USA}
\author{R.~Abbott}
\affiliation{LIGO, California Institute of Technology, Pasadena, CA 91125, USA}
\author{T.~D.~Abbott}
\affiliation{Louisiana State University, Baton Rouge, LA 70803, USA}
\author{S.~Abraham}
\affiliation{Inter-University Centre for Astronomy and Astrophysics, Pune 411007, India}
\author{F.~Acernese}
\affiliation{Universit\`a di Salerno, Fisciano, I-84084 Salerno, Italy}
\affiliation{INFN, Sezione di Napoli, Complesso Universitario di Monte S.Angelo, I-80126 Napoli, Italy}
\author{K.~Ackley}
\affiliation{OzGrav, School of Physics \& Astronomy, Monash University, Clayton 3800, Victoria, Australia}
\author{C.~Adams}
\affiliation{LIGO Livingston Observatory, Livingston, LA 70754, USA}
\author{R.~X.~Adhikari}
\affiliation{LIGO, California Institute of Technology, Pasadena, CA 91125, USA}
\author{V.~B.~Adya}
\affiliation{Max Planck Institute for Gravitational Physics (Albert Einstein Institute), D-30167 Hannover, Germany}
\affiliation{Leibniz Universit\"at Hannover, D-30167 Hannover, Germany}
\author{C.~Affeldt}
\affiliation{Max Planck Institute for Gravitational Physics (Albert Einstein Institute), D-30167 Hannover, Germany}
\affiliation{Leibniz Universit\"at Hannover, D-30167 Hannover, Germany}
% ...

\collaboration{The LIGO Scientific Collaboration and the Virgo Collaboration}
%
%

 }{
 \author{The LIGO Scientific Collaboration and the Virgo Collaboration}
}
}

\date[\relax]{compiled \today}

\begin{abstract}
The detection of gravitational waves by Advanced LIGO and Advanced Virgo provides an opportunity to test general relativity in a regime that is inaccessible to traditional astronomical observations and laboratory tests. We present four tests of the consistency of the data with binary black hole gravitational waveforms predicted by general relativity. One test subtracts the best-fit waveform from the data and checks the consistency of the residual with detector noise. The second test checks the consistency of the low- and high-frequency parts of the observed signals. The third test checks that phenomenological deviations introduced in the waveform model (including in the post-Newtonian coefficients) are consistent with zero. The fourth test constrains modifications to the propagation of gravitational waves due to a modified dispersion relation, including that from a massive graviton. We present results both for individual events and also results obtained by combining together particularly strong events from the first and second observing runs of Advanced LIGO and Advanced Virgo, as collected in the catalog GWTC-1. We do not find any inconsistency of the data with the predictions of general relativity and improve our previously presented combined constraints by factors of $\MinimumImprovementFromPreviousBoundsOTwoTGR$ to $\MaximumImprovementFromPreviousBoundsOTwoTGR$. In particular, we bound the mass of the graviton to be $m_g \leq \NinetyPercentCombinedGravitonMassBoundScaledOTwoTGR \times 10^{-23} \text{ eV}/c^2$ ($90\%$ credible level), an improvement of a factor of $\NinetyPercentCombinedGravitonMassBoundOTwoTGRImprovementFromGWOneSevenZeroOneZeroFourPaper$ over our previously presented results.
Additionally, we check that the four gravitational-wave events published for the first time in GWTC-1 do not lead to stronger constraints on alternative polarizations than those published previously.
\end{abstract}

\maketitle

\section{Introduction}
\label{sec:intro}
Einstein's theory of gravity, general relativity (GR), has withstood a 
large number of experimental tests~\cite{lrr-2014-4}. 
With the advent of gravitational-wave (GW) astronomy and the observations 
by the Advanced LIGO~\cite{TheLIGOScientific:2014jea} and Advanced Virgo~\cite{TheVirgo:2014hva} 
detectors, a range of new tests of GR have become possible. 
These include both weak field tests of the propagation of 
GWs, as well as tests of the strong field regime of 
compact binary sources. See~\cite{GW150914:TGR,O1:BBH,GW170104,GW170814paper,bns-tgr} for previous applications of such tests to GW data.

We report results from tests of GR on all the confident binary black hole
GW events in the catalog GWTC-1~\cite{GWOSC:GWTC}, i.e., those from the first and second observing runs of the advanced generation
of detectors. Besides all of the events previously announced
(GW150914, GW151012, GW151226, GW170104, GW170608, and GW170814)~\detections,
this includes the four new GW events reported in~\cite{O2:Catalog} 
(GW170729, GW170809, GW170818, and GW170823). 
We do not investigate any of the marginal triggers in GWTC-1, which have a false-alarm rate (FAR) greater than
one per year.
Table~\ref{tab:events} displays a complete list of the events we consider.
Tests of GR on the binary neutron star event GW170817 are described in~\cite{bns-tgr}.

The search results in~\cite{O2:Catalog} originate from two
modeled searches and one weakly modeled search~\cite{TheLIGOScientific:2016qqj, 
TheLIGOScientific:2016uux,O1:BBH,O2:Catalog}. 
The modeled searches use templates based on GR to find candidate events and 
to assess their significance. However, detection by such searches 
does not in itself imply full compatibility of the signal with 
GR \cite{Vallisneri:2013rc,Vitale:2013bma}.
The weakly modeled search relies on coherence of signals between multiple detectors, 
as expected for an astrophysical source. While it assumes that the morphology 
of the signal resembles a chirp (whose frequency increases with time), as 
expected for a compact binary coalescence, it does 
not assume that the detailed waveform shape agrees with GR. 
A transient signal strongly deviating from GR would likely be found by the 
weakly modeled search, even if missed by the modeled searches.
So far, however, all significant [$\mathrm{FAR} < (1\, \mathrm{yr})^{-1}$] transient 
signals found by the weakly modeled
search were also found
by at least one of the modeled searches~\cite{O2:Catalog}.

At present, there are no complete theories of gravity other than GR that 
are mathematically and physically viable and provide well defined alternative
predictions for the waveforms arising from the coalescence of two black holes 
(if, indeed, these theories even admit black holes).\footnote{There are very preliminary simulations of
scalar waveforms from binary black holes in the effective field theory (EFT)
framework in alternative theories~\cite{Okounkova:2017yby,Witek:2018dmd}, and the leading corrections to the gravitational
waveforms in head-on collisions~\cite{Okounkova:2019dfo}, but these simulations require much more development before their
results can be used in gravitational wave data analysis. There are also concerns about the mathematical viability of the theories
considered when they are not treated in the EFT framework.
}
Thus, we cannot test GR by direct comparison with other specific theories. Instead, we can 
\begin{inparaenum}[(i)]
\item check the consistency of the GR predictions with the data and 
\item introduce \textit{ad hoc} modifications in GR waveforms to determine the degree to which the values of the deviation parameters agree with GR.
\end{inparaenum}
These methods are agnostic to any particular choice of alternative theory.
For the most part, our results should therefore be interpreted as observational constraints on possible GW phenomenologies, independent of the overall suitability or well-posedness of any specific alternative to GR.
These limits are useful in providing a quantitative indication of the degree to which the data is described by GR; they may also be interpreted more specifically in the context of any given alternative to produce constraints, if applicable.

In particular, with regard to the consistency of the GR predictions (i), we 
\begin{inparaenum}[(a)]
\item look for residual power after subtracting the best-fitting GR waveform from the data, and  
\item evaluate the consistency of the high and low frequency components of the observed signal. 
\end{inparaenum}
With regard to deviations from GR (ii), we separately introduce parametrizations for 
\begin{inparaenum}[(a)]
\item the emitted waveform, and 
\item its propagation. 
\end{inparaenum}
The former could be viewed as representing possible GR modifications in the 
strong-field region close to the binary, while the latter would correspond to 
weak-field modifications away from the source.
Although we consider these independently, modifications to GW propagation 
would most likely be accompanied by modifications to GW generation 
in any given extension of GR.
We have also checked that none of the events discussed here provide stronger constraints on models with purely vector and purely scalar GW polarizations than those previously published in~\cite{GW170814paper,bns-tgr}. 
Our analyses do not reveal any inconsistency of the data with the predictions of GR.
These results supersede all our previous testing GR results on the binary black hole signals found in O1 and O2~\cite{GW150914:TGR,O1:BBH,GW170104,GW170814paper}. In particular, the previously published residuals and propagation test results were affected by a slight normalization issue.

Limits on deviations from GR for individual events are dominated 
by statistical errors due to detector noise. These errors can be 
reduced by appropriately combining results from multiple events.
Sources of systematic errors, on the other hand, 
include uncertainties in the detector calibration and power spectral density (PSD) 
estimation and errors in the modeling of waveforms in GR. 
Detector calibration uncertainties are modeled as corrections to the measured 
detector response function and are marginalized over. Studies on the effect of PSD uncertainties are currently ongoing.
A full characterization of the systematic errors due to the GR waveform models that we employ is beyond the scope of this study; some investigations can be found in~\cite{Bohe:2016gbl,Khan:2015jqa,Blackman:2017dfb,Khan:2018fmp,Abbott:2016wiq}. 

This paper is organized as follows.
Section~\ref{sec:data} provides an overview of the data sets employed here, 
while Sec.~\ref{sec:events} details which GW events are used to produce the individual and combined results presented in this paper. 
In Sec.~\ref{sec:inference} we explain the gravitational waveforms and data analysis formalisms which our tests of GR are based on, before we present the results in the following sections.
Section~\ref{sec:waveforms} contains two signal consistency tests: the 
residuals test in~\ref{sec:residuals} and the inspiral-merger-ringdown consistency test
 in~\ref{sec:imr-test}. Results from parameterized tests are given 
in Sec.~\ref{sec:gwnature} for GW generation,
and in Sec.~\ref{sec:propagation} for GW propagation.
We briefly discuss the study of GW polarizations in Sec.~\ref{sec:pols}.
Finally, we conclude in Sec.~\ref{sec:conclusions}. We give results for individual events and some checks on waveform systematics in the Appendix.

The results of each test and associated data products can be found in Ref.~\cite{GWOSC:O2TGR}. The GW strain data for all the events considered are available at the Gravitational Wave Open Science Center~\cite{GWOSC}.

\section{Data, calibration and cleaning}
\label{sec:data}

The first observing run of Advanced LIGO (O1) lasted from
September 12{$^{\rm th}$}, 2015 to January 19{$^{\rm th}$}, 2016.
The second observing run (O2) lasted from 
November 30{$^{\rm th}$}, 2016 to August 25{$^{\rm th}$}, 2017, 
with the Advanced Virgo observatory joining on August 1{$^{\rm st}$}, 2017.
This paper includes all GW events originating from the coalescence
of two black holes found in these two data sets and
published in~\cite{O1:BBH,O2:Catalog}.

The GW detector's response to changes 
in the differential arm length (the interferometer's degree of 
freedom most sensitive to GWs) must be calibrated 
using independent, accurate, absolute references. 
The LIGO detectors use photon recoil (radiation pressure) from 
auxiliary laser systems to induce mirror motions that change 
the arm cavity lengths, 
allowing a direct measure of the detector response
~\cite{Karki:2016pht, Cahillane:2017vkb, Viets:2017yvy}. 
Calibration of Virgo relies on measurements of Michelson interference
fringes as the main optics swing freely, using the primary laser 
wavelength as a fiducial length. Subsequent measurements 
propagate the calibration to arrive at the final
detector response~\cite{VIR-0362A-18, Acernese:2018bfl}.
These complex-valued, frequency-dependent measurements of the LIGO and Virgo
detectors' response yield the uncertainty in their respective estimated 
amplitude and phase of the GW strain output.
The amplitude and phase correction factors are modeled as cubic splines
and marginalized over in the estimation of astrophysical source parameters
~\cite{SplineCalMarg-T1400682, Veitch:2014wba, Abbott:2018wiz, O2:Catalog}. 
Additionally, the uncertainty in the time stamping 
of Virgo data (much larger than the LIGO timing uncertainty, which is included
in the phase correction factor) is also accounted for in the analysis.

Post-processing techniques to subtract noise contributions and frequency
lines from the data around gravitational-wave events were developed in O2 and 
introduced in~\cite{GW170608, GW170814paper, DetectionPaper}, for the astrophysical 
parameter estimation of GW170608, GW170814, and GW170817. 
This noise subtraction was achieved using optimal Wiener filters to calculate 
coupling transfer functions from auxiliary sensors~\cite{Driggers:2018gii}.
A new, optimized parallelizable method in the frequency domain~\cite{Davis:2018yrz} 
allows large scale noise subtraction on LIGO data. 
All of the O2 analyses presented in this manuscript use the noise-subtracted 
data set with the latest calibration available. The O1 data set is the same used in 
previous publications, as the effect of noise subtraction is expected to be negligible.
Reanalysis of the O1 events is motivated by improvements 
in the parameter estimation pipeline, an improved frequency-dependent calibration, 
and the availability of new waveform models.

\section{Events and significance}
\label{sec:events}
We present results for all confident detections of binary black hole events in 
GTWC-1~\cite{GWOSC:GWTC}, i.e., all such events detected during 
O1 and O2 with a FAR lower than one per year, as published in~\cite{O2:Catalog}.
The central columns of Table~\ref{tab:events} list the FARs of each event as evaluated by the three search pipelines used in~\cite{O2:Catalog}.
Two of these pipelines (\pycbc{} and \gstlal{}) rely on waveform templates computed from binary black hole coalescences in GR.
Making use of a measure of significance that assumes the validity of GR 
could potentially lead to biases in the selection of events to be tested, 
systematically disfavoring signals in which a GR violation would be most evident. 
Therefore, it is important to consider the possibilities that 
\begin{inparaenum}[(1)]
\item there were GW signals with such large deviations from GR 
that they were missed entirely by the modeled searches, and 
\item there were events that were picked up by the modeled searches but classified 
as marginal (and thus excluded from our analysis) because of their significant 
deviations from GR. 
\end{inparaenum}

These worries can largely be dispelled by considering
the third GW search pipeline, the coherent WaveBurst (\cwb{}) weakly modeled 
search presented in~\cite{O2:Catalog}. 
This \cwb{} search~\cite{Klimenko:2008fu, Klimenko:2015ypf, TheLIGOScientific:2016uux}
was tuned to detect chirping signals---like those that would be expected from compact 
binary coalescences---but was not tuned to any specific GR predictions.\footnote{Chirping
signals from compact binary coalescences are a feature of many theories of gravity. All that is
required is that the orbital frequency increases as the binary
radiates energy and angular momentum in GWs.}
\cwb{} is most sensitive to short signals from high-mass binary black holes. 
It is still able to detect signals from lower-mass binaries (e.g., GW151226), 
though with reduced significance compared to the modeled searches. Thus, a signal from
a low-mass binary, or a marginal event, with a significant departure from the GR predictions 
(hence not detected by the GR modeled searches) would not necessarily be detected by the \cwb{} 
search with a $\mathrm{FAR} < (1\, \mathrm{yr})^{-1}$. However, if there is a population of 
such signals, they will not all be weak and/or from low-mass binaries. Thus, one would expect 
some of the signals in the population to be detected by \cwb{}, even if they evade detection by the modeled searches.

All signals detected by the \cwb{} search with $\mathrm{FAR} < (1\, \mathrm{yr})^{-1}$ 
were also found by at least one modeled search with $\mathrm{FAR} < (1\, \mathrm{yr})^{-1}$. 
Given the above considerations, this is evidence that our analysis does not exclude chirping GW signals that were missed in the modeled searches because of drastic departures from GR. 
Similarly, this is also evidence against the possibility of marginal events representing a population of GR-deviating signals, as none of them show high significance [$\mathrm{FAR} < (1\, \mathrm{yr})^{-1}$] in the \cwb{} search only. 
Thus, we believe that we have not biased our analysis by considering only the ten events with $\mathrm{FAR} < (1\, \mathrm{yr})^{-1}$, as published in~\cite{O2:Catalog}.

We consider each of the GW events individually, carrying out different analyses on a case-by-case basis.
Some of the tests presented here, such as the inspiral-merger-ringdown (IMR) 
consistency test in Sec.~\ref{sec:imr-test} and the parameterized tests in 
Sec.~\ref{sec:gwnature}, distinguish between the inspiral and the post-inspiral 
regimes of the signal.
The separation between these two regimes is performed in the frequency-domain, 
choosing a particular cutoff frequency determined by the parameters of the event.
Larger-mass systems merge at lower frequencies, presenting a short inspiral signal in band; 
lower mass systems have longer observable inspiral signals, but the detector's sensitivity
decreases at higher frequencies and hence the post-inspiral signal becomes less informative. 
Therefore, depending on the total mass of the system, a particular signal might not provide 
enough information within the sensitive frequency band of the GW detectors for all analyses. 

As a proxy for the amount of information that can be extracted from each part of the signal, 
we calculate the signal-to-noise ratio (SNR) of the inspiral and the post-inspiral 
parts of the signals separately.
We only apply inspiral (post-inspiral) tests if the inspiral (post-inspiral) SNR is greater than 6.
Each test uses a different inspiral-cutoff frequency, and hence they assign different SNRs to the two regimes (details provided in the relevant section for each test).
In Table~\ref{tab:events} we indicate which analyses have been performed on which event, based on this frequency and the corresponding SNR.%
\footnote{\label{footnote:snr_thres} While we perform these tests on all events
with SNR $>6$ in the appropriate regime, in a few cases the results appear uninformative and the posterior
distribution extends across the entire prior considered.
Since the results are prior dependent, upper limits should not be set from 
these individual analyses. 
See Sec.~\ref{app:generation} of the Appendix for details.}

In addition to the individual analysis of each event, we derive combined constraints on departures from GR using multiple signals simultaneously.
Constraints from individual events are largely dominated by statistical uncertainties due to detector noise.
Combining events together can reduce such statistical errors on parameters that take consistent values across all events.
However, it is impossible to make joint probabilistic statements from multiple events without prior assumptions about the nature of each observation and how it relates to others in the set.
This means that, although there are well-defined statistical procedures for producing joint results, there is no unique way of doing so.

In light of this, we adopt what we take to be the most straightforward strategy, although future studies may follow different criteria.
First, in combining events we assume that deviations from GR are manifested equally across events, independent of source properties.
This is justified for studies of modified GW propagation, since those effects should not depend on the source.%
\footnote{Propagation effects do depend critically on source distance. However, this dependence is factored out explicitly, in a way that allows for combining events as we do here (see Sec.~\ref{sec:propagation}).}
For other analyses, it is quite a strong assumption to take all deviations from GR to be independent of source properties.
Such combined tests should not be expected to necessarily reveal generic source-dependent deviations, although they might if the departures from GR are large enough (see, e.g., \cite{Ghosh:2017gfp}).
Future work may circumvent this issue by combining marginalized likelihood ratios (Bayes factors), instead of posterior probability distributions \cite{Agathos:2013upa}.
More general ways of combining results are discussed and implemented in Refs.~\cite{Zimmerman:2019wzo,Isi:2019asy}.

Second, we choose to produce combined constraints only from events 
that were found in both modeled searches 
(\pycbc{}~\cite{pycbc-github, Canton:2014ena, Usman:2015kfa} and 
\gstlal{}~\cite{Sachdev:2019vvd,Messick:2016aqy})
with a FAR of at most one per one-thousand years.
This ensures that there is a very small probability of inclusion of a 
non-astrophysical event.
The events used for the combined results are indicated with bold names in Table~\ref{tab:events}.
The events thus excluded from the combined analysis have low SNR and would therefore contribute only marginally to tightened constraints.
Excluding marginal events from our analyses amounts to assigning a null \textit{a priori} probability to the possibility that those data contain any information about the tests in question.
This is, in a sense, the most conservative choice.

In summary, we enforce two significance thresholds:  $\mathrm{FAR} < (1\, \mathrm{yr})^{-1}$, for single-event analyses,  and $\mathrm{FAR} < (1000\, \mathrm{yr})^{-1}$, for combined results.
This two-tiered setup allows us to produce conservative joint results  by including only the most significant events, while also providing  information about a broader (less significant) set of triggers.
This is intended to enable the interested reader to combine individual  results with less stringent criteria and under different statistical assumptions,  according to their specific needs and tolerance for false positives.
In the future, we may adapt our thresholds depending on the rate of detections.

\begin{table*}
\caption{\label{tab:events}
The GW events considered in this paper, separated by observing run. The first block of columns gives the names of the events and lists some of their relevant
properties obtained using GR waveforms (luminosity distance $D_\text{L}$, source frame total mass $M_\text{tot}$ 
and final mass $M_\text{f}$, and dimensionless final spin $a_\text{f}$). The next block of columns gives the significance, measured by the false-alarm-rate (FAR),
with which each event was detected by each of the three searches employed, as well as the matched filter signal-to-noise
ratio from the stochastic sampling analyses with GR waveforms. A dash indicates that an event was not identified by a search.
The parameters and SNR values give the medians and 90\% credible intervals. All the events except for GW151226 and GW170729 are consistent with
a binary of nonspinning black holes (when analyzed assuming GR). See~\cite{O2:Catalog} for more details about all the events.
The last block of columns indicates which GR tests are performed on a given event: RT = residuals test (Sec.~\ref{sec:residuals}); IMR = inspiral-merger-ringdown consistency test (Sec.~\ref{sec:imr-test}); PI \& PPI = parameterized tests of GW generation for inspiral and post-inspiral phases (Sec.~\ref{sec:gwnature}); MDR = modified GW dispersion relation (Sec.~\ref{sec:propagation}).
The events with bold names are used to obtain the 
combined results for each test.
}
\begin{tabular}{ c c c c c c c c c c c c c c c c c}
\toprule
\multirow{2}{*}{Event} & \multicolumn{4}{c}{Properties} & \hphantom{X} & \multicolumn{3}{c}{FAR} & & \multirow{2}{*}{SNR} & \hphantom{X} & \multicolumn{5}{c}{GR tests performed}\\
\cline{2-5}
\cline{7-9}
\cline{13-17}
& $D_\text{L}$ & $M_\text{tot}$ & $M_\text{f}$ & $a_\text{f}$ & & \pycbc & \gstlal & \cwb & & & & RT & IMR & PI \ & PPI \ & MDR\\
 & [Mpc] & [$M_\odot$] & [$M_\odot$] & & & [yr$^{-1}$] & [yr$^{-1}$] & [yr$^{-1}$]  & & & & & & & &\\
\midrule
\textbf{GW150914}\footnote{\label{footnote:O1FARs} The FARs for these events 
differ from those in~\cite{O1:BBH} because the data were 
re-analyzed with the new pipeline statistics used in O2
(see~\cite{O2:Catalog} for more details).}\hphantom{,,} & \DISTANCECOMPACTOneFiveZeroNineOneFourCat & \MTOTSCOMPACTOneFiveZeroNineOneFourCat & \MFINALSavgCOMPACTOneFiveZeroNineOneFourCat & \SPINFINALCOMPACTOneFiveZeroNineOneFourCat & & \FARperYearPyCBCGWOneFiveZeroNineOneFour & \FARperYearGstLALGWOneFiveZeroNineOneFour & \FARperYearcWBGWOneFiveZeroNineOneFour & & \PEMATCHSNRCOMPACTOneFiveZeroNineOneFourCat & & \cmark & \cmark & \cmark & \cmark & \cmark\\[0.075cm]
GW151012\textsuperscript{\ref{footnote:O1FARs}}\hphantom{,,} & \DISTANCECOMPACTOneFiveOneZeroOneTwoCat & \MTOTSCOMPACTOneFiveOneZeroOneTwoCat & \MFINALSavgCOMPACTOneFiveOneZeroOneTwoCat & \SPINFINALCOMPACTOneFiveOneZeroOneTwoCat & & \FARperYearPyCBCGWOneFiveOneZeroOneTwo & \FARperYearGstLALGWOneFiveOneZeroOneTwo & \FARperYearcWBGWOneFiveOneZeroOneTwo & & \PEMATCHSNRCOMPACTOneFiveOneZeroOneTwoCat & & \cmark & -- & -- & \cmark & \cmark\\[0.075cm]
\textbf{GW151226}\textsuperscript{\ref{footnote:O1FARs},}\footnote{At least one black hole has dimensionless spin $> \SPINMAXMINOneFiveOneTwoTwoSixCat$ ($99\%$ credible level).} &\DISTANCECOMPACTOneFiveOneTwoTwoSixCat & \MTOTSCOMPACTOneFiveOneTwoTwoSixCat & \MFINALSavgCOMPACTOneFiveOneTwoTwoSixCat & \SPINFINALCOMPACTOneFiveOneTwoTwoSixCat & & \FARperYearPyCBCGWOneFiveOneTwoTwoSix & \FARperYearGstLALGWOneFiveOneTwoTwoSix & \FARperYearcWBGWOneFiveOneTwoTwoSix & & \PEMATCHSNRCOMPACTOneFiveOneTwoTwoSixCat & & \cmark & -- & \cmark & -- & \cmark\\
\midrule
\textbf{GW170104}\hphantom{,,,,} &\DISTANCECOMPACTOneSevenZeroOneZeroFourCat & \MTOTSCOMPACTOneSevenZeroOneZeroFourCat & \MFINALSavgCOMPACTOneSevenZeroOneZeroFourCat & \SPINFINALCOMPACTOneSevenZeroOneZeroFourCat & & \FARperYearPyCBCGWOneSevenZeroOneZeroFour & \FARperYearGstLALGWOneSevenZeroOneZeroFour & \FARperYearcWBGWOneSevenZeroOneZeroFour & & \PEMATCHSNRCOMPACTOneSevenZeroOneZeroFourCat & & \cmark & \cmark & \cmark & \cmark & \cmark\\[0.075cm]
\textbf{GW170608}\hphantom{,,,,} &  \DISTANCECOMPACTOneSevenZeroSixZeroEightCat & \MTOTSCOMPACTOneSevenZeroSixZeroEightCat & \MFINALSavgCOMPACTOneSevenZeroSixZeroEightCat & \SPINFINALCOMPACTOneSevenZeroSixZeroEightCat & & \FARperYearPyCBCGWOneSevenZeroSixZeroEight & \FARperYearGstLALGWOneSevenZeroSixZeroEight & \FARperYearcWBGWOneSevenZeroSixZeroEight & & \PEMATCHSNRCOMPACTOneSevenZeroSixZeroEightCat & & \cmark & -- & \cmark & \cmark & \cmark\\[0.075cm]
GW170729\footnote{This event has a higher significance in the unmodeled search than in the modeled searches. Additionally, at least one black hole has dimensionless spin $> \SPINMAXMINOneSevenZeroSevenTwoNineCat$ ($99\%$ credible level).}\hphantom{,,} & \DISTANCECOMPACTOneSevenZeroSevenTwoNineCat & \MTOTSCOMPACTOneSevenZeroSevenTwoNineCat & \MFINALSavgCOMPACTOneSevenZeroSevenTwoNineCat & \SPINFINALCOMPACTOneSevenZeroSevenTwoNineCat & & \FARperYearPyCBCGWOneSevenZeroSevenTwoNine & \FARperYearGstLALGWOneSevenZeroSevenTwoNine & \FARperYearcWBGWOneSevenZeroSevenTwoNine & & \PEMATCHSNRCOMPACTOneSevenZeroSevenTwoNineCat & & \cmark & \cmark & -- & \cmark & \cmark\\[0.075cm]
\textbf{GW170809}\hphantom{,,,,} & \DISTANCECOMPACTOneSevenZeroEightZeroNineCat & \MTOTSCOMPACTOneSevenZeroEightZeroNineCat & \MFINALSavgCOMPACTOneSevenZeroEightZeroNineCat & \SPINFINALCOMPACTOneSevenZeroEightZeroNineCat & & \FARperYearPyCBCGWOneSevenZeroEightZeroNine & \FARperYearGstLALGWOneSevenZeroEightZeroNine & \FARperYearcWBGWOneSevenZeroEightZeroNine & & \PEMATCHSNRCOMPACTOneSevenZeroEightZeroNineCat & & \cmark & \cmark & -- & \cmark & \cmark\\[0.075cm]
\textbf{GW170814}\hphantom{,,,,} & \DISTANCECOMPACTOneSevenZeroEightOneFourCat & \MTOTSCOMPACTOneSevenZeroEightOneFourCat & \MFINALSavgCOMPACTOneSevenZeroEightOneFourCat & \SPINFINALCOMPACTOneSevenZeroEightOneFourCat & & \FARperYearPyCBCGWOneSevenZeroEightOneFour & \FARperYearGstLALGWOneSevenZeroEightOneFour & \FARperYearcWBGWOneSevenZeroEightOneFour & & \PEMATCHSNRCOMPACTOneSevenZeroEightOneFourCatLI & & \cmark & \cmark & \cmark & \cmark & \cmark\\[0.075cm]
GW170818\hphantom{,,,,}
& \DISTANCECOMPACTOneSevenZeroEightOneEightCat & \MTOTSCOMPACTOneSevenZeroEightOneEightCat & \MFINALSavgCOMPACTOneSevenZeroEightOneEightCat & \SPINFINALCOMPACTOneSevenZeroEightOneEightCat & & \FARperYearPyCBCGWOneSevenZeroEightOneEight & \FARperYearGstLALGWOneSevenZeroEightOneEight & \FARperYearcWBGWOneSevenZeroEightOneEight & & \PEMATCHSNRCOMPACTOneSevenZeroEightOneEightCat & & \cmark & \cmark & -- & \cmark & \cmark\\[0.075cm]
\textbf{GW170823}\hphantom{,,,,} & \DISTANCECOMPACTOneSevenZeroEightTwoThreeCat & \MTOTSCOMPACTOneSevenZeroEightTwoThreeCat & \MFINALSavgCOMPACTOneSevenZeroEightTwoThreeCat & \SPINFINALCOMPACTOneSevenZeroEightTwoThreeCat & & \FARperYearPyCBCGWOneSevenZeroEightTwoThree & \FARperYearGstLALGWOneSevenZeroEightTwoThree & \FARperYearcWBGWOneSevenZeroEightTwoThree & & \PEMATCHSNRCOMPACTOneSevenZeroEightTwoThreeCat & & \cmark & \cmark & -- & \cmark & \cmark\\
\bottomrule
\end{tabular}
\end{table*}

\section{Parameter inference}
\label{sec:inference}
The starting point for all the analyses presented here are
waveform models that describe the GWs emitted
by coalescing black-hole binaries. The GW signature depends on
the intrinsic parameters describing the binary as well as the 
extrinsic parameters specifying the location and orientation 
of the source with respect to the detector network.
The intrinsic parameters for circularized black-hole binaries in GR
are the two masses $m_i$ of the black holes and the 
two spin vectors $\vec S_i$ defining the rotation of each black hole,
where $i\in\{1,2\}$ labels the two black holes.
We assume that the binary has negligible orbital eccentricity, as is expected
to be the case when the binary enters the band of ground-based
detectors~\cite{PhysRev.131.435, Peters:1964}
(except in some more extreme formation scenarios,\footnote{These scenarios
could occur often enough, compared to the expected rate of detections,
that the inclusion of eccentricity in waveform models is a necessity
for tests of GR in future observing runs; see, 
e.g.,~\cite{Hinderer:2017jcs, Cao:2017ndf, 
Hinder:2017sxy, Huerta:2017kez, Klein:2018ybm, Moore:2018kvz,Moore:2019xkm,Tiwari:2019jtz}
for recent work on developing such waveform models.}
e.g.,~\cite{Samsing:2017xmd,Rodriguez:2017pec,Zevin:2018kzq,Rodriguez:2018pss}).
The extrinsic parameters comprise four parameters that specify the space-time 
location of the binary black-hole, namely the sky location (right ascension 
and declination), the luminosity distance, and the time of coalescence.
In addition, there are three extrinsic parameters that determine the 
orientation of the binary with respect to Earth, namely the inclination angle 
of the orbit with respect to the observer, the polarization angle,
and the orbital phase at coalescence. 

We employ two waveform families that model binary
black holes in GR: the effective-one-body based
\SEOB~\cite{Bohe:2016gbl}
waveform family that assumes non-precessing spins
for the black holes (we use the frequency domain reduced order model
\SEOBROM{} for reasons of computational efficiency), and the phenomenological waveform family
\IMRP~\cite{Husa:2015iqa,Khan:2015jqa,Hannam:2013oca}
that models the effects of precessing spins
using two effective parameters by twisting up the underlying aligned-spin model.
We use \IMRP{} to obtain all the main results
given in this paper, and use \SEOB{} to check 
the robustness of these results, whenever possible. 
When we use \IMRP{}, we impose a prior $m_1/m_2 \leq 18$ on the mass ratio, as the
waveform family is not calibrated against numerical relativity simulations for $m_1/m_2 > 18$.
We do not impose a similar prior when using \SEOB{}, since it includes information about the
extreme mass ratio limit. Neither of these
waveform models includes the full spin dynamics 
(which requires 6 spin parameters). Fully precessing waveform models have been recently 
developed~\cite{Taracchini:2013, Babak:2016tgq, Blackman:2017pcm, Khan:2018fmp, Varma:2019csw}
and will be used in future applications of these tests.

The waveform models used in this paper do not include
the effects of subdominant (non-quadrupole) modes, which are expected to be
small for comparable-mass binaries~\cite{PhysRevD.85.024018, PhysRevD.90.124032}.
The first generation of binary black hole waveform models including spin and higher order modes 
has recently been developed~\cite{Blackman:2017pcm, London:2017bcn, Cotesta:2018fcv, Varma:2018mmi, Varma:2019csw}.
Preliminary results in~\cite{O2:Catalog}, using NR simulations supplemented by 
NR surrogate waveforms, indicate that the higher mode content of the 
GW signals detected by Advanced LIGO and Virgo is weak enough that models
without the effect of subdominant modes do not introduce substantial biases 
in the intrinsic parameters of the binary.
For unequal-mass binaries, the effect of the non-quadrupole modes is more 
pronounced~\cite{Varma:2016dnf}, 
particularly when the binary's orientation is close to edge-on. 
In these cases, the presence of non-quadrupole modes can show up as
a deviation from GR when using waveforms that only include the quadrupole modes, 
as was shown in~\cite{Pang:2018hjb}. 
Applications of tests of GR with the new waveform models that include non-quadrupole modes
will be carried out in the future.

We believe that our simplifying assumptions on the waveform models
(zero eccentricity, simplified treatment of spins, and neglect of subdominant modes)
are justified by astrophysical considerations and previous
studies. Indeed, as we show in the remainder of the paper, the
observed signals are consistent with the waveform models. Of course,
had our analyses resulted in evidence for violations of GR, we
would have had to revisit these simplifications very carefully.

The tests described in this paper are performed within the framework 
of Bayesian inference, by means of the \textsc{LALInference} 
code~\cite{Veitch:2014wba} in the LIGO Scientific Collaboration Algorithm 
Library Suite (LALSuite)~\cite{lalsuite}. We estimate the PSD using the \textsc{BayesWave}
code~\cite{Cornish:2014kda, Littenberg:2014oda}, as described in Appendix~B of~\cite{O2:Catalog}.
Except for the residuals test described in Sec.~\ref{sec:residuals},
we use the waveform models described in this section to estimate from the data 
the posterior distributions of the parameters of the binary.
These include not only the intrinsic and extrinsic parameters
mentioned above, but also other parameters that describe
possible departures from the GR predictions. Specifically, for the parameterized 
tests in Secs.~\ref{sec:gwnature} and~\ref{sec:propagation}, we modify the
phase $\Phi(f)$ of the frequency-domain waveform
\begin{equation}
\label{eq:h_tilde}
\tilde{h}(f) = A(f)e^{i\Phi(f)}.
\end{equation}
For the GR parameters, we use the same prior distributions as the main 
parameter estimation analysis described in~\cite{O2:Catalog}, though 
for a number of the tests we need to extend the ranges of these priors 
to account for correlations with the non-GR parameters, 
or for the fact that only a portion of the signal is analyzed 
(as in Sec.~\ref{sec:imr-test}). We also use the same low-frequency cutoffs
for the likelihood integral as in~\cite{O2:Catalog}, i.e., $20$~Hz for all events
except for GW170608, where $30$~Hz is used for LIGO Hanford, as discussed
in~\cite{GW170608}, and GW170818, where $16$~Hz is used for all three detectors.
For the model agnostic residual test described in Sec.~\ref{sec:residuals},
we use the \textsc{BayesWave} code~\cite{Cornish:2014kda} which describes 
the GW signals in terms of a number of Morlet-Gabor wavelets.

\section{Consistency tests}
\label{sec:waveforms}
\subsection{Residuals test}
\label{sec:residuals}

\newcommand{\bsn}{{\cal B}^{\rm S}_{\rm N}}
\newcommand{\bsg}{{\cal B}^{\rm S}_{\rm G}}
\newcommand{\rhores}{\mathrm{SNR}_{90}}
\newcommand{\rhosig}{\mathrm{SNR}_\mathrm{GR}}
\newcommand{\rhodiff}{\mathrm{SNR}_\mathrm{res}}
\newcommand{\rhonoise}{\rhores^\mathrm{n}}

\newcommand{\ff}{\mathrm{FF}_{90}}

One way to evaluate the ability of GR to describe GW signals is to subtract the best-fit template from the data and make sure the residuals have the statistical properties expected of instrumental noise.
This largely model-independent test is sensitive to a wide range of possible disagreements between the data and our waveform models, including those caused by deviations from GR and by modeling systematics.
This analysis can look for GR violations without relying on specific parametrizations of the deviations, making it a versatile tool.
Results from a similar study on our first detection were already presented in \cite{GW150914:TGR}.

In order to establish whether the residuals agree with noise 
(Gaussian or otherwise), we proceed as follows.
For each event in our set, we use \linf{} and the \IMRP{}
waveform family to obtain an estimate of the best-fit
(i.e., maximum likelihood) binary black-hole waveform based on GR.
This waveform incorporates factors that account for uncertainty in the detector calibration, as described in Sec.~\ref{sec:data}.
This best-fit waveform is then subtracted from the data to obtain residuals for a 1 second window centered on the trigger time reported in \cite{O2:Catalog}.%
\footnote{The analysis is sensitive only to residual power in that 1 s window due to technicalities related to how \bw{} handles its sine-Gaussian basis elements \cite{Cornish:2014kda, Littenberg:2014oda}.}
If the GR-based model provides a good description of the signal, we expect the resulting residuals at each detector to lack any significant coherent SNR 
beyond what is expected from noise fluctuations.
We compute the coherent SNR using \bw{}, which models the multi-detector residuals as a superposition of incoherent Gaussian noise and an elliptically-polarized coherent signal.
The residual network SNR reported by \bw{} is the SNR that would correspond to such a coherent signal.

In particular, for each event, \bw{} produces a distribution of possible residual signals consistent with the data, together with corresponding \textit{a posteriori} probabilities.
This is trivially translated into a probability distribution over the coherent residual SNR.
We summarize each of these distributions by computing the corresponding 90\%-credible upper limit ($\rhores$).
This produces one number per event that represents an upper bound on the coherent power that could be present in its residuals.

We may translate the $\rhores$ into a measure of how well the best-fit templates describe the signals in our data.
We do this through the fitting factor \cite{Apostolatos:1995}, $\mathrm{FF} \coloneqq \rhosig\, / (\rhodiff^2 + \rhosig^2)^{1/2}$, where $\rhodiff$ is the coherent residual SNR and $\rhosig$ is the network SNR of the best-fit template (see~Table~\ref{tab:events} for network SNRs).
By setting $\rhodiff=\rhores$, we produce a 90\%-credible lower limit on the fitting factor ($\ff$).
Because FF is itself a lower limit on the overlap between the true and best-fit templates, so is $\ff$.
As in \cite{GW150914:TGR}, we may then assert
that the disagreement between the true waveform and our GR-based template is at most $(1-\ff)\times100\%$. 
This is interesting as a measure of the sensitivity of our test, but does not tell us about the consistency of the residuals with instrumental noise.

\begin{table}
\caption{Results of the residuals analysis.
For each event, this table presents the 90\%-credible upper limit on the reconstructed network SNR after subtraction of the best-fit GR waveform ($\rhores$), a corresponding lower limit on the fitting factor ($\ff$ in the text), and the $\rhores$ $p$-value.
$\rhores$ is a measure of the maximum possible coherent signal power not captured by the best-fit GR template, while the $p$-value is an estimate of the probability that instrumental noise produced such $\rhores$ or higher. We also indicate which interferometers (IFOs) were used in the analysis of a given event, either the two Advanced LIGO detectors (HL) or the two Advanced LIGO detectors plus Advanced Virgo (HLV).
See Sec.~\ref{sec:residuals} in the main text for details.}
\label{tab:residuals}
\centering
\begin{tabular}{l@{\quad} c@{\quad} c@{\quad} c@{\quad} c@{\quad}}
\toprule
Event    & IFOs & Residual $\rhores$ & Fitting factor & $p$-value \\
\midrule                                              
GW150914 & HL   & 6.1               & $\geq0.97$     & 0.46             \\
GW151012 & HL   & 7.3               & $\geq0.79$     & 0.11             \\
GW151226 & HL   & 5.6               & $\geq0.91$     & 0.81             \\
\midrule
GW170104 & HL   & 5.1               & $\geq0.94$     & 0.99             \\
GW170608 & HL   & 7.9               & $\geq0.89$     & 0.05             \\
GW170729 & HLV  & 6.5               & $\geq0.85$     & 0.74             \\
GW170809 & HLV  & 6.5               & $\geq0.88$     & 0.78             \\
GW170814 & HLV  & 8.9               & $\geq0.88$     & 0.16             \\
GW170818 & HLV  & 9.2               & $\geq0.78$     & 0.19             \\
GW170823 & HL   & 5.5               & $\geq0.90$     & 0.86             \\
\bottomrule
\end{tabular}
\end{table}

To assess whether the obtained residual $\rhores$ values are consistent with detector noise, we run an identical \bw{} analysis on 200 different sets of noise-only detector data near each event.
This allows us to estimate the $p$-value for the null hypothesis that the residuals are consistent with noise.
The $p$-value gives the probability of noise producing coherent power with $\rhonoise$ greater than or equal to the residual value $\rhores$, i.e., $p \coloneqq P(\rhonoise \ge \rhores \mid \mathrm{noise})$.
In that sense, a smaller $p$-value indicates a smaller chance that the residual power arose from instrumental noise only.
For each event, our estimate of $p$ is produced from the fraction of noise instantiations that yielded $\rhonoise\ge \rhores$ (that is, from the empirical survival function).%
\footnote{Computing $p$-values would not be necessary if the noise was perfectly Gaussian, in which case we could predict the noise-only distribution of $\rhonoise$ from first principles.}

\begin{figure}[t] 
	\centering
	\includegraphics[width=\columnwidth]{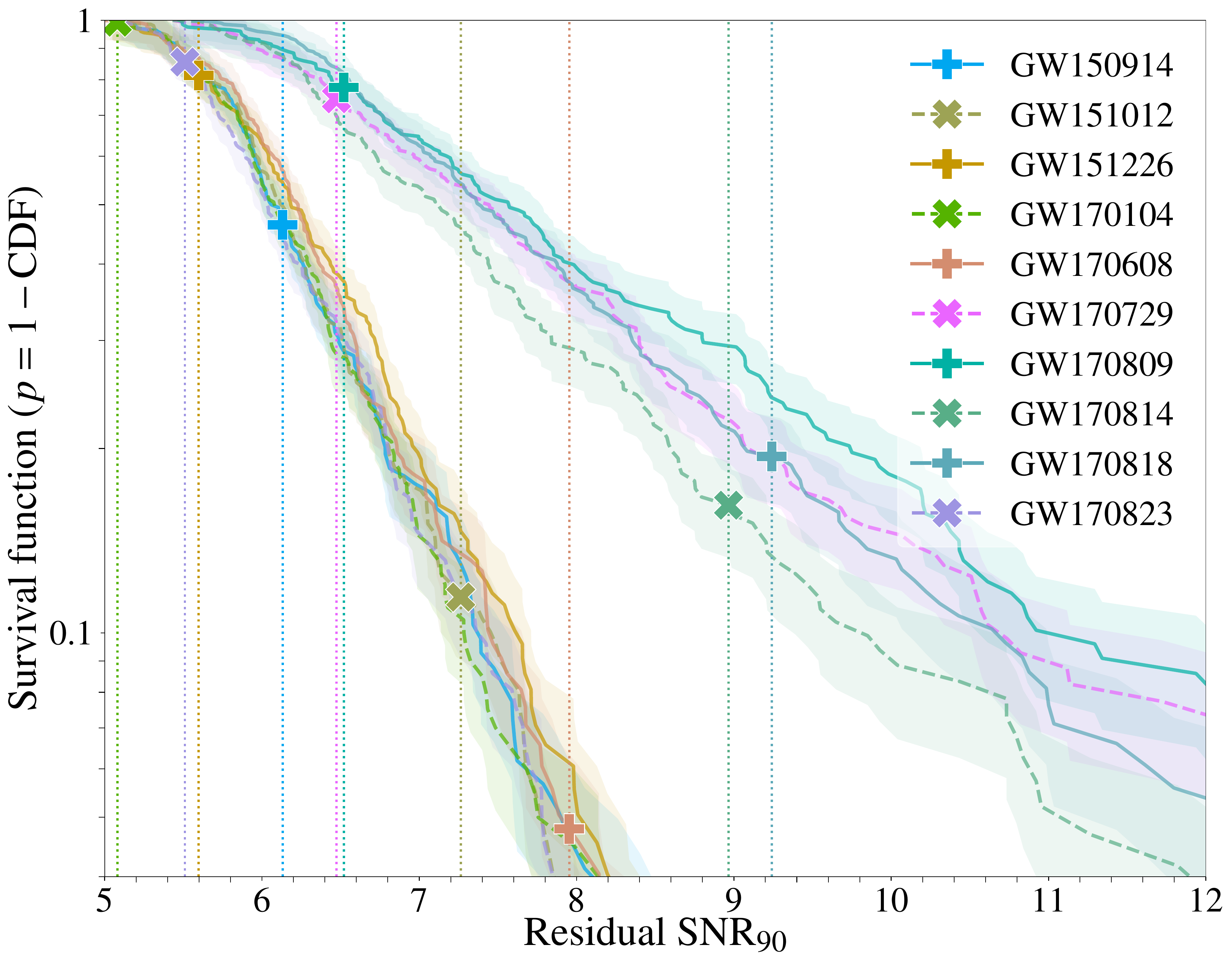}
	\caption{Survival function ($p = 1-\mathrm{CDF}$) of the 90\%-credible upper limit on the network SNR ($\rhores$) for each event ({solid or dashed} curves), compared to the measured residual values (vertical dotted lines).
For each event, the value of the survival function at the measured $\rhores$ gives the $p$-value reported in Table~\ref{tab:residuals} (markers). 
The colored bands correspond to uncertainty regions for a Poisson process and have half width $\pm p/\sqrt{N}$, with $N$ the number of noise-only instantiations that yielded $\rhonoise$ greater than the abscissa value.}
	\label{fig:residuals_snr}
\end{figure}

Our results are summarized in Table \ref{tab:residuals}.
For each event, we present the values of the residual $\rhores$, the lower limit on the fitting factor ($\ff$), and the $\rhores$ $p$-value.
The background distributions that resulted in those $p$-values are shown in Fig.~\ref{fig:residuals_snr}.
In Fig.~\ref{fig:residuals_snr} we represent these distributions through the empirical estimate of their survival functions, i.e., $p(\rhores) = 1-\mathrm{CDF}(\rhores)$, with ``CDF'' the cumulative distribution function.
Fig.~\ref{fig:residuals_snr} also displays the actual value of $\rhores$ measured from the residuals of each event (dotted vertical line).
In each case, the height of the curve evaluated at the $\rhores$ measured for the corresponding detection yields the $p$-value reported in Table \ref{tab:residuals} (markers in Fig.~\ref{fig:residuals_snr}).

The values of residual $\rhores$ vary widely among events because they depend on the specific state of the instruments at the time of detection: segments of data with elevated noise levels are more likely to result in spurious coherent residual power, even if the signal agrees with GR.
In particular, the background distributions for events seen by three detectors are qualitatively different from those seen by only two.
This is both due to 
\begin{inparaenum}[(i)]
\item the fact that \bw{} is configured to expect the SNR to increase with the number of detectors and
\item the fact that Virgo data present a higher rate of non-Gaussianities than LIGO.
\end{inparaenum}
We have confirmed both these factors play a role by studying the background $\rhores$ distributions for real data from each possible pair of detectors, as well as distributions over simulated Gaussian noise.
Specifically, removing Virgo from the analysis results in a reduction in the coherent residual power for GW170729 ($\rhores^{\rm HL}=6.5$), GW170809 ($\rhores^{\rm HL}=6.3$), GW170814 ($\rhores^{\rm HL}=6.0$), and GW170818 ($\rhores^{\rm HL}=7.2$).

The event-by-event variation of $\rhores$ is also reflected in the values of $\ff$.
GW150914 provides the strongest result with $\ff = {0.97}$, which corresponds
to an upper limit of {3\%} on the magnitude of potential deviations from our GR-based template,%
\footnote{This value is better than the one quoted in \cite{GW150914:TGR} by 1 percentage point. The small difference is explained by several factors, including that paper's use of the maximum {\it a posteriori} waveform (instead of maximum likelihood) and 95\% (instead of 90\%) credible intervals, as well as improvements in data calibration.}
in the specific sense defined in \cite{GW150914:TGR} and discussed above.
On the other hand, GW170818 yields the weakest result with $\ff = 0.78$
and a corresponding upper limit on waveform disagreement of $1-\ff=22\%$.
The average $\ff$ over all events is 0.88.

The set of $p$-values shown in Table \ref{tab:residuals} is consistent with all coherent residual power being due to instrumental noise.
Assuming that this is indeed the case, we expect the $p$-values to be uniformly distributed over $[0, 1]$, which explains the variation in Table \ref{tab:residuals}.
With only ten events, however, it is difficult to obtain strong quantitative evidence of the uniformity of this distribution.
Nevertheless, we follow Fisher's method \cite{Fisher1948} to compute a meta $p$-value for the null hypothesis that the individual $p$-values in Table \ref{tab:residuals} are uniformly distributed.
We obtain a meta $p$-value of 0.4, implying that there is no evidence for coherent residual power that cannot be explained by noise alone.
All in all, this means that there is no statistically significant evidence for deviations from GR.

\begin{figure}[tbh] 
	\begin{center}
	\includegraphics[width=3.5in]{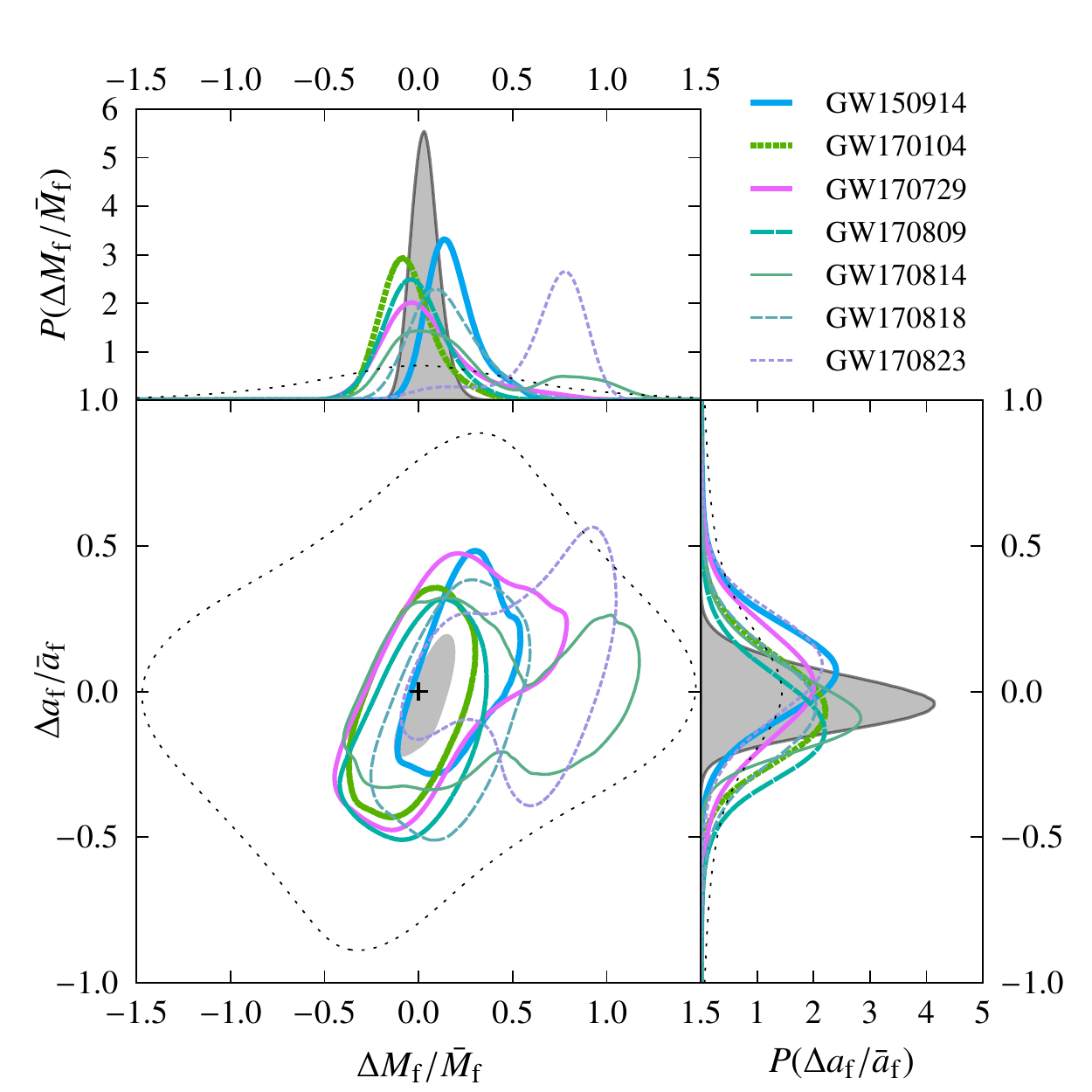}
	\end{center}
	\caption{Results of the inspiral-merger-ringdown consistency test for the selected BBH events (see Table~\ref{tab:events}). The main panel shows 90\% credible regions of the posterior distributions of $(\Delta M_\mathrm{f}/\bar{M}_\mathrm{f}, \Delta a_\mathrm{f}/\bar{a}_\mathrm{f})$, with the cross marking the expected value for GR. The side panels show the marginalized posteriors for $\Delta M_\mathrm{f}/\bar{M}_\mathrm{f}$ and $\Delta a_\mathrm{f}/\bar{a}_\mathrm{f}$. The thin black dashed curve represents the prior distribution, and the grey shaded areas correspond to the combined posteriors from the five most significant events (as outlined in Sec.~\ref{sec:events} and Table~\ref{tab:events}).}
	\label{fig:imr_test_posteriors}
\end{figure}

\begin{table}
\caption{Results from the inspiral-merger-ringdown consistency test for selected binary black hole events. $f_\text{c}$ denotes the cutoff frequency used to demarcate the division between the inspiral and post-inspiral regimes; $\rho_\mathrm{IMR}$, $\rho_\mathrm{insp}$, and $\rho_\mathrm{post-insp}$ are the median values of the SNR in the full signal, the inspiral part, and the post-inspiral part, respectively; and the GR quantile denotes the fraction of the posterior enclosed by the isoprobability contour that passes through the GR value, with smaller values indicating better consistency with GR.
(Note, however, that the posterior distribution will be broader for smaller SNRs, and hence the GR quantile will typically be smaller in such cases. This effect is further complicated by the randomness of the noise.)}
\label{tab:imr_test_params}
\centering
\begin{tabular}{c@{\quad} c@{\quad}c@{\quad}c@{\quad} c@{\quad} c}
\toprule
Event 		& $f_\text{c}$ [Hz]  & $\rho_\mathrm{IMR}$ & $\rho_\mathrm{insp}$ & $\rho_\mathrm{post-insp}$ & GR quantile [\%] \\
\midrule
GW150914	& 132	  & 25.3	& 19.4	& 16.1	& 55.5 \\	
GW170104	& 143	  & 13.7	& 10.9	& 8.5	& 24.4 \\	
GW170729	& 91	  & 10.7	& 8.6	& 6.9	& 10.4 \\	
GW170809	& 136	  & 12.7	& 10.6	& 7.1	& 14.7 \\
GW170814	& 161	  & 16.8	& 15.3	& 7.2	& 7.8 \\
GW170818	& 128	  & 12.0	& 9.3	& 7.2	& 25.5 \\
GW170823	& 102	  & 11.9	& 7.9	& 8.5	& 80.4 \\
\bottomrule
\end{tabular}
\end{table}

\subsection{Inspiral-merger-ringdown consistency test}
\label{sec:imr-test}

The inspiral-merger-ringdown consistency test for binary black 
holes~\cite{Ghosh:2016qgn,Ghosh:2017gfp} checks the consistency of the 
low-frequency part of the observed signal (roughly corresponding to
the inspiral of the black holes) with the high-frequency part 
(to a good approximation, produced by the post-inspiral stages).%
\footnote{Note that this is not exactly equal to testing the consistency between the early and 
late part of the waveform in time domain, because the low-frequency part of the signal could
 be ``contaminated'' by power from late times and vice versa. In practice, this effect is negligible with our 
choice of cutoff frequencies. See~\cite{Ghosh:2017gfp} for a discussion.}
The cutoff frequency $f_\text{c}$ between 
the two regimes is chosen as the frequency of the innermost stable 
circular orbit of a Kerr black hole~\cite{Bardeen:1972fi}, with mass and 
dimensionless spin computed by applying NR fits~\cite{Healy:2016lce, 
2041-8205-825-2-L19, PhysRevD.95.064024,spinfit-T1600168} to the median 
values of the posterior distributions of the initial masses and spherical coordinate
components of the spins.
This determination of $f_\text{c}$ is performed separately for each event and 
based on parameter inference of the full signal (see 
Table~\ref{tab:imr_test_params} for values of $f_\text{c}$).\footnote{
The frequency $f_\text{c}$ was determined using preliminary parameter 
inference results, so the values in Table~\ref{tab:imr_test_params} are 
slightly different than those that would be obtained using the posterior 
samples in GWTC-1~\cite{GWOSC:GWTC}. However, the test is robust against 
small changes in the cutoff frequency~\cite{Ghosh:2017gfp}.}
The binary's parameters are then estimated independently from 
the low (high) frequency parts of the data by restricting the noise-weighted 
integral in the likelihood calculation to frequencies below (above) 
this frequency cutoff $f_\text{c}$. 
For each of these independent estimates of the source parameters, 
we make use of fits to numerical-relativity simulations given 
in~\cite{Healy:2016lce, 2041-8205-825-2-L19, PhysRevD.95.064024} 
to infer the mass $M_\mathrm{f}$ and dimensionless spin magnitude
$a_\mathrm{f} = c |\vec S_\mathrm{f}| / (G M_\mathrm{f}^2)$ of the remnant black hole.%
\footnote{As in~\cite{GW170104}, we average the $M_\mathrm{f}, a_\mathrm{f}$ posteriors obtained by different fits~\cite{Healy:2016lce, 2041-8205-825-2-L19, PhysRevD.95.064024} after augmenting the fitting formulae for aligned-spin binaries by adding the contribution from in-plane spins~\cite{spinfit-T1600168}. However, unlike in~\cite{GW170104,spinfit-T1600168}, we do not evolve the spins before applying the fits, due to technical reasons.}
If the data are consistent with GR, these two independent estimates have to be consistent with 
each other~\cite{Ghosh:2016qgn,Ghosh:2017gfp}. 
Because this consistency test ultimately compares between the inspiral 
and the post-inspiral results, posteriors of both parts must be informative. 
In the case of low-mass binaries, the SNR in the part $f>f_\text{c}$ is insufficient 
to perform this test, so that we only analyze seven events as indicated in 
Tables~\ref{tab:events} and~\ref{tab:imr_test_params}.

In order to quantify the consistency of the two different estimates of the final black hole's mass and spin we define two dimensionless quantities that quantify the fractional difference between them: 
$\Delta M_\mathrm{f}/\bar{M}_\mathrm{f} \coloneqq 2\, (M_\mathrm{f}^{\text{insp}} - M_\mathrm{f}^{\text{post-insp}})/(M_\mathrm{f}^{\text{insp}} + M_\mathrm{f}^{\text{post-insp}})$ and $\Delta a_\mathrm{f}/\bar{a}_\mathrm{f} \coloneqq 2\, (a_\mathrm{f}^{\text{insp}} - a_\mathrm{f}^{\text{post-insp}})/(a_\mathrm{f}^{\text{insp}} + a_\mathrm{f}^{\text{post-insp}})$, where the superscripts indicate the estimates of the mass and spin from the inspiral and post-inspiral parts of the signal.%
\footnote{For black hole binaries with comparable masses and moderate spins, as we consider here, the remnant black hole is expected to have $a_\mathrm{f} \gtrsim 0.5$; see, e.g., \cite{Healy:2016lce, 2041-8205-825-2-L19, PhysRevD.95.064024} for fitting formulae derived from numerical simulations, or Table~\ref{tab:events} for values of the remnant's spins obtained from GW events. Hence, $\Delta a_\mathrm{f}/\bar{a}_\mathrm{f}$ is expected to yield finite values.}
The posteriors of these dimensionless parameters, estimated from different events, are shown in Fig.~\ref{fig:imr_test_posteriors}. For all events, the posteriors are consistent with the GR value ($\Delta M_\mathrm{f}/\bar{M}_\mathrm{f} = 0, \Delta a_\mathrm{f}/\bar{a}_\mathrm{f} = 0)$. The fraction of the posterior enclosed by the isoprobability contour that passes through the GR value (i.e., the GR quantile) for each event is shown in Table~\ref{tab:imr_test_params}. Figure~\ref{fig:imr_test_posteriors} also shows the posteriors obtained by combining all the events that pass the stronger significance threshold $\mathrm{FAR} < (1000\, \mathrm{yr})^{-1}$, as outlined in Sec.~\ref{sec:events} (see the same section for a discussion of caveats).

The parameter estimation is performed employing uniform priors in component masses and spin magnitudes and isotropic priors in spin directions~\cite{O2:Catalog}. This will introduce a non-flat prior in the deviation parameters $\Delta M_\mathrm{f}/\bar{M}_\mathrm{f}$ and $\Delta a_\mathrm{f}/\bar{a}_\mathrm{f}$, which is shown as a thin, dashed contour in Fig.~\ref{fig:imr_test_posteriors}. Posteriors are estimated employing the precessing spin phenomenological waveform family \IMRP{}. To assess the systematic errors due to imperfect waveform modeling, we also estimate the posteriors using the effective-one-body based waveform family \SEOB{} that models binary black holes with non-precessing spins. There is no qualitative difference between the results derived using the two different waveforms families (see Sec.~\ref{appendix:imr_test} of the Appendix).

We see additional peaks in the posteriors estimated from GW170814 and GW170823. Detailed follow-up investigations did not show any evidence of the presence of a coherent signal in multiple detectors that differs from the GR prediction. The second peak in GW170814 is introduced by the posterior of $M_\mathrm{f}^{\text{post-insp}}$, while the extra peak in GW170823 is introduced by the posterior of $M_\mathrm{f}^{\text{insp}}$.
Injection studies in real data around the time of these events, using 
simulated GR waveforms with parameters consistent with GW170814 and GW170823, 
suggest that such secondary peaks occur for $\sim 10\%$ of injections. 
Features in the posteriors of GW170814 and GW170823 are thus consistent with expected noise fluctuations.

\section{Parameterized tests of gravitational wave generation}
\label{sec:gwnature}

A deviation from GR could manifest itself as
a modification of the dynamics of two orbiting compact objects, and in
particular, the evolution of the orbital (and hence, GW) phase. 
In an analytical waveform model like {\IMRP}, the details of the GW phase evolution
are controlled by coefficients that are either 
analytically calculated or determined by fits to numerical-relativity (NR) simulations, 
always under the assumption that GR is the underlying theory.
In this section we investigate deviations from the GR binary dynamics by 
introducing shifts in each of the individual GW phase coefficients of {\IMRP}.
Such shifts correspond to deviations in the waveforms from the predictions of GR.
We then treat these shifts as additional unconstrained GR-violating parameters, which we
measure in addition to the standard parameters describing the binary.

The early inspiral of compact binaries is well modeled by the post-Newtonian (PN) 
approximation~\cite{Blanchet:1995ez,Blanchet:2004ek,Blanchet:2005tk,Blanchet:2013haa} to GR, which is based 
on the expansion of the orbital quantities in terms of a small velocity parameter $v/c$. 
For a given set of intrinsic parameters, coefficients for the different 
orders in $v/c$ in the PN series are uniquely determined. 
A consistency test of GR using measurements of the inspiral 
PN phase coefficients was first proposed in~\cite{Arun:2006hn,Arun:2006yw,Mishra:2010tp},
while a generalized parametrization was motivated in~\cite{Yunes:2009ke}. 
Bayesian implementations based on such parametrized methods were
presented and tested in~\cite{LiEtAl:2012a,LiEtAl:2012b,Agathos:2013upa,Cornish:2011ys} and 
were also extended to the post-inspiral part of the gravitational-wave signal~\cite{Sampson:2013jpa,Meidam:2017dgf}. 
These ideas were applied to the first GW observation, GW150914~\cite{GW150914_paper}, 
yielding the first bounds on higher-order PN coefficients~\cite{GW150914:TGR}.
Since then, the constraints have been revised with the binary black 
hole events that followed, GW151226 in O1~\cite{O1:BBH} and GW170104 in O2~\cite{GW170104}.
More recently, the first such constraints from a binary neutron star merger were placed with the 
detection of GW170817~\cite{bns-tgr}.
Bounds on parametrized violations of GR from GW detections have been mapped, to leading order, to
constraints on specific alternative theories of gravity (see, e.g.,~\cite{Yunes:2016jcc}).
In this paper, we present individual constraints on parametrized deviations from 
GR for each of the GW sources in O1 and O2 listed in Table~\ref{tab:events}, 
as well as the tightest combined constraints obtained to date by combining information 
from all the significant binary black holes events observed so far,
as described in Sec.~\ref{sec:events}.

The frequency-domain GW phase evolution $\Phi(f)$ in the early-inspiral stage of {\IMRP} is described by a PN expansion, augmented with higher-order phenomenological coefficients. 
The PN phase evolution is analytically expressed in closed form by employing the stationary phase approximation.
The late-inspiral and post-inspiral (intermediate and merger-ringdown) 
stages are described by phenomenological analytical expressions.
The transition frequency\footnote{This frequency is different than the cutoff frequency used in the inspiral-merger-ringdown 
consistency test, as was briefly mentioned in Sec.~\ref{sec:events}.} from inspiral to intermediate regime is given by the condition
$G M f/c^3 = 0.018$, with $M$ the total mass of the binary in the detector frame, since this is the lowest
frequency above which this model was calibrated with NR data~\cite{Khan:2015jqa}. 
Let us use $p_i$ to collectively denote all of the inspiral and post-inspiral parameters $\varphi_i$, $\alpha_i$, $\beta_i$, that will be introduced below.
Deviations from GR in all stages are expressed by means of relative shifts $\delta\hat{p}_i$ in the corresponding waveform coefficients: $p_i \rightarrow (1+\delta\hat p_i) \ p_i $, which are used as additional free parameters in our extended waveform models.

We denote the testing parameters corresponding to PN phase coefficients by $\delta\hat\varphi_i$, 
where $i$ indicates the power of $v/c$ beyond leading (Newtonian or $0$PN) order in $\Phi(f)$.
The frequency dependence of the corresponding phase term is $f^{(i-5)/3}$.
In the parametrized model, $i$ varies from 0 to 7, including the terms with logarithmic dependence 
at $2.5$PN and $3$PN. The non-logarithmic term at $2.5$PN (i.e., $i = 5$) cannot be constrained, because of its degeneracy
with a constant reference phase (e.g., the phase at coalescence).
These coefficients were introduced in their current form in Eq.~(19) of~\cite{LiEtAl:2012a}.
In addition, we also test for $i=-2$, representing an effective $-1$PN term, which is motivated below.
The full set of inspiral parameters are thus
\[ \{ \delta\hat\varphi_{-2}, \delta\hat\varphi_0, \delta\hat\varphi_1, 
\delta\hat\varphi_2, \delta\hat\varphi_3, \delta\hat\varphi_4, \delta\hat\varphi_{5l}, \delta\hat\varphi_6, 
\delta\hat\varphi_{6l}, \delta\hat\varphi_7\}. \]
Since the $-1$PN term and the $0.5$PN term are absent in the GR phasing,
we parametrize $\delta\hat\varphi_{-2}$ and $\delta\hat\varphi_1$ as \emph{absolute} deviations, with a pre-factor equal to the $0$PN coefficient.

\begin{figure*}[tbh!]
\includegraphics[keepaspectratio, width=\textwidth]{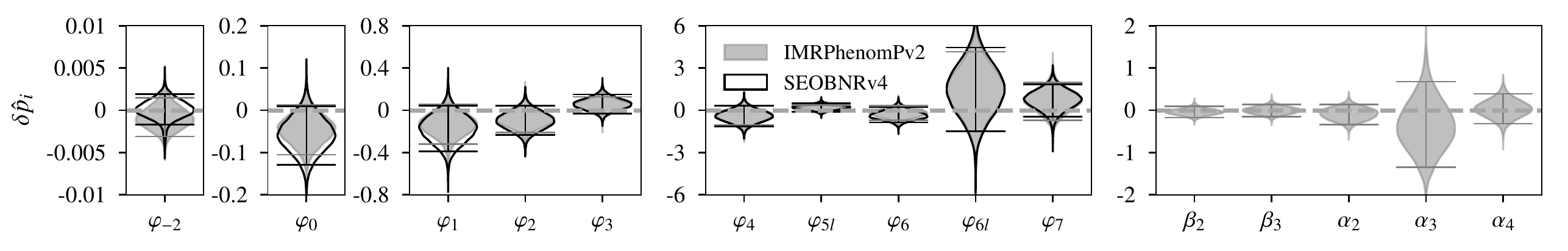}
\caption{Combined posteriors for parametrized violations of GR, obtained from all events in Table~\ref{tab:events} with a significance of $\mathrm{FAR < (1000~yr)}^{-1}$ in both modeled searches.
The horizontal lines indicate the 90\% credible intervals, and the dashed horizontal line at zero corresponds to the expected GR values.
  The combined posteriors on $\varphi_i$ in the inspiral regime are obtained from the events which in addition exceed the SNR threshold in the inspiral regime (GW150914, GW151226, GW170104, GW170608, and GW170814), analyzed with {\IMRP} (grey shaded region) and {\SEOB} (black outline).
  The combined posteriors on the intermediate and merger-ringdown parameters $\beta_i$ and $\alpha_i$ are obtained from events which exceed the SNR threshold in the post-inspiral regime (GW150914, GW170104, GW170608, GW170809, GW170814, and GW170823), analyzed with {\IMRP}.}
\label{fig:violin_plot}
\end{figure*}

The $-1$PN term of $\delta\hat{\varphi}_{2}$ can be interpreted as arising
from the emission of dipolar radiation. For binary black holes, this could occur in, e.g., alternative theories of gravity where
an additional scalar charge is sourced by terms related to curvature~\cite{Yagi:2015oca,Yagi:2016jml}.
At leading order, this introduces a deviation in the  $-1$PN coefficient of the 
waveform~\cite{Barausse:2016eii, Arun:2012hf}. 
This effectively introduces a term in the inspiral GW 
phase, varying with frequency as $f^{-7/3}$,
while the gravitational flux is modified 
as $\mathcal{F_\mathrm{GR}}\rightarrow\mathcal{F_\mathrm{GR}}(1+Bc^2/v^2)$.
The first bound on $\delta\hat\varphi_{-2}$ was published in~\cite{bns-tgr}.
The higher-order terms in the above expansion also lead to a
modification in the higher-order PN coefficients.
Unlike the case of GW170817 (which we study separately in~\cite{bns-tgr}),
where the higher-order terms in the expansion of the flux are negligible,
the contribution of higher-order terms can be significant 
in the binary black-hole signals that we study here.
This prohibits an exact interpretation of the $-1$PN term as the strength of dipolar radiation.  
Hence, this analysis only serves as a test of the presence of an effective $-1$PN term in the inspiral phasing, 
which is absent in GR. 

To measure the above GR violations in the post-Newtonian inspiral, we employ 
two waveform models: (i) the analytical frequency-domain model {\IMRP} 
which also provided the natural parametrization 
for our tests and (ii) \SEOB{}, which we use in the form of \textsc{SEOBNRv4\_ROM}, a frequency-domain,
reduced-order-model of the {\SEOB} model.
The inspiral part of {\SEOB} is based on a numerical evolution of
the aligned-spin effective-one-body dynamics of the binary and its
post-inspiral model is phenomenological. The entire {\SEOB} model
is calibrated against NR simulations.
Despite its non-analytical nature, \textsc{SEOBNRv4\_ROM} can also be used to test the 
parametrized modifications of the early inspiral defined above.
Using the method presented in~\cite{bns-tgr}, we add deviations to the waveform phase 
corresponding to a given $\delta\hat\varphi_i$ at low frequencies and then taper 
the corrections to zero at a frequency consistent with the transition frequency 
between early-inspiral and intermediate phases used by \mbox{\textsc{IMRPhenomPv2}}.
The same procedure cannot be applied to the later stages of the waveform, thus the 
analysis performed with {\SEOB} is restricted to the post-Newtonian inspiral, cf. Fig.~\ref{fig:violin_plot}.

The analytical descriptions of the intermediate and merger-ringdown stages in the {\IMRP} model allow for a 
straightforward way of parametrizing deviations from GR,
denoted by $\{ \delta\hat\beta_2, \delta\hat\beta_3 \}$ and 
$\{ \delta\hat\alpha_2, \delta\hat\alpha_3, \delta\hat\alpha_4 \}$ respectively, following~\cite{Meidam:2017dgf}.
Here the parameters $\delta\hat\beta_i$ correspond to deviations from the NR-calibrated phenomenological coefficients
$\beta_i$ of the intermediate stage, while the parameters $\delta\hat\alpha_i$ refer to modifications of the
merger-ringdown coefficients $\alpha_i$ obtained from a combination of phenomenological fits and analytical black-hole
perturbation theory calculations~\cite{Khan:2015jqa}. 

Using \textsc{LALInference}, we calculate posterior distributions of the
parameters characterizing the waveform (including those that 
describe the binary in GR). 
Our parametrization recovers GR at $\delta\hat{p}_i = 0$, so 
consistency with GR is verified if the posteriors of 
$\delta\hat p_i$ have support at zero. 
We perform the analyses by varying one $\delta\hat p_i$ at a time; 
as shown in Ref.~\cite{Sampson:2013lpa}, this is fully robust to detecting 
deviations present in multiple PN-orders. In addition, allowing for a larger parameter 
space by varying multiple coefficients simultaneously would not improve our efficiency in identifying
violations of GR, as it would yield less informative posteriors.
A specific alternative theory of gravity would likely yield correlated deviations 
in many parameters, including modifications that we have not considered here.
This would be the target of an exact comparison of an alternative theory with GR, which would only be possible if
a complete, accurate description of the inspiral-merger-ringdown signal in that theory was available.

We use priors uniform on $\delta\hat p_i$ and symmetric around zero.
Figure~\ref{fig:violin_plot} shows the combined posteriors 
of $\delta\hat p_i$ (marginalized over all other parameters) 
estimated from the combination of all the events that cross the significance 
threshold of $\mathrm{FAR < (1000~yr)}^{-1}$ in both modeled searches; see Table~\ref{tab:events}. 
Events with SNR$<6$ in the inspiral regime (parameters $\delta\hat\varphi_i$) or 
in the post-inspiral regime ($\delta\hat\beta_i$ and $\delta\hat\alpha_i$
for the intermediate and merger-ringdown parameters respectively) are not 
included in the results, since the
data from those instances failed to provide useful constraints
(see Sec.~\ref{sec:events} for more details).
This SNR threshold, however, is not equally effective in ensuring informative results for all cases;
see Sec.~\ref{app:generation} in the Appendix for a detailed discussion. 
In all cases considered, the posteriors are consistent with $\delta\hat p_i = 0$ within statistical fluctuations.
Bounds on the inspiral coefficients obtained with the two different waveform models are found to be in good agreement with each other.
Finally, we note that the event-combining analyses on $\delta\hat{p}_i$ assume that these parametrized violations are constant across all events considered.
This assumption should not be made when testing a specific theory that predicts violations that depend on the binary's parameters.
Posterior distributions of $\delta\hat p_i$ for the individual-event analysis, also showing full consistency with GR, are provided in
Sec.~\ref{app:generation} of the Appendix.

\begin{figure}[tb]
\includegraphics[keepaspectratio, width=0.47\textwidth]{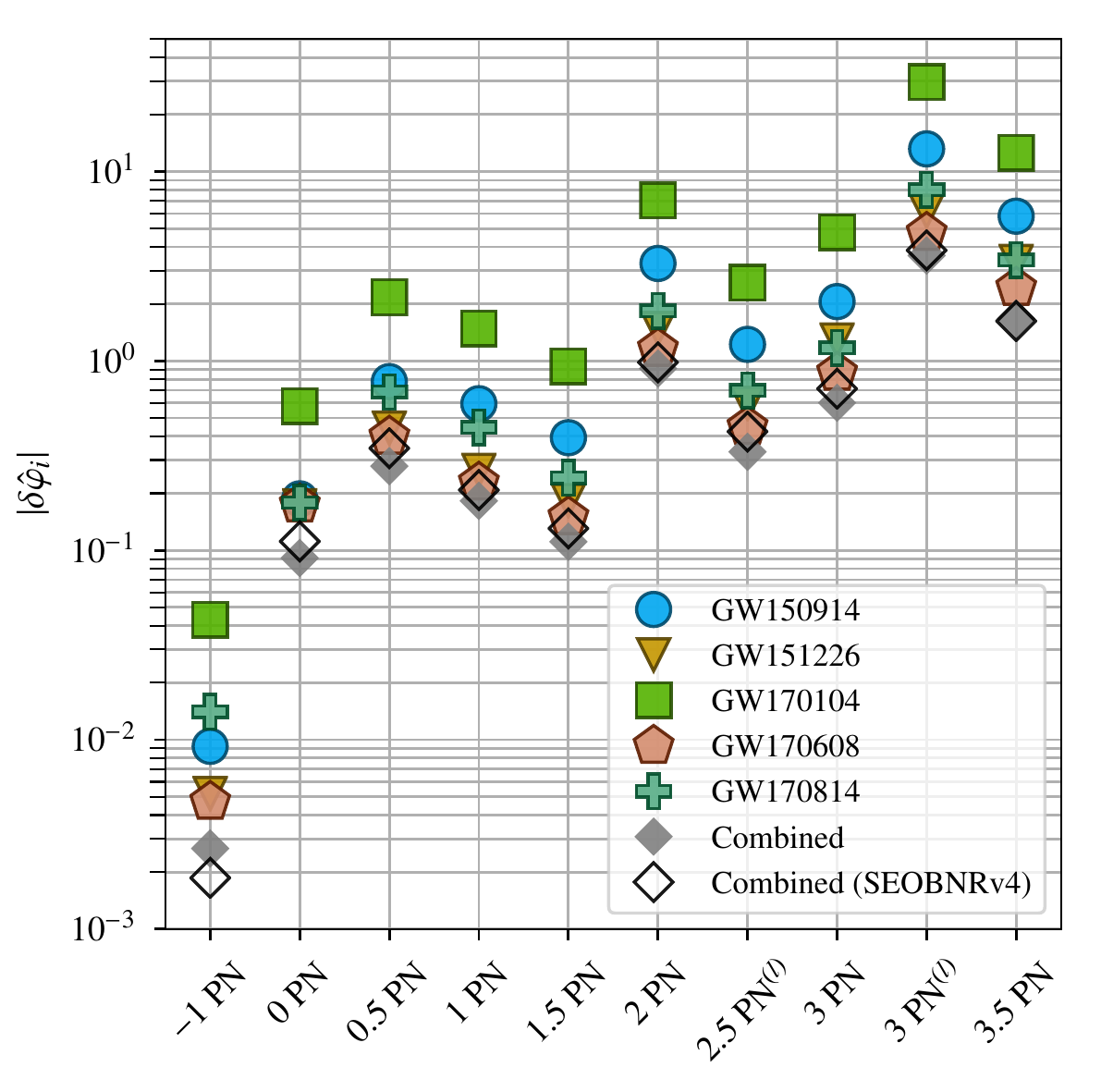}
\caption{90\% upper bounds on the absolute magnitude of the GR-violating parameters $\delta\hat\varphi_n$, from $-1$PN through 3.5PN in the inspiral phase.
  At each PN order, we show results obtained from each of the events listed in Table~\ref{tab:events} that cross the SNR threshold in the inspiral regime, analyzed with {\IMRP}.
  Bounds obtained from combining posteriors of events detected with a significance that exceeds a threshold of $\mathrm{FAR < (1000~yr)}^{-1}$ in both modelled searches are shown for both analyses, using {\IMRP} (filled diamonds) and {\SEOB} (empty diamonds).}
\label{fig:pn_bounds}
\end{figure}

Figure~\ref{fig:pn_bounds} shows the $90\%$ upper bounds on $|\delta\hat\varphi_i|$ for all the individual events 
which cross the SNR threshold (SNR $>$ 6) in the inspiral regime (the most massive of which is GW150914). 
The bounds from the combined posteriors are also shown; these include the events which exceed both the 
SNR threshold in the inspiral regime as well as the significance threshold, namely GW150914, GW151226, GW170104, GW170608, 
and GW170814.
The bound from the likely lightest mass binary black hole event GW170608 at 1.5PN is currently the strongest constraint obtained 
on a positive PN coefficient from a single binary black hole event, as shown in Fig.~\ref{fig:pn_bounds}.
However, the constraint at this order is about five times worse than that obtained 
from the binary neutron star event GW170817 alone~\cite{bns-tgr}.
The $-1$PN bound is two orders of magnitude better for GW170817 than the best bound obtained here (from GW170608). The corresponding best $-1$PN bound coming from the double pulsar PSR J0737$-$3039, is a few orders of magnitude tighter still, at $|\delta\hat\varphi_{-2}| \lesssim 10^{-7}$~\cite{Barausse:2016eii,Kramer:2006nb}.
At $0$PN we find that the bound from GW170608 beats the one from GW170817, but remains weaker than the one from the double pulsar by one order of magnitude~\cite{YunesHughes2010,Kramer:2006nb}.
For all other PN orders, GW170608 also provides the best bounds, which at high PN orders are 
of the same order of magnitude as the ones from GW170817.
Our results can be compared statistically to those obtained by performing the same tests on simulated GR and non-GR
waveforms given in~\cite{Meidam:2017dgf}.
The results presented here are consistent with those of GR waveforms injected into realistic detector data.
The combined bounds are the tightest obtained so far, improving on the bounds obtained in~\cite{O1:BBH} by
factors between 1.1 and 1.8.

\section{Parameterized tests of gravitational wave propagation}
\label{sec:propagation}
We now place constraints on a phenomenological modification of the GW dispersion relation, i.e., on a possible frequency dependence of the speed of GWs. This modification, introduced in~\cite{Mirshekari:2011yq} and first applied to LIGO data in~\cite{GW170104}, is obtained by adding a power-law term in the momentum to the dispersion relation $E^2 = p^2c^2$ of GWs in GR, giving
\begin{equation}
\label{eq:dispersion}
E^2 = p^2c^2 + A_\alpha p^\alpha c^\alpha.
\end{equation}
Here, $c$ is the speed of light, $E$ and $p$ are the energy and momentum of the GWs, 
and $A_\alpha$ and $\alpha$ are phenomenological parameters. 
We consider $\alpha$ values from $0$ to $4$ in steps of $0.5$. However,
we exclude $\alpha = 2$, where the speed of the GWs is modified in a frequency-independent manner, and therefore gives no observable dephasing.\footnote{For a source with
an electromagnetic counterpart, $A_2$ can be constrained by comparison with the arrival time 
of the photons, as was done with GW170817/GRB170817A~\cite{Monitor:2017mdv}.} 
Thus, in all cases except for $\alpha = 0$, we are considering a Lorentz-violating dispersion relation.
The group velocity associated with this dispersion relation is $v_g/c = (dE/dp)/c = 1 + (\alpha - 1)A_\alpha E^{\alpha - 2}/2 + O(A_\alpha^2)$.
The associated length scale is $\lambda_A \coloneqq hc |A_\alpha|^{1/(\alpha - 2)}$, where $h$ is Planck's constant. $\lambda_A$ gives the scale of modifications to the Newtonian potential (the Yukawa potential for $\alpha = 0$) associated with this dispersion relation.

While Eq.~\eqref{eq:dispersion} is a purely phenomenological model, it encompasses a variety of more fundamental predictions (at least to leading order)~\cite{Mirshekari:2011yq,Yunes:2016jcc}. In particular, $A_0 > 0$ corresponds to a massive graviton, i.e., the same dispersion as for a massive particle in vacuo~\cite{Will:1997bb}, with a graviton mass given by $m_g = A_0^{1/2}/c^2$.\footnote{Thus, the Yukawa screening length is $\lambda_0 = h/(m_g c)$.} Furthermore, $\alpha$ values of $2.5$, $3$, and $4$ correspond to the leading predictions of multi-fractal spacetime~\cite{Calcagni:2009kc}; doubly special relativity~\cite{AmelinoCamelia:2002wr}; and Ho{\v{r}}ava-Lifshitz~\cite{Horava:2009uw} and extra dimensional~\cite{Sefiedgar:2010we} theories, respectively. The standard model extension also gives a leading contribution with $\alpha = 4$~\cite{Kostelecky:2016kfm}, only considering the non-birefringent terms; our analysis does not allow for birefringence.

In order to obtain a waveform model with which to constrain these propagation effects, we start by assuming that the waveform extracted in the binary's local wave zone (i.e., near to the binary compared to the distance from the binary to Earth, but far from the binary compared to its own size) is well-described by a waveform in GR.\footnote{This is likely to be a good assumption for $\alpha < 2$, where we constrain $\lambda_A$ to be much larger than the size of the binary. For $\alpha > 2$, where we constrain $\lambda_A$ to be much \emph{smaller} than the size of the binary, one has to posit a screening mechanism in order to be able to assume that the waveform in the binary's local wave zone is well-described by GR, as well as for this model to evade Solar System constraints.} Since we are able to bound these propagation effects to be very small, we can work to linear order in $A_\alpha$ when computing the effects of this dispersion on the frequency-domain GW phasing,\footnote{The dimensionless parameter controlling the size of the linear correction is $A_\alpha f^{\alpha - 2}$, which is $\lesssim 10^{-18}$ at the 90\% credible level for the events we consider and frequencies up to $1$~kHz.} thus obtaining a correction~\cite{Mirshekari:2011yq} that is added to $\Phi(f)$ in Eq.~\eqref{eq:h_tilde}:
\begin{equation} \label{eq:phase_corr}
\delta\Phi_\alpha(f) = \sign(A_\alpha) ~ \begin{dcases}
\frac{\pi D_\text{L}}{\alpha - 1}\lambda_{A, \text{eff}}^{\alpha - 2}\left(\frac{f}{c}\right)^{\alpha - 1}, & \alpha\neq 1\\
\frac{\pi D_\text{L}}{\lambda_{A, \text{eff}}}\ln\left(\frac{\pi G\mathcal{M}^\text{det}f}{c^3}\right), & \alpha = 1
\end{dcases}.
\end{equation}
Here, $D_\text{L}$ is the binary's luminosity distance, $\mathcal{M}^\text{det}$ is the binary's detector-frame (i.e., redshifted) chirp mass, and $\lambda_{A, \text{eff}}$ is the effective wavelength parameter used in the sampling, defined as
\begin{equation}
\lambda_{A, \text{eff}} \coloneqq \left[\frac{(1 + z)^{1-\alpha}D_\text{L}}{D_\alpha}\right]^{1/(\alpha - 2)}\lambda_A \, .
\end{equation}
The parameter $z$ is the binary's redshift, and $D_\alpha$ is a distance parameter given by
\begin{equation}
D_\alpha = \frac{(1 + z)^{1-\alpha}}{H_0}\int_0^z\frac{(1 + \bar{z})^{\alpha - 2}}{\sqrt{\Omega_\text{m}(1 + \bar{z})^3 + \Omega_\Lambda}}d\bar{z} \, ,
\end{equation}
where $H_0 = 67.90 \text{ km s}^{-1} \text{ Mpc}^{-1}$ is the Hubble constant, and $\Omega_\text{m} = 0.3065$ and $\Omega_\Lambda = 0.6935$ are the matter and dark energy density parameters; these are the TT+lowP+lensing+ext values from~\cite{Ade:2015xua}.\footnote{We use these values for consistency with the results presented in~\cite{O2:Catalog}. If we instead use the more recent results from~\cite{Aghanim:2018eyx}, specifically the TT,TE,EE+lowE+lensing+BAO values used for comparison in~\cite{O2:Catalog}, then there are very minor changes to the results presented in this section. For instance, the upper bounds in Table~\ref{tab:prop_results_summary} change by at most $\sim 0.1\%$.}

The dephasing in Eq.~\eqref{eq:phase_corr} is obtained by treating the gravitational wave as a stream of particles (gravitons), which travel at the particle velocity $v_p/c = pc/E = 1 - A_\alpha E^{\alpha - 2}/2 + O(A_\alpha^2)$. There are suggestions to use the particle velocity when considering doubly special relativity, though there are also suggestions to use the group velocity $v_g$ in that case (see, e.g.,~\cite{AmelinoCamelia:2005ik} and references therein for both arguments). However, the group velocity is appropriate for, e.g., multi-fractal spacetime theories (see, e.g.,~\cite{Calcagni:2016zqv}).
To convert the bounds presented here to the case where the particles travel at the group velocity, scale the $A_\alpha$ bounds for $\alpha\neq 1$ by factors of $1/(1 - \alpha)$.
The group velocity calculation gives an unobservable constant phase shift for $\alpha = 1$.

\begin{table*}
\caption{\label{tab:prop_results_summary}
90\% credible level upper bounds on the graviton mass $m_g$ and the absolute value of the modified dispersion relation parameter $A_\alpha$, as well as the GR quantiles $Q_\text{GR}$. The $<$ and $>$ labels denote the bounds for $A_\alpha < 0$ and $>0$, respectively,
and we have defined the dimensionless quantity $\bar{A}_\alpha \coloneqq A_\alpha/\text{eV}^{2-\alpha}$. Events with names in boldface are used to obtain the combined results.
}
\centering
\scalebox{0.93}{
\begin{tabular}{ccc*{7}{cccc}ccc}
\toprule
& $m_g$ & & \threec{$|\bar{A}_0|$} & & \threec{$|\bar{A}_{0.5}|$} & & \threec{$|\bar{A}_1|$} & & \threec{$|\bar{A}_{1.5}|$} & & \threec{$|\bar{A}_{2.5}|$} & & \threec{$|\bar{A}_3|$} & & \threec{$|\bar{A}_{3.5}|$} & & \threec{$|\bar{A}_4|$}\\
\cline{4-6}
\cline{8-10}
\cline{12-14}
\cline{16-18}
\cline{20-22}
\cline{24-26}
\cline{28-30}
\cline{32-34}
Event & [$10^{-23}$ & & $<$ & $>$ & $Q_\text{GR}$ & & $<$ & $>$ & $Q_\text{GR}$ & & $<$ & $>$ & $Q_\text{GR}$ & & $<$ & $>$ & $Q_\text{GR}$ & & $<$ & $>$ & $Q_\text{GR}$ & & $<$ & $>$ & $Q_\text{GR}$ & & $<$ & $>$ & $Q_\text{GR}$ & & $<$ & $>$ & $Q_\text{GR}$\\
&  eV$/c^2$] & & \twoc{[$10^{-44}$]} & [\%] & & \twoc{[$10^{-38}$]} & [\%] & & \twoc{[$10^{-32}$]} & [\%] & & \twoc{[$10^{-25}$]} & [\%] & & \twoc{[$10^{-13}$]} & [\%] & & \twoc{[$10^{-8}$]} & [\%] & & \twoc{[$10^{-2}$]} & [\%] & & \twoc{[$10^{4}$]} & [\%]\\
\midrule
\textbf{GW150914} & $9.9$ & & $1.4$ & $1.1$ & $71$ & & $5.9$ & $4.9$ & $57$ & & $5.5$ & $3.5$ & $74$ & & $3.2$ & $2.1$ & $74$ & & $2.4$ & $2.1$ & $50$ & & $17$ & $19$ & $37$ & & $11$ & $20$ & $42$ & & $7.7$ & $9.8$ & $52$\\
GW151012 & $17$ & & $3.8$ & $3.5$ & $40$ & & $3.6$ & $11$ & $35$ & & $6.6$ & $9.5$ & $41$ & & $1.9$ & $2.5$ & $46$ & & $2.8$ & $1.5$ & $56$ & & $21$ & $9.7$ & $56$ & & $18$ & $15$ & $61$ & & $14$ & $6.7$ & $58$\\
\textbf{GW151226} & $29$ & & $7.1$ & $9.3$ & $21$ & & $8.1$ & $21$ & $10$ & & $13$ & $13$ & $26$ & & $3.4$ & $3.3$ & $28$ & & $3.4$ & $2.1$ & $68$ & & $22$ & $9.7$ & $61$ & & $14$ & $6.1$ & $58$ & & $19$ & $5.4$ & $72$\\
\midrule
\textbf{GW170104} & $9.2$ & & $2.5$ & $0.99$ & $62$ & & $4.2$ & $2.4$ & $72$ & & $7.0$ & $2.9$ & $76$ & & $1.9$ & $0.80$ & $83$ & & $1.1$ & $2.8$ & $23$ & & $10$ & $13$ & $35$ & & $6.4$ & $10$ & $37$ & & $6.5$ & $8.7$ & $43$\\
\textbf{GW170608} & $30$ & & $13$ & $9.2$ & $49$ & & $22$ & $8.8$ & $68$ & & $15$ & $28$ & $49$ & & $3.1$ & $3.8$ & $64$ & & $3.3$ & $2.1$ & $49$ & & $9.8$ & $8.3$ & $46$ & & $87$ & $8.0$ & $46$ & & $30$ & $3.7$ & $38$\\
GW170729 & $7.4$ & & $0.29$ & $0.64$ & $17$ & & $0.93$ & $1.1$ & $26$ & & $2.1$ & $4.5$ & $16$ & & $0.79$ & $1.5$ & $18$ & & $4.2$ & $1.4$ & $94$ & & $36$ & $8.6$ & $96$ & & $26$ & $8.9$ & $94$ & & $43$ & $7.1$ & $95$\\
\textbf{GW170809} & $9.3$ & & $1.4$ & $1.2$ & $64$ & & $2.5$ & $2.5$ & $49$ & & $11$ & $5.8$ & $49$ & & $1.4$ & $2.3$ & $37$ & & $3.8$ & $1.3$ & $79$ & & $22$ & $6.9$ & $81$ & & $18$ & $6.2$ & $78$ & & $13$ & $6.5$ & $82$\\
\textbf{GW170814} & $8.5$ & & $4.0$ & $1.1$ & $94$ & & $5.2$ & $1.8$ & $92$ & & $15$ & $5.2$ & $93$ & & $3.0$ & $1.3$ & $96$ & & $1.3$ & $3.9$ & $5.7$ & & $7.4$ & $25$ & $6.2$ & & $5.4$ & $29$ & $7.5$ & & $4.2$ & $13$ & $9.6$\\
GW170818 & $7.2$ & & $1.4$ & $0.66$ & $74$ & & $2.5$ & $1.9$ & $80$ & & $4.9$ & $4.3$ & $73$ & & $1.7$ & $0.67$ & $79$ & & $1.4$ & $3.4$ & $28$ & & $17$ & $15$ & $41$ & & $19$ & $7.9$ & $73$ & & $12$ & $9.0$ & $49$\\
\textbf{GW170823} & $6.3$ & & $1.2$ & $0.49$ & $61$ & & $1.1$ & $1.6$ & $51$ & & $2.7$ & $2.3$ & $49$ & & $0.99$ & $1.2$ & $46$ & & $2.5$ & $1.3$ & $53$ & & $11$ & $16$ & $41$ & & $8.8$ & $11$ & $37$ & & $14$ & $11$ & $46$\\
\midrule
\textbf{Combined} & $\NinetyPercentCombinedGravitonMassBoundScaledOTwoTGR$ & & $0.80$ & $0.34$ & $79$ & & $1.2$ & $0.70$ & $73$ & & $2.5$ & $1.2$ & $70$ & & $0.70$ & $0.37$ & $86$ & & $0.50$ & $0.80$ & $28$ & & $2.9$ & $3.7$ & $25$ & & $2.0$ & $3.7$ & $35$ & & $1.4$ & $2.3$ & $34$\\
\bottomrule
\end{tabular}}
\end{table*}

\begin{figure}[tb]
\includegraphics[width=0.48\textwidth]{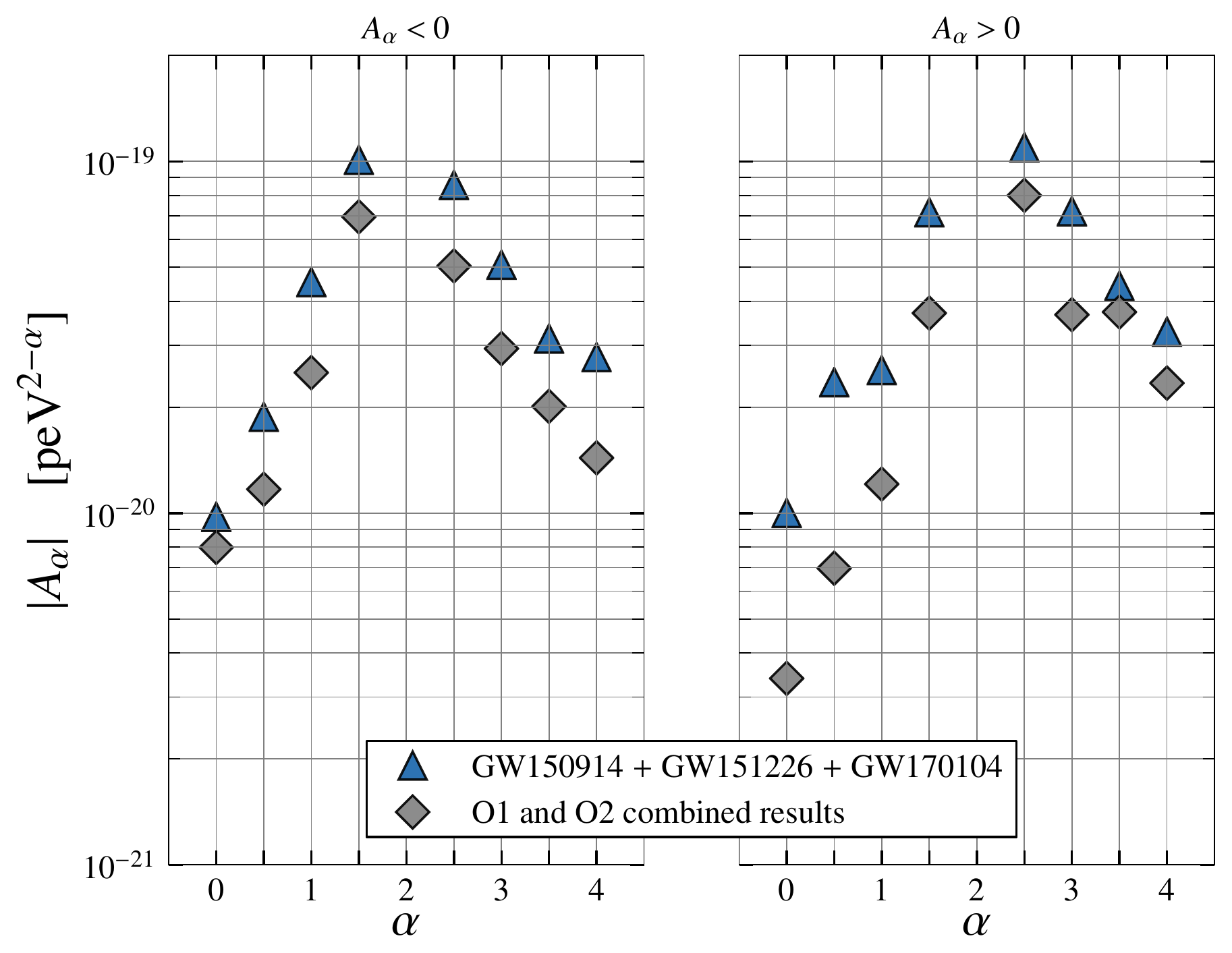}
\caption{90\% credible upper bounds on the absolute value of the modified dispersion relation parameter $A_\alpha$. We show results for positive and negative values of $A_\alpha$ separately. Specifically, we give the updated versions of the results from combining together GW150914, GW151226, and GW170104 (first given in~\cite{GW170104}), as well as the results from combining together all the events meeting our significance threshold for combined results (see Table~\ref{tab:events}). Picoelectronvolts (peV) provide a convenient scale, because $1 \text{ peV} \simeq h\times 250$~Hz, where $250$~Hz is roughly around the most sensitive frequencies of the LIGO and Virgo instruments.}
\label{fig:prop_results_summary}
\end{figure}

\begin{figure}[tb]
\includegraphics[width=0.48\textwidth]{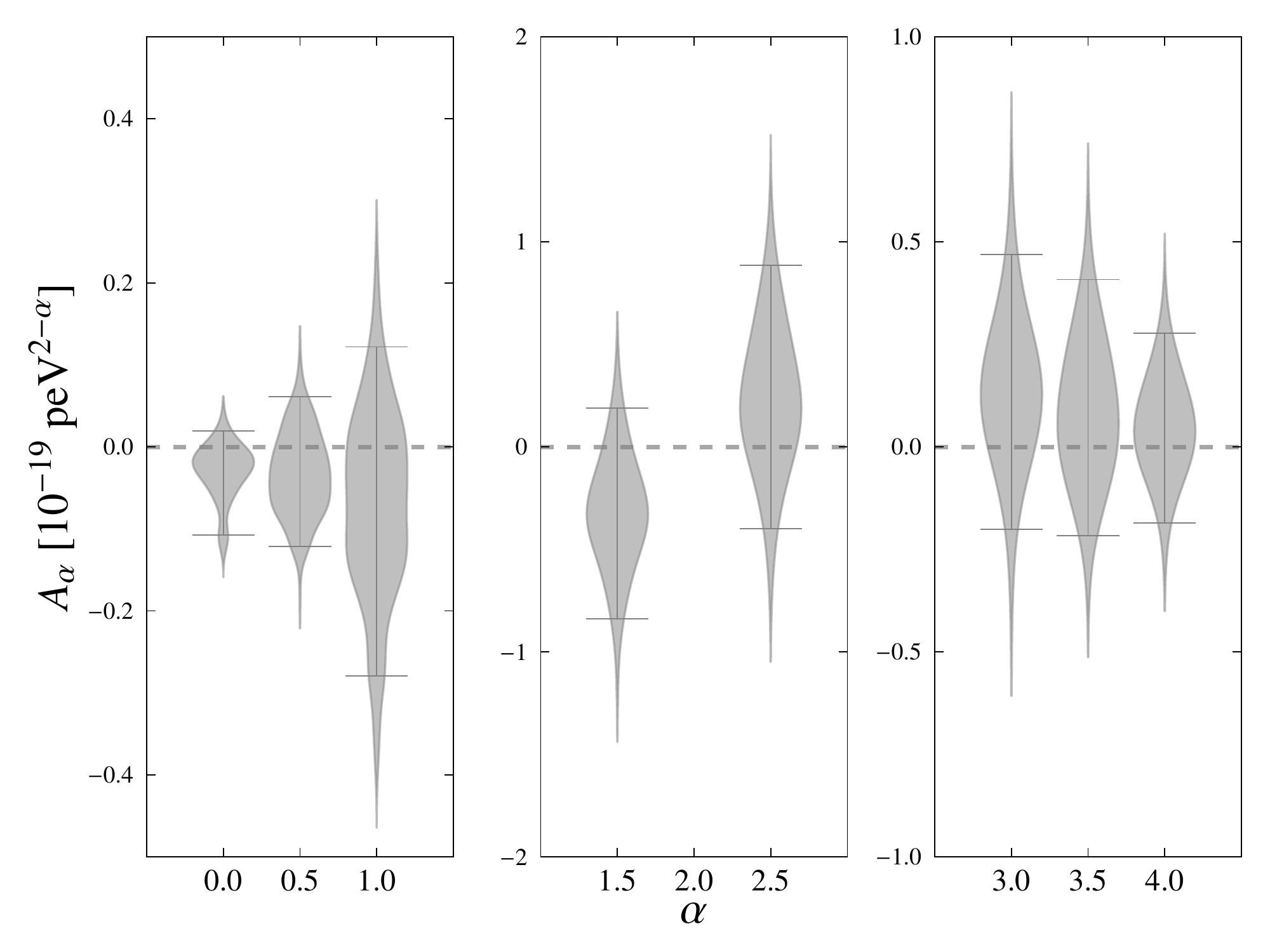}
\caption{Violin plots of the full posteriors on the modified dispersion relation parameter $A_\alpha$ calculated from the combined events, with the $90\%$ credible interval around the median indicated.}
\label{fig:prop_results_violin}
\end{figure}

We consider the cases of positive and negative $A_\alpha$ separately, and obtain the results shown in Table~\ref{tab:prop_results_summary} and Fig.~\ref{fig:prop_results_summary} when applying this analysis to the GW events under consideration. While we sample with a flat prior in $\log\lambda_{A, \text{eff}}$, our bounds are given using priors flat in $A_\alpha$ for all results except for the mass of the graviton, where we use a prior flat in the graviton mass. We also show the results from combining together all the signals that satisfy our selection criterion. We are able to combine together the results from different signals with no ambiguity, since the known distance dependence is accounted for in the waveforms.

Figure~\ref{fig:prop_results_violin} displays the full $A_\alpha$ posteriors obtained by combining all selected events (using \IMRP{} waveforms). To obtain the full $A_\alpha$ posteriors, we combine together the positive and negative $A_\alpha$ results for individual events by weighting by their Bayesian evidences; we then combine the posteriors from individual events. We give the analogous plots for the individual events in Sec.~\ref{app:propagation} of the Appendix. The combined positive and negative $A_\alpha$ posteriors are also used to compute the GR quantiles given in Table~\ref{tab:prop_results_summary}, which give the probability to have $A_\alpha < 0$, where $A_\alpha = 0$ is the GR value. Thus, large or small values of the GR quantile indicate that the distribution is not peaked close to the GR value. For a GR signal, the GR quantile will be distributed uniformly in $[0,1]$ for different noise realizations. The GR quantiles we find are consistent with such a uniform distribution. In particular, the (two-tailed) meta $p$-value for all events and $\alpha$ values obtained using Fisher's method~\cite{Fisher1948} (as in Sec.~\ref{sec:residuals}) is $0.9995$.

We find that the combined bounds overall improve on those quoted in~\cite{GW170104} by roughly the factor of $\sqrt{7/3} \simeq 1.5$ expected from including more events, with the bounds for some quantities improving by up to a factor of \MaximumImprovementFromPreviousBoundsOTwoTGR, due to the inclusion of several more massive and distant systems in the sample.\footnote{While the results in~\cite{GW170104} were affected by a slight normalization issue, and also had insufficiently fine binning in the computation of the upper bounds, we find improvements of up to a factor of $3.4$ when comparing to the combined GW150914 + GW151226 + GW170104 bounds we compute here.} These massive and distant systems, notably GW170823 (and GW170729, which is not included in the combined results), generally give the best individual bounds, particularly for small values of $\alpha$, where the dephasing is largest at lower frequencies. Closer and less massive systems such as GW151226 and GW170608 provide weaker bounds, overall. However, their bounds can be comparable to those of the more massive, distant events for larger values of $\alpha$. The lighter systems have more power at higher frequencies where the dephasing from the modified dispersion is larger for larger values of $\alpha$.

The new combined bound on the mass of the graviton of $m_g \leq \NinetyPercentCombinedGravitonMassBoundScaledOTwoTGR \times 10^{-23} \text{ eV}/c^2$ is a factor of $\NinetyPercentCombinedGravitonMassBoundOTwoTGRImprovementFromGWOneSevenZeroOneZeroFourPaper$ improvement on the one presented in~\cite{GW170104}. It is also a small improvement on the bound of $m_g \leq 6.76 \times 10^{-23} \text{ eV}/c^2$ ($90\%$ confidence level) obtained from Solar System ephemerides in~\cite{Bernus:2019rgl}.\footnote{The much stronger bound in~\cite{Will:2018gku} is deduced from a post-fit analysis (i.e., using the residuals of a fit to Solar System ephemerides performed without including the effects of a massive graviton). It may therefore overestimate Solar System constraints, as is indeed seen to be the case in~\cite{Bernus:2019rgl}.} However, these bounds are complementary, since the GW bound comes from the radiative sector, while the Solar System bound considers the static modification to the Newtonian potential. See, e.g.,~\cite{deRham:2016nuf} for a review of bounds on the mass of the graviton.

We find that the posterior on $A_\alpha$ peaks away from $0$ in some cases (illustrated in Sec.~\ref{app:propagation} of the Appendix), and the GR quantile is in one of the tails of the distribution. This feature is expected for a few out of 10 events, simply from Gaussian noise fluctuations. We have performed simulations of 100 GR sources with source-frame component masses lying between 25 and 45 $M_\odot$, isotropically distributed spins with dimensionless magnitudes up to $0.99$, and at luminosity distances between 500 and 800 Mpc. These simulations used the waveform model {\IMRP} and considered the Advanced LIGO and Virgo network, using Gaussian noise with the detectors' design sensitivity power spectral densities. We found that in about 20 -- 30\% of cases, the GR quantile lies in the tails of the distribution (i.e., $< 10\%$ or $> 90\%$), when the sources injected are analyzed using the same waveform model (\IMRP{}).

In order to assess the impact of waveform systematics, we also analyze some events using the aligned-spin \SEOB{} model. We consider GW170729 and GW170814 in depth in this study because the GR quantiles of the \IMRP{} results lie in the tails of the distributions, and find that the $90\%$ upper bounds and GR quantiles presented in Table~\ref{tab:prop_results_summary} differ by at most a factor of $2.3$ for GW170729 and $1.5$ for GW170814 when computed using the \SEOB{} model. These results are presented in Sec.~\ref{app:propagation} of the Appendix.

There are also uncertainties in the determination of the 90\% bounds due to the finite number of samples and the long tails of the distributions. As in Ref.~\cite{GW170104}, we quantify this uncertainty using Bayesian bootstrapping~\cite{rubin1981}. We use 1000 bootstrap realizations for each value of $\alpha$ and sign of $A_\alpha$, obtaining a distribution of 90\% bounds on $A_\alpha$. We consider the 90\% credible interval of this distribution and find that its width is $< 30\%$ of the values for the 90\% bounds on $A_\alpha$ given in Table~\ref{tab:prop_results_summary} for all but $10$ of the $160$ cases we consider (counting positive and negative $A_\alpha$ cases separately). For GW170608, $A_4 < 0$, the width of the 90\% credible interval from bootstrapping is $91\%$ of the value in Table~\ref{tab:prop_results_summary}. This ratio is $\leq 47\%$ for all the remaining cases. Thus, there are a few cases where the bootstrapping uncertainty in the bound on $A_\alpha$ is large, but for most cases, this is not a substantial uncertainty.

\section{Polarizations}
\label{sec:pols}
Generic metric theories of gravity may allow up to six polarizations of gravitational waves \cite{Eardley:1974nw}: two tensor modes (helicity $\pm 2$), two vector modes (helicity $\pm1$), and two scalar modes (helicity 0).
Of these, only the two tensor modes ($+$ and $\times$) are permitted in GR.
We may attempt to reconstruct the polarization content of a passing GW using a network of detectors \cite{lrr-2014-4,Chatziioannou:2012rf,Isi:2017equ,Callister:2017ocg,Isi:2017fbj}.
This is possible because instruments with different orientations will respond differently to signals from a given sky location depending on their polarization.
In particular, the strain signal in detector $I$ can be written as $h_I (t)  = \sum_A F^A{}_I h_A (t)$, with $F^A{}_I$ the detector's response function and $h_A(t)$ the $A$-polarized part of the signal \cite{lrr-2014-4,Blaut:2012zz}.

In order to fully disentangle the polarization content of a transient signal, at least 5 detectors are needed to break all degeneracies \cite{Chatziioannou:2012rf}.%
\footnote{Differential-arm detectors are only sensitive to the traceless scalar mode, meaning we can only hope to distinguish five, not six, polarizations.}
This limits the polarization measurements that are currently feasible.
In spite of this, we may extract some polarization information from signals detected with both  LIGO detectors and Virgo \cite{Isi:2017fbj}.
This was done previously with GW170814 and GW170817 to provide evidence that GWs are tensor polarized, instead of fully vector or fully scalar \cite{GW170814paper,bns-tgr}.
Besides GW170814, there are three binary black hole events that were detected with the full network (GW170729, GW170809, and GW170818).
Of these events, only GW170818 has enough SNR and is sufficiently well localized to provide any relevant information (cf.~Fig.~8 in \cite{O2:Catalog}).
The Bayes factors (marginalized likelihood ratios) obtained in this case are $12\pm3$ for tensor vs vector and $407\pm100$ for tensor vs scalar, where the error corresponds to the uncertainty due to discrete sampling in the evidence computations.
These values are comparable to those from GW170814, for which the latest recalibrated and cleaned data (cf.~Sec.~\ref{sec:data}) yield Bayes factors of $30\pm 4$ and $220\pm 27$ for tensor vs vector and scalar respectively.%
\footnote{These values are less stringent than those previously published in \cite{GW170814paper}. This is solely due to the change in data, which impacted the sky locations inferred under the non-GR hypotheses.}
Values from these binary black holes are many orders of magnitude weaker than those obtained from GW170817, where we benefited from the precise sky-localization provided by an electromagnetic counterpart \cite{bns-tgr}.

\section{Conclusions and outlook}
\label{sec:conclusions}
We have presented the results from various tests of GR performed using 
the binary black hole signals from the catalog GWTC-1~\cite{GWOSC:GWTC}, 
i.e., those observed by Advanced LIGO and Advanced Virgo during the first 
two observing runs of the advanced detector era. These tests, which are 
among the first tests of GR in the highly relativistic, nonlinear regime 
of strong gravity, do not reveal any inconsistency of our data with the 
predictions of GR. We have presented full results on four tests of the 
consistency of the data with gravitational waveforms from binary black hole 
systems as predicted by GR. The first two of these tests check the self-consistency 
of our analysis. One checks that the residual remaining after subtracting the 
best-fit waveform is consistent with detector noise. The other checks that the 
final mass and spin inferred from the low- and high-frequency parts of the signal 
are consistent. The third and fourth tests introduce parameterized deviations in 
the waveform model and check that these deviations are consistent with their GR 
value of zero. In one test, these deviations are completely phenomenological 
modifications of the coefficients in a waveform model, including the post-Newtonian 
coefficients. In the other test, the deviations are those arising from the propagation 
of GWs with a modified dispersion relation, which includes the dispersion due to a 
massive graviton as a special case. In addition, we also check whether the observed 
polarizations are consistent with being purely tensor modes (as expected in GR) 
as opposed to purely scalar modes or vector modes.

We present results from all binary black hole events that are detected with a false 
alarm rate better than $(1\, \mathrm{yr})^{-1}$. This includes results from the 
re-analysis of some of the events which were published 
earlier~\cite{GW150914:TGR,O1:BBH,GW170104,GW170814paper}, with better calibration 
and data quality. Assuming that the parameters that describe deviations from GR take 
values that are independent of source properties, we can combine results from 
multiple events. We choose to combine results only from highly significant events, 
detected with a false alarm rate better than $(1000\, \mathrm{yr})^{-1}$ in both modeled 
searches. Combining together these results has allowed us to significantly reduce the 
statistical errors on constraints on deviations from GR predictions, as compared to 
those from individual events. The combined constraints presented here improve our 
previously presented constraints by factors of $\MinimumImprovementFromPreviousBoundsOTwoTGR$ 
to $\MaximumImprovementFromPreviousBoundsOTwoTGR$, with the largest improvements obtained 
for certain cases of the modified dispersion test. Notable constraints include that on the 
graviton's mass 
$m_g \leq \NinetyPercentCombinedGravitonMassBoundScaledOTwoTGR \times 10^{-23} \text{ eV}/c^2$ 
(an improvement of a factor of 
$\NinetyPercentCombinedGravitonMassBoundOTwoTGRImprovementFromGWOneSevenZeroOneZeroFourPaper$ 
over previously presented constraints) and the first constraint on the $-1$PN coefficient 
obtained from binary black holes.

With the expected observations of additional binary black hole merger events in the 
upcoming LIGO/Virgo observing runs~\cite{Aasi:2013wya,O2:Catalog}, the statistical errors 
of the combined results will soon decrease significantly. A number of potential sources 
of systematic errors (due to imperfect modeling of GR waveforms, calibration uncertainties, noise artifacts, etc.) need to be understood for future high-precision tests of strong gravity 
using GW observations. However, work to improve the analysis on all these fronts is well 
underway, for instance the inclusion of full spin-precession 
dynamics~\cite{Taracchini:2013, Babak:2016tgq, Blackman:2017pcm, Khan:2018fmp, Varma:2019csw},
non-quadrupolar modes~\cite{Blackman:2017pcm, London:2017bcn, Cotesta:2018fcv, Varma:2018mmi, Varma:2019csw}, and
eccentricity~\cite{Hinderer:2017jcs, Cao:2017ndf, Hinder:2017sxy, Huerta:2017kez, Klein:2018ybm, Moore:2018kvz, Moore:2019xkm, Tiwari:2019jtz} 
in waveform models, as well as analyses that compare directly with numerical relativity 
waveforms~\cite{Abbott:2016apu, Lange:2017wki}. Additionally, a new, more flexible parameter
estimation infrastructure is currently being developed~\cite{Ashton:2018jfp}, and this will 
allow for improvements in, e.g., the treatment of calibration uncertainties or PSD estimation to be incorporated 
more easily. We thus expect that tests of general relativity using the data from upcoming 
observing runs will be able to take full advantage of the increased sensitivity of the 
detectors.

\acknowledgments
The authors gratefully acknowledge the support of the United States
National Science Foundation (NSF) for the construction and operation of the
LIGO Laboratory and Advanced LIGO as well as the Science and Technology Facilities Council (STFC) of the
United Kingdom, the Max-Planck-Society (MPS), and the State of
Niedersachsen/Germany for support of the construction of Advanced LIGO 
and construction and operation of the GEO600 detector. 
Additional support for Advanced LIGO was provided by the Australian Research Council.
The authors gratefully acknowledge the Italian Istituto Nazionale di Fisica Nucleare (INFN),  
the French Centre National de la Recherche Scientifique (CNRS) and
the Foundation for Fundamental Research on Matter supported by the Netherlands Organisation for Scientific Research, 
for the construction and operation of the Virgo detector
and the creation and support  of the EGO consortium. 
The authors also gratefully acknowledge research support from these agencies as well as by 
the Council of Scientific and Industrial Research of India, 
the Department of Science and Technology, India,
the Science \& Engineering Research Board (SERB), India,
the Ministry of Human Resource Development, India,
the Spanish  Agencia Estatal de Investigaci\'on,
the Vicepresid\`encia i Conselleria d'Innovaci\'o, Recerca i Turisme and the Conselleria d'Educaci\'o i Universitat del Govern de les Illes Balears,
the Conselleria d'Educaci\'o, Investigaci\'o, Cultura i Esport de la Generalitat Valenciana,
the National Science Centre of Poland,
the Swiss National Science Foundation (SNSF),
the Russian Foundation for Basic Research, 
the Russian Science Foundation,
the European Commission,
the European Regional Development Funds (ERDF),
the Royal Society, 
the Scottish Funding Council, 
the Scottish Universities Physics Alliance, 
the Hungarian Scientific Research Fund (OTKA),
the Lyon Institute of Origins (LIO),
the Paris \^{I}le-de-France Region, 
the National Research, Development and Innovation Office Hungary (NKFIH), 
the National Research Foundation of Korea,
Industry Canada and the Province of Ontario through the Ministry of Economic Development and Innovation, 
the Natural Science and Engineering Research Council Canada,
the Canadian Institute for Advanced Research,
the Brazilian Ministry of Science, Technology, Innovations, and Communications,
the International Center for Theoretical Physics South American Institute for Fundamental Research (ICTP-SAIFR), 
the Research Grants Council of Hong Kong,
the National Natural Science Foundation of China (NSFC),
the Leverhulme Trust, 
the Research Corporation, 
the Ministry of Science and Technology (MOST), Taiwan
and
the Kavli Foundation.
The authors gratefully acknowledge the support of the NSF, STFC, MPS, INFN, CNRS and the
State of Niedersachsen/Germany for provision of computational resources.
The authors would like to thank Clifford Will and Nicol{\'a}s Yunes for useful discussions.

\vspace{5mm}

\appendix*
\section{Individual results and systematics studies}
In the main body of the paper, for most analyses, we present only the combined results from all events. Here we present the posteriors from various tests obtained from individual events. In addition, we offer a limited discussion on systematic errors in the analysis, due to the specific choice of a GR waveform approximant.

\subsection{Residuals test}

As mentioned in Sec.~\ref{sec:residuals}, the residuals test is sensitive to all kinds of disagreement between the best-fit GR-based waveform and the data.
This is true whether the disagreement is due to actual deviations from GR or more mundane reasons, like physics missing from our waveform models (e.g., higher-order modes).
Had we found compelling evidence of coherent power in the residuals that could not be explained by instrumental noise, further investigations would be required to determine its origin.
However, given our null result, we can simply state that we find no evidence for shortcomings in the best-fit waveform, neither from deviations from GR nor modeling systematics.

As the sensitivity of the detectors improves, the issue of systematics will become increasingly more important.
To address this, future versions of this test will be carried out by subtracting a best-fit waveform produced with more accurate GR-based models, including numerical relativity.

\subsection{Inspiral-merger-ringdown consistency test}
\label{appendix:imr_test}

\begin{figure}[tbh] 
	\begin{center}
	\includegraphics[width=3.5in]{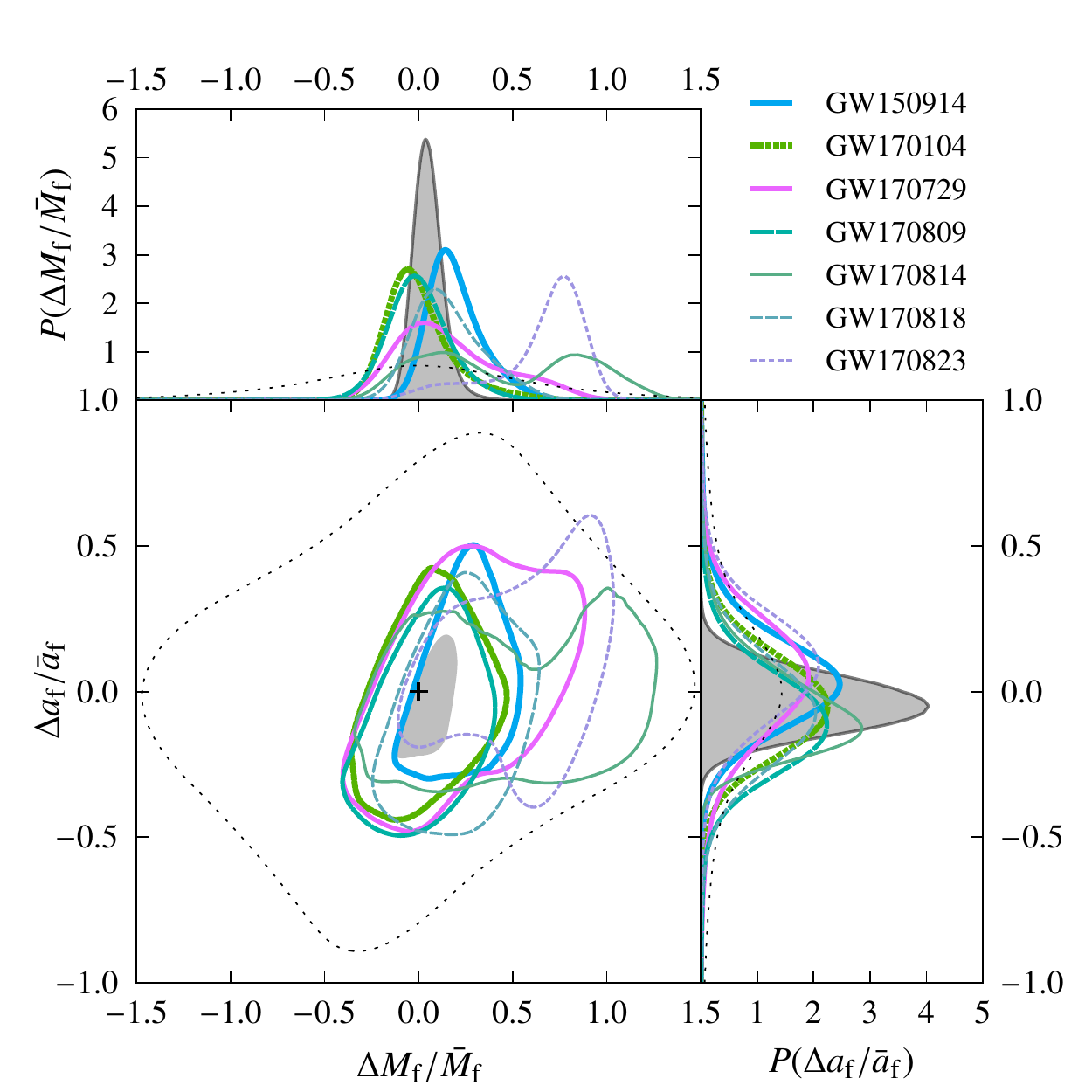}
	\end{center}
	\caption{Same as Fig.~\ref{fig:imr_test_posteriors}, except that the posteriors are computed using the nonprecessing-spin \SEOB{} waveforms.} 
	\label{fig:imr_test_posteriors_seobnr}
\end{figure}

In order to gauge the systematic errors in the IMR consistency test results due to imperfect waveform modeling, we have also estimated the posteriors of the deviation parameters $\Delta M_\mathrm{f}/\bar{M}_\mathrm{f}$ and $\Delta a_\mathrm{f}/\bar{a}_\mathrm{f}$ using the effective-one-body based waveform family \SEOB{} that models binary black holes with non-precessing spins. This analysis uses the same priors as used in the main analysis presented in Sec.\ref{sec:imr-test}, except that spins are assumed to be aligned/antialigned with the orbital angular momentum of the binary. The resulting posteriors are presented in Fig.~\ref{fig:imr_test_posteriors_seobnr} and are broadly consistent with the posteriors using \textsc{IMRPhenomPv2} presented in Fig.~\ref{fig:imr_test_posteriors}. The differences in the posteriors of some of the individual events are not surprising, due to the different assumptions on the spins. For all events, the GR value is recovered in the 90\% credible region of the posteriors.

\subsection{Parametrized tests of gravitational wave generation}
\label{app:generation}
\begin{figure*}[tb]
\includegraphics[keepaspectratio, width=\textwidth]{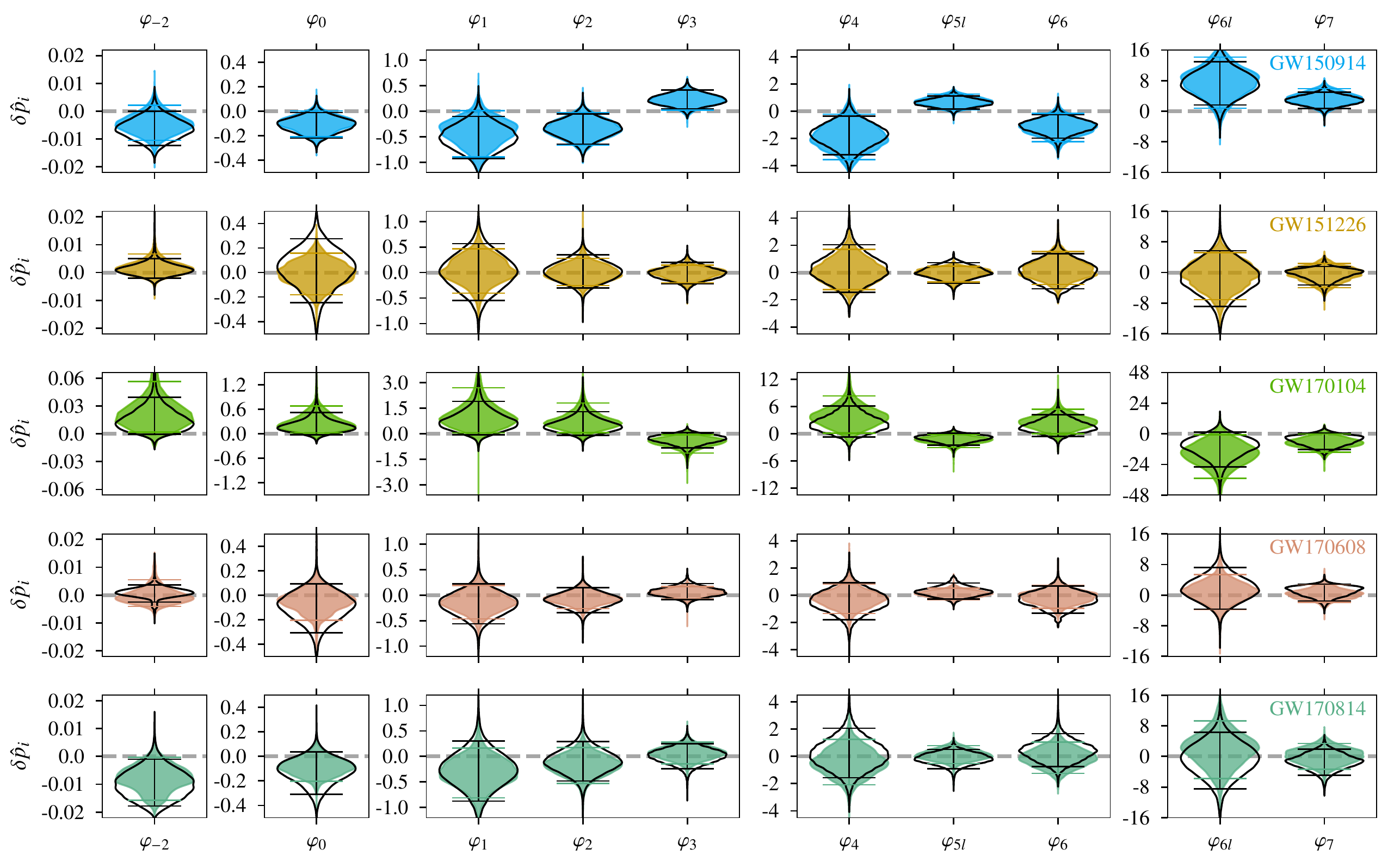}
\caption{Violin plots showing inspiral $\delta\hat{p}_{i}$ posteriors for the individual binary black-hole events of GWTC-1~\cite{O2:Catalog} outlined in Sec.~\ref{sec:events} (see ``PI'' column of Table~\ref{tab:events}), using {\IMRP} (shaded regions) and {\SEOB} (black solid lines).
Thin horizontal lines indicate the 90\% credible intervals, which show an overall statistical consistency with GR (dashed grey line).}
\label{fig:violin_all_PI}
\end{figure*}

\begin{figure*}[tb]
\includegraphics[keepaspectratio, width=\textwidth]{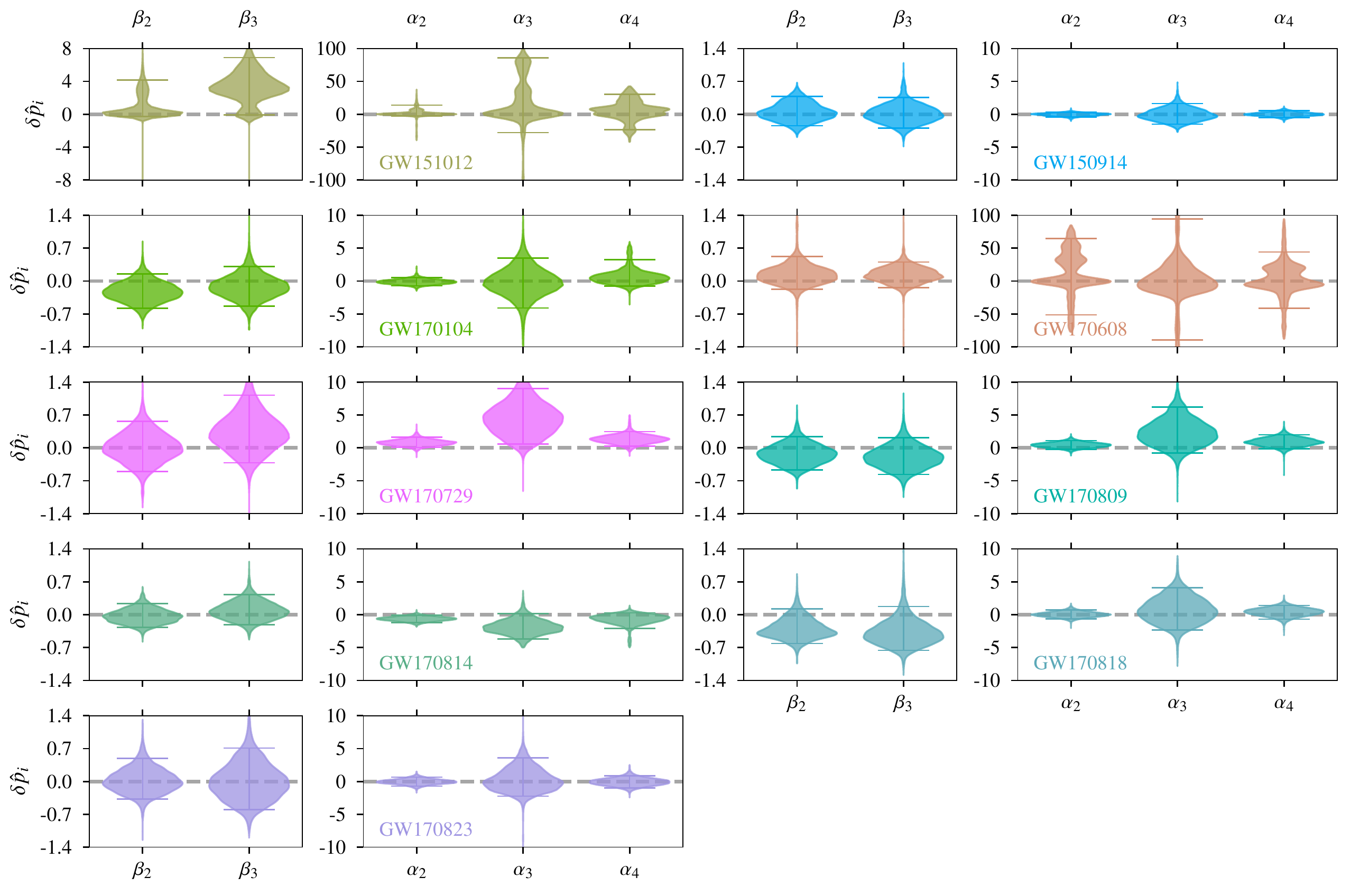}
\caption{Violin plots showing post-inspiral $\delta\hat{p}_{i}$ posteriors for the individual binary black-hole events of GWTC-1~\cite{O2:Catalog} outlined in Sec.~\ref{sec:events} (see ``PPI'' column of Table~\ref{tab:events}), using {\IMRP}.
Thin horizontal lines indicate the 90\% credible intervals, which show an overall statistical consistency with GR (dashed grey line).}
\label{fig:violin_all_PPI}
\end{figure*}

Figures~\ref{fig:violin_all_PI} and~\ref{fig:violin_all_PPI} report the parameterized
tests of waveform deviations for the individual events, augmenting the results shown 
in Fig.~\ref{fig:violin_plot}.
A statistical summary of the posterior PDFs, showing median and symmetric 90\% credible level bounds for the measured parameters is given in Table~\ref{tab:gen_results_wf_comparison}.
Sources with low SNR in the inspiral regime yield uninformative posterior distributions on 
$\delta\hat\varphi_i$. These sources are the ones further away and with higher mass, 
which merge at lower frequencies.
For instance, although GW170823 has a total mass close to that of GW150914, being 
much further away (and redshifted to lower frequencies) makes it a low-SNR event, leaving
very little information content in the inspiral regime.
The same holds true for GW170729, which has a larger mass. 
Conversely, low-mass events like GW170608, having a significantly larger SNR in the 
inspiral regime and many more cycles in the frequency band provide very strong constraints 
in the $\delta\hat\varphi_i$ parameters (especially the low-order ones) while providing
no useful constraints in the merger-ringdown parameters $\delta\hat\alpha_i$.

The choice of the $\mathrm{SNR} > 6$ threshold explained in Sec.~\ref{sec:events} 
ensures that most analyses are informative. 
However, this is not true in all cases, as not all parameters are as easily determined 
from the data (cf.\ the good constraints one obtains on the chirp mass with the much weaker constraints on the
mass ratio). 
The two events for which the SNR threshold is insufficient are GW151012 and GW170608, 
where some post-inspiral parameters are largely unconstrained. 
The post-inspiral regime is itself divided into the intermediate and merger-ringdown regimes, and 
for both these events we find the intermediate regime parameters ($\delta\hat\beta_i$) to be 
informative; however, the merger--ringdown $\delta\hat\alpha_3$ for GW170608 and all $\delta\hat\alpha_i$ 
for GW151012 extend across the entire prior range considered in the analyses, far beyond the range constrained by other events (as can be seen in Fig.~\ref{fig:violin_all_PPI}). 
Although we use the results from GW170608 $\delta\hat\alpha_i$ 
for combining posteriors in Fig.~\ref{fig:violin_plot}, the combined bounds remain unaffected by adding these results. 
For future tests, a more discerning  threshold than a simple SNR cut, 
for example including information like the number of cycles of the signal in band, 
may be used to select which events will provide useful constraints.

Both here and in Sec.~\ref{sec:gwnature} we report results on the parametrized deviations in the PN regime using two waveform models, {\IMRP} and {\SEOB}.
There is a subtle difference between the ways deviations from GR are introduced and parametrized in the two models.
With IMRPhenomPv2, we directly constrain $\delta \hat{\varphi}_i$, which represent fractional deviations in the non-spinning portion of the $(i/2)$PN phase coefficients. 
The SEOBNRv4 analysis instead uses a parameterization that also applies the fractional deviations to spin contributions, as described in~\cite{bns-tgr}. The results are then mapped post-hoc from this native parameterization to posteriors on $\delta \hat{\varphi}_i$, shown in Figs.~\ref{fig:violin_plot} and~\ref{fig:violin_all_PI} (black solid lines).

In the {\SEOB} analysis at $3.5$PN, the native (spin-inclusive) posteriors contain tails that extend to the edge of the prior range.
This is due to a zero-crossing of the $3.5$PN term in the $(\eta,a_1,a_2)$ parameter space, which makes the corresponding relative deviation ill-defined.
After the post-hoc mapping to posteriors on $\delta\hat{\varphi}_7$, no tails appear and we find good agreement with the {\IMRP} analysis, as expected.
By varying the prior range, we estimate a systematic uncertainty of at most a few percent on the quoted $90\%$ bounds due to the truncation of tails.

\begin{table*}
\caption{\label{tab:gen_results_wf_comparison}
Median value and symmetric 90\% credible level bounds of the waveform parameters $\delta\hat{p}_{i}$, as well as the GR quantiles $Q_\text{GR}$. For the inspiral parameters, we show results in pairs of rows for when the data from individual events are analyzed using \IMRP{} (P) and \SEOB{} (S), while for the post-inspiral phenomenological parameters, results are obtained only for \IMRP{}.}
\centering
\scalebox{0.88}{

\begin{tabular}{*{18}{c}}
\toprule
 & & & & \twoc{GW150914}& & \twoc{GW151226}& & \twoc{GW170104}& & \twoc{GW170608}& & \twoc{GW170814}\\
\cline{5-6}
\cline{8-9}
\cline{11-12}
\cline{14-15}
\cline{17-18}
\twoc{Parameter} & Model & & $\tilde{X}^{+}_{-}$ & $Q_\text{GR}$ [\%] & & $\tilde{X}^{+}_{-}$ & $Q_\text{GR}$ [\%] & & $\tilde{X}^{+}_{-}$ & $Q_\text{GR}$ [\%] & & $\tilde{X}^{+}_{-}$ & $Q_\text{GR}$ [\%] & & $\tilde{X}^{+}_{-}$ & $Q_\text{GR}$ [\%]\\
\midrule
\vspace{1pt}\multirow{2}{*}{$\delta\hat\phi_{-2}$} & \multirow{2}{*}{$[10^{-2}]$} & P & & $-0.46^{+0.68}_{-0.59}$ & $88$& & $0.14^{+0.52}_{-0.35}$ & $27$& & $2.0^{+3.6}_{-1.9}$ & $3$& & $-0.06^{+0.61}_{-0.35}$ & $60$& & $-0.76^{+0.74}_{-0.81}$ & $96$\\
\vspace{4pt} &  & S & & $-0.63^{+0.62}_{-0.61}$ & $95$& & $0.10^{+0.40}_{-0.31}$ & $30$& & $1.4^{+2.6}_{-1.4}$ & $5.9$& & $0.10^{+0.27}_{-0.34}$ & $26$& & $-0.99^{+0.89}_{-0.79}$ & $97$\\
\vspace{1pt}\multirow{2}{*}{$\delta\hat\phi_{0}$} & \multirow{2}{*}{$[10^{-1}]$} & P & & $-1.0^{+1.0}_{-1.1}$ & $95$& & $-0.1^{+1.7}_{-1.7}$ & $53$& & $2.5^{+4.3}_{-2.3}$ & $3.1$& & $-0.5^{+1.4}_{-1.6}$ & $73$& & $-0.8^{+1.2}_{-1.2}$ & $88$\\
\vspace{4pt} &  & S & & $-1.1^{+1.0}_{-1.1}$ & $97$& & $0.3^{+2.4}_{-2.8}$ & $41$& & $1.8^{+3.5}_{-2.0}$ & $7.3$& & $-0.9^{+1.8}_{-2.2}$ & $77$& & $-1.1^{+1.5}_{-2.0}$ & $89$\\
\vspace{1pt}\multirow{2}{*}{$\delta\hat\phi_{1}$} & \multirow{2}{*}{$[10^{0}]$} & P & & $-0.39^{+0.40}_{-0.50}$ & $95$& & $0.03^{+0.43}_{-0.44}$ & $45$& & $1.0^{+1.7}_{-0.9}$ & $4.1$& & $-0.14^{+0.33}_{-0.33}$ & $76$& & $-0.31^{+0.47}_{-0.50}$ & $86$\\
\vspace{4pt} &  & S & & $-0.51^{+0.41}_{-0.41}$ & $98$& & $0.02^{+0.55}_{-0.56}$ & $48$& & $0.7^{+1.2}_{-0.8}$ & $6.6$& & $-0.16^{+0.38}_{-0.40}$ & $77$& & $-0.27^{+0.57}_{-0.61}$ & $79$\\
\vspace{1pt}\multirow{2}{*}{$\delta\hat\phi_{2}$} & \multirow{2}{*}{$[10^{0}]$} & P & & $-0.35^{+0.32}_{-0.31}$ & $97$& & $-0.01^{+0.29}_{-0.24}$ & $52$& & $0.7^{+1.1}_{-0.6}$ & $3.5$& & $-0.07^{+0.21}_{-0.20}$ & $72$& & $-0.17^{+0.34}_{-0.36}$ & $80$\\
\vspace{4pt} &  & S & & $-0.34^{+0.28}_{-0.30}$ & $97$& & $0.02^{+0.33}_{-0.32}$ & $47$& & $0.47^{+0.83}_{-0.56}$ & $8.6$& & $-0.09^{+0.24}_{-0.25}$ & $75$& & $-0.10^{+0.39}_{-0.38}$ & $67$\\
\vspace{1pt}\multirow{2}{*}{$\delta\hat\phi_{3}$} & \multirow{2}{*}{$[10^{-1}]$} & P & & $2.2^{+2.0}_{-1.9}$ & $3.4$& & $-0.1^{+1.5}_{-2.0}$ & $54$& & $-4.8^{+4.2}_{-6.5}$ & $97$& & $0.5^{+1.2}_{-1.2}$ & $26$& & $0.7^{+2.1}_{-2.2}$ & $30$\\
\vspace{4pt} &  & S & & $2.2^{+2.0}_{-1.8}$ & $1.8$& & $-0.2^{+2.0}_{-2.0}$ & $55$& & $-3.2^{+3.8}_{-5.2}$ & $91$& & $0.6^{+1.6}_{-1.4}$ & $26$& & $0.1^{+2.4}_{-2.4}$ & $48$\\
\vspace{1pt}\multirow{2}{*}{$\delta\hat\phi_{4}$} & \multirow{2}{*}{$[10^{0}]$} & P & & $-1.9^{+1.7}_{-1.6}$ & $97$& & $0.1^{+1.6}_{-1.3}$ & $47$& & $3.7^{+4.6}_{-3.6}$ & $4.2$& & $-0.3^{+1.1}_{-1.1}$ & $67$& & $-0.5^{+1.7}_{-1.6}$ & $67$\\
\vspace{4pt} &  & S & & $-1.7^{+1.3}_{-1.5}$ & $98$& & $0.2^{+1.8}_{-1.6}$ & $41$& & $2.2^{+3.9}_{-2.9}$ & $10$& & $-0.4^{+1.3}_{-1.4}$ & $70$& & $0.1^{+1.9}_{-1.7}$ & $45$\\
\vspace{1pt}\multirow{2}{*}{$\delta\hat\phi_{5}^{(l)}$} & \multirow{2}{*}{$[10^{0}]$} & P & & $0.70^{+0.56}_{-0.58}$ & $2.2$& & $-0.03^{+0.49}_{-0.67}$ & $54$& & $-1.4^{+1.3}_{-1.6}$ & $97$& & $0.09^{+0.42}_{-0.42}$ & $36$& & $0.10^{+0.69}_{-0.62}$ & $40$\\
\vspace{4pt} &  & S & & $0.61^{+0.49}_{-0.45}$ & $1.5$& & $0.00^{+0.71}_{-0.77}$ & $50$& & $-1.0^{+1.1}_{-1.5}$ & $92$& & $0.25^{+0.62}_{-0.54}$ & $23$& & $-0.16^{+0.66}_{-0.72}$ & $64$\\
\vspace{1pt}\multirow{2}{*}{$\delta\hat\phi_{6}$} & \multirow{2}{*}{$[10^{0}]$} & P & & $-1.2^{+1.0}_{-1.1}$ & $97$& & $0.2^{+1.3}_{-1.1}$ & $40$& & $2.6^{+2.9}_{-2.5}$ & $4.7$& & $-0.11^{+0.84}_{-0.84}$ & $59$& & $-0.1^{+1.2}_{-1.1}$ & $56$\\
\vspace{4pt} &  & S & & $-1.07^{+0.80}_{-0.92}$ & $99$& & $0.1^{+1.3}_{-1.3}$ & $45$& & $1.6^{+2.7}_{-2.1}$ & $11$& & $-0.3^{+1.0}_{-1.0}$ & $68$& & $0.4^{+1.3}_{-1.1}$ & $29$\\
\vspace{1pt}\multirow{2}{*}{$\delta\hat\phi_{6}^{(l)}$} & \multirow{2}{*}{$[10^{1}]$} & P & & $0.78^{+0.63}_{-0.70}$ & $3.5$& & $-0.05^{+0.56}_{-0.66}$ & $55$& & $-1.6^{+1.5}_{-1.9}$ & $96$& & $0.09^{+0.44}_{-0.47}$ & $36$& & $0.20^{+0.73}_{-0.78}$ & $32$\\
\vspace{4pt} &  & S & & $0.72^{+0.57}_{-0.55}$ & $1.8$& & $-0.13^{+0.70}_{-0.76}$ & $61$& & $-1.0^{+1.1}_{-1.6}$ & $92$& & $0.15^{+0.57}_{-0.52}$ & $31$& & $-0.07^{+0.69}_{-0.78}$ & $56$\\
\vspace{1pt}\multirow{2}{*}{$\delta\hat\phi_{7}$} & \multirow{2}{*}{$[10^{0}]$} & P & & $3.2^{+2.7}_{-2.6}$ & $2.2$& & $-0.4^{+2.8}_{-3.5}$ & $58$& & $-7.0^{+6.9}_{-7.4}$ & $95$& & $0.3^{+2.5}_{-2.2}$ & $42$& & $0.2^{+3.2}_{-3.5}$ & $47$\\
\vspace{4pt} &  & S & & $2.8^{+2.1}_{-2.0}$ & $1.3$& & $-0.1^{+1.7}_{-3.1}$ & $54$& & $-4.8^{+5.4}_{-7.2}$ & $93$& & $1.1^{+1.8}_{-2.7}$ & $23$& & $-1.2^{+3.1}_{-3.2}$ & $73$\\
\vspace{4pt}$\delta\hat\beta_{2}$ & $[10^{0}]$ & P & & $0.03^{+0.35}_{-0.28}$ & $43$& & $-^{}_{}$ & $-$& & $-0.23^{+0.38}_{-0.35}$ & $85$& & $0.13^{+0.40}_{-0.30}$ & $25$& & $-0.04^{+0.28}_{-0.23}$ & $61$\\
\vspace{4pt}$\delta\hat\beta_{3}$ & $[10^{0}]$ & P & & $0.00^{+0.36}_{-0.30}$ & $50$& & $-^{}_{}$ & $-$& & $-0.13^{+0.44}_{-0.41}$ & $71$& & $0.12^{+0.28}_{-0.26}$ & $21$& & $0.07^{+0.35}_{-0.29}$ & $35$\\
\vspace{4pt}$\delta\hat\alpha_{2}$ & $[10^{0}]$ & P & & $-0.02^{+0.34}_{-0.42}$ & $54$& & $-^{}_{}$ & $-$& & $-0.13^{+0.67}_{-0.58}$ & $64$& & $3^{+62}_{-55}$ & $40$& & $-0.64^{+0.57}_{-0.59}$ & $97$\\
\vspace{4pt}$\delta\hat\alpha_{3}$ & $[10^{0}]$ & P & & $-0.1^{+1.7}_{-1.4}$ & $55$& & $-^{}_{}$ & $-$& & $-0.3^{+3.8}_{-3.8}$ & $55$& & $-1^{+95}_{-89}$ & $52$& & $-1.9^{+2.1}_{-1.8}$ & $93$\\
\vspace{4pt}$\delta\hat\alpha_{4}$ & $[10^{0}]$ & P & & $-0.01^{+0.54}_{-0.51}$ & $52$& & $-^{}_{}$ & $-$& & $0.5^{+2.8}_{-1.3}$ & $28$& & $-1^{+45}_{-41}$ & $51$& & $-0.6^{+0.9}_{-1.5}$ & $85$\\
\midrule
 & & & & \twoc{GW151012}& & \twoc{GW170729}& & \twoc{GW170809}& & \twoc{GW170818}& & \twoc{GW170823}\\
\cline{5-6}
\cline{8-9}
\cline{11-12}
\cline{14-15}
\cline{17-18}
\twoc{Parameter} & Model  & & $\tilde{X}^{+}_{-}$ & $Q_\text{GR}$ [\%] & & $\tilde{X}^{+}_{-}$ & $Q_\text{GR}$ [\%] & & $\tilde{X}^{+}_{-}$ & $Q_\text{GR}$ [\%] & & $\tilde{X}^{+}_{-}$ & $Q_\text{GR}$ [\%] & & $\tilde{X}^{+}_{-}$ & $Q_\text{GR}$ [\%]\\
\midrule
\vspace{4pt}$\delta\hat\beta_{2}$ & $[10^{0}]$ & P & & $0.3^{+3.8}_{-0.6}$ & $24$& & $0.01^{+0.56}_{-0.51}$ & $49$& & $-0.13^{+0.37}_{-0.33}$ & $74$& & $-0.32^{+0.44}_{-0.29}$ & $89$& & $0.00^{+0.49}_{-0.38}$ & $50$\\
\vspace{4pt}$\delta\hat\beta_{3}$ & $[10^{0}]$ & P & & $3.3^{+3.6}_{-3.4}$ & $5.6$& & $0.31^{+0.81}_{-0.63}$ & $21$& & $-0.20^{+0.42}_{-0.36}$ & $80$& & $-0.41^{+0.58}_{-0.36}$ & $90$& & $-0.02^{+0.74}_{-0.57}$ & $52$\\
\vspace{4pt}$\delta\hat\alpha_{2}$ & $[10^{0}]$ & P & & $0^{+14}_{-3}$ & $37$& & $0.78^{+0.85}_{-0.73}$ & $4$& & $0.42^{+0.69}_{-0.64}$ & $14$& & $0.01^{+0.71}_{-0.65}$ & $49$& & $-0.04^{+0.70}_{-0.66}$ & $55$\\
\vspace{4pt}$\delta\hat\alpha_{3}$ & $[10^{0}]$ & P & & $5^{+81}_{-33}$ & $29$& & $4.4^{+4.6}_{-3.8}$ & $2.1$& & $2.2^{+4.0}_{-3.0}$ & $12$& & $0.6^{+3.5}_{-2.9}$ & $38$& & $0.1^{+3.5}_{-2.3}$ & $49$\\
\vspace{4pt}$\delta\hat\alpha_{4}$ & $[10^{0}]$ & P & & $6^{+24}_{-30}$ & $33$& & $1.3^{+1.2}_{-1.1}$ & $2.2$& & $0.9^{+1.1}_{-1.0}$ & $7.6$& & $0.4^{+1.0}_{-1.1}$ & $26$& & $-0.09^{+0.97}_{-0.92}$ & $56$\\

\bottomrule
\end{tabular}
}
\end{table*}

\subsection{Parameterized tests of gravitational wave propagation}
\label{app:propagation}

\begin{figure*}[tb]
\includegraphics[width=0.95\textwidth]{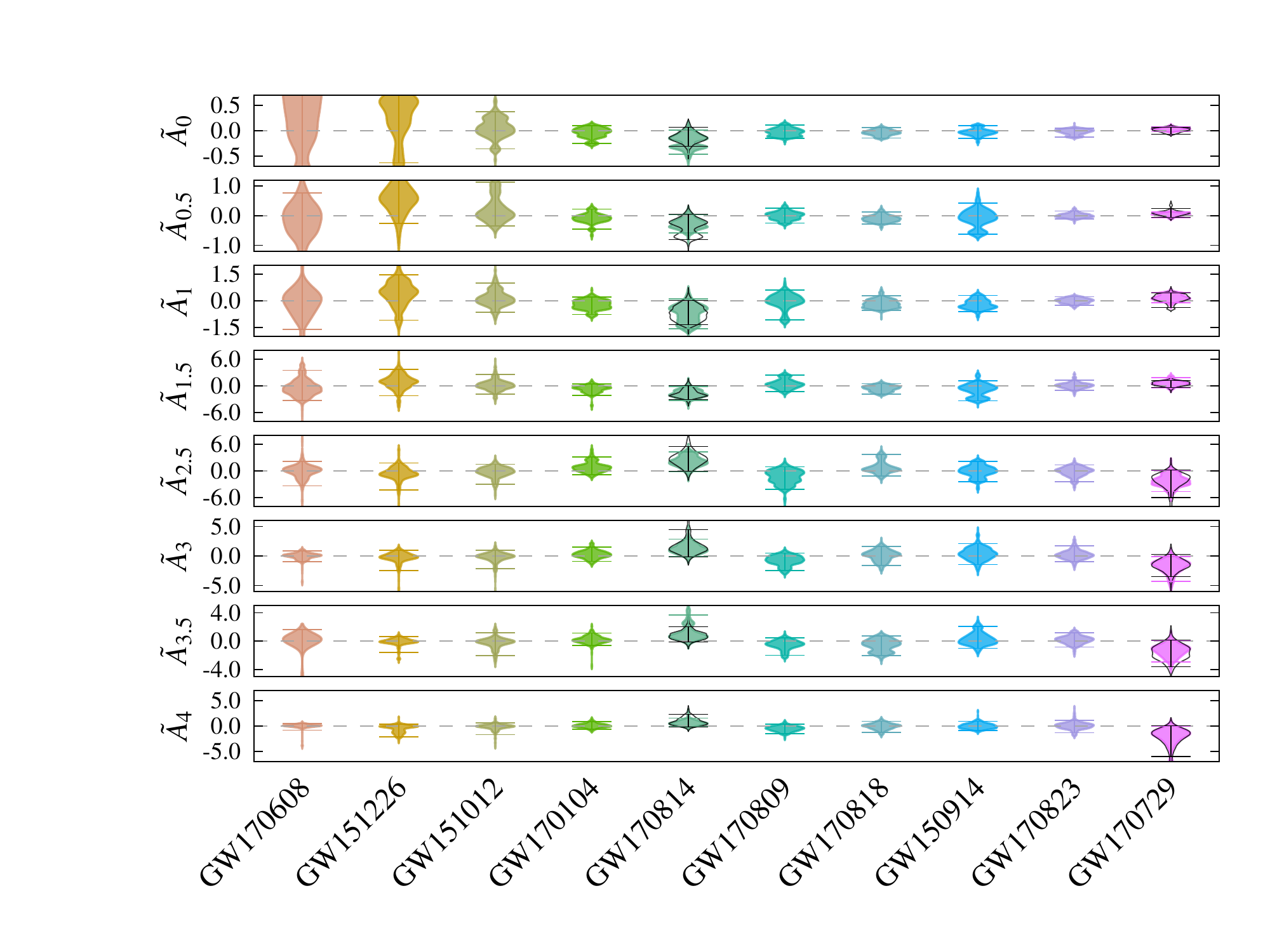}
\caption{Violin plots of the full posteriors for the modified dispersion relation parameter $A_\alpha$ for the individual binary black-hole events of GWTC-1~\cite{O2:Catalog}, with the $90\%$ credible interval around the median indicated. The events are ordered left-to-right by increasing median total mass. The filled violins are the \IMRP{} results, while the unfilled violins give the \SEOB{} results for GW170729 and GW170814. Here $\tilde{A}_\alpha \coloneqq A_\alpha/(10^{-19} \text{ peV}^{2-\alpha})$. We have let the much less constraining posteriors extend off the edges of the plots in order to show the more constraining posteriors in detail. The violin plots are scaled so that the maximum value of the posterior always has the same width, and these maximum values never occur off the plot.}
\label{fig:prop_indiv_results_violin}
\end{figure*}

Posteriors on $A_\alpha$ for individual events are shown in 
Fig.~\ref{fig:prop_indiv_results_violin}, with data for positive and 
negative $A_\alpha$ combined into one violin plot.  We provide results 
for all events with the \IMRP{} waveform model and also show results of 
the analysis with the \SEOB{} waveform model for GW170729 and GW170814.
In Table~\ref{tab:prop_results_wf_comparison} we compare the 90\% bounds on 
$A_\alpha$ and GR quantiles obtained with \IMRP{} and \SEOB{} for GW170729 and 
GW170814. We focus on these two events
because the GR quantiles obtained with \IMRP{} lie in the tails of the 
distributions, and we find that this remains true for most $\alpha$ values 
in the analysis with \SEOB. For GW170729 and $\alpha \in \{0, 0.5, 1\}$, the 
GR quantiles obtained using the two waveforms differ by factors of $\sim 2$; 
the two waveforms give values that are in much closer agreement for the other 
cases.

Additionally, for the GW151012 event and certain $\alpha$ values, a technical issue with our
computation of the likelihood meant that specific points with relatively large
values of $A_\alpha$ had to be manually removed from the posterior distribution.
In particular, for computational efficiency, the likelihood is calculated on as
short a segment of data as practical, with duration set by the longest waveform
to be sampled. Large values of $A_\alpha$ yield highly dispersed waveforms that
are pushed beyond the confines of the segment we use, causing the waveform
templates to wrap around the boundaries. This invalidates the assumptions
underlying our likelihood computation and causes an artificial enhancement of
the SNR as reported by the analysis. As expected, recomputing the SNRs for
these points on a segment that properly fits the waveform results in smaller
values that are consistent with noise. Therefore, we exclude from our analysis
parameter values yielding waveforms that would not be contained by the data
segment used, which is equivalent to using a stricter prior on $A_\alpha$.
Failure to do this may result in the appearance of outliers with spuriously
high likelihood for large values of $A_\alpha$, as we have seen in our own
analysis.

\begin{table}
\caption{\label{tab:prop_results_wf_comparison}
90\% credible level upper bounds on the absolute value of the modified dispersion relation parameter $A_\alpha$, as well as the GR quantiles $Q_\text{GR}$ for \IMRP{} (P) and \SEOB{} (S) runs. The $<$ and $>$ labels denote the bounds for $A_\alpha < 0$ and $>0$, respectively, with the given scalings and $\bar{A}_\alpha \coloneqq A_\alpha/\text{eV}^{2-\alpha}$.}
\centering
\scalebox{0.88}{
\begin{tabular}{*{16}{c}}
\toprule
 & & & \sixc{GW170729} & & \sixc{GW170814}\\
\cline{4-9}
\cline{11-16}
\twoc{Quantity} & & \twoc{$<$} & \twoc{$>$} & \twoc{$Q_\text{GR}$ [\%]} & & \twoc{$<$} & \twoc{$>$} & \twoc{$Q_\text{GR}$ [\%]}\\
& & & P & S & P & S & P & S & & P & S & P & S & P & S\\
\midrule
$\bar{A}_{0}$ & [$10^{-44}$] & & $0.29$ & $0.67$ & $0.64$ & $0.67$ & $17$ & $39$ & $\phantom{\Bigl(}$ & $4.0$ & $2.7$ & $1.1$ & $1.4$ & $94$ & $87$\\
$\bar{A}_{0.5}$ & [$10^{-38}$] & & $0.93$ & $0.86$ & $1.1$ & $1.8$ & $26$ & $16$ & $\phantom{\Bigl(}$ & $5.2$ & $7.6$ & $1.8$ & $1.8$ & $92$ & $92$\\
$\bar{A}_{1}$ & [$10^{-32}$] & & $2.1$ & $4.6$ & $4.5$ & $4.2$ & $16$ & $26$ & $\phantom{\Bigl(}$ & $15$ & $12$ & $5.2$ & $4.0$ & $93$ & $95$\\
$\bar{A}_{1.5}$ & [$10^{-25}$] & & $0.79$ & $0.74$ & $1.5$ & $1.2$ & $18$ & $17$ & $\phantom{\Bigl(}$ & $3.0$ & $2.8$ & $1.3$ & $1.0$ & $96$ & $94$\\
$\bar{A}_{2.5}$ & [$10^{-13}$] & & $4.2$ & $4.7$ & $1.4$ & $1.6$ & $94$ & $92$ & $\phantom{\Bigl(}$ & $1.3$ & $1.2$ & $3.9$ & $4.7$ & $5.7$ & $5.1$\\
$\bar{A}_{3}$ & [$10^{-8}$] & & $36$ & $30$ & $8.6$ & $16$ & $96$ & $92$ & $\phantom{\Bigl(}$ & $7.4$ & $8.3$ & $25$ & $36$ & $6.2$ & $6.4$\\
$\bar{A}_{3.5}$ & [$10^{-2}$] & & $26$ & $32$ & $8.9$ & $12$ & $94$ & $94$ & $\phantom{\Bigl(}$ & $5.4$ & $5.9$ & $29$ & $17$ & $7.5$ & $8.5$\\
$\bar{A}_{4}$ & [$10^{4}$] & & $43$ & $47$ & $7.1$ & $9.1$ & $95$ & $95$ & $\phantom{\Bigl(}$ & $4.2$ & $4.7$ & $13$ & $20$ & $9.6$ & $11$\\
\bottomrule
\end{tabular}}
\end{table}

\bibliography{cbc-group}

\begin{thebibliography}{134}%
\makeatletter
\providecommand \@ifxundefined [1]{%
 \@ifx{#1\undefined}
}%
\providecommand \@ifnum [1]{%
 \ifnum #1\expandafter \@firstoftwo
 \else \expandafter \@secondoftwo
 \fi
}%
\providecommand \@ifx [1]{%
 \ifx #1\expandafter \@firstoftwo
 \else \expandafter \@secondoftwo
 \fi
}%
\providecommand \natexlab [1]{#1}%
\providecommand \enquote  [1]{``#1''}%
\providecommand \bibnamefont  [1]{#1}%
\providecommand \bibfnamefont [1]{#1}%
\providecommand \citenamefont [1]{#1}%
\providecommand \href@noop [0]{\@secondoftwo}%
\providecommand \href [0]{\begingroup \@sanitize@url \@href}%
\providecommand \@href[1]{\@@startlink{#1}\@@href}%
\providecommand \@@href[1]{\endgroup#1\@@endlink}%
\providecommand \@sanitize@url [0]{\catcode `\\12\catcode `\$12\catcode
  `\&12\catcode `\#12\catcode `\^12\catcode `\_12\catcode `\%12\relax}%
\providecommand \@@startlink[1]{}%
\providecommand \@@endlink[0]{}%
\providecommand \url  [0]{\begingroup\@sanitize@url \@url }%
\providecommand \@url [1]{\endgroup\@href {#1}{\urlprefix }}%
\providecommand \urlprefix  [0]{URL }%
\providecommand \Eprint [0]{\href }%
\@ifxundefined \urlstyle {%
  \providecommand \doi  [0]{\begingroup \@sanitize@url \@doi}%
  \providecommand \@doi [1]{\endgroup \@@startlink {\doibase
  #1}doi:\discretionary {}{}{}#1\@@endlink }%
}{%
  \providecommand \doi  [0]{doi:\discretionary{}{}{}\begingroup
  \urlstyle{rm}\Url }%
}%
\providecommand \doibase [0]{http://dx.doi.org/}%
\providecommand \Doi [0]{\begingroup \@sanitize@url \@Doi }%
\providecommand \@Doi  [1]{\endgroup\@@startlink{\doibase#1}\@@Doi}%
\providecommand \@@Doi [1]{#1\@@endlink}%
\providecommand \selectlanguage [0]{\@gobble}%
\providecommand \bibinfo  [0]{\@secondoftwo}%
\providecommand \bibfield  [0]{\@secondoftwo}%
\providecommand \translation [1]{[#1]}%
\providecommand \BibitemOpen [0]{}%
\providecommand \bibitemStop [0]{}%
\providecommand \bibitemNoStop [0]{.\EOS\space}%
\providecommand \EOS [0]{\spacefactor3000\relax}%
\providecommand \BibitemShut  [1]{\csname bibitem#1\endcsname}%
\bibitem [{\citenamefont {Will}(2014)}]{lrr-2014-4}%
  \BibitemOpen
  \bibfield  {author} {\bibinfo {author} {\bibfnamefont {C.~M.}\ \bibnamefont
  {Will}},\ }\bibfield  {title} {\enquote {\bibinfo {title} {{The Confrontation
  between General Relativity and Experiment}},}\ }\Doi {10.12942/lrr-2014-4}
  {\bibfield  {journal} {\bibinfo  {journal} {Living Rev. Relativity},\
  }\textbf {\bibinfo {volume} {17}},\ \bibinfo {pages} {4} (\bibinfo {year}
  {2014})},\ \Eprint {http://arxiv.org/abs/1403.7377} {arXiv:1403.7377 [gr-qc]}
  \BibitemShut {NoStop}%
\bibitem [{\citenamefont {Aasi}\ \emph {et~al.}(2015)\citenamefont {Aasi} \emph
  {et~al.}}]{TheLIGOScientific:2014jea}%
  \BibitemOpen
  \bibfield  {author} {\bibinfo {author} {\bibfnamefont {J.}~\bibnamefont
  {Aasi}} \emph {et~al.} (\bibinfo {collaboration} {LIGO Scientific
  Collaboration}),\ }\bibfield  {title} {\enquote {\bibinfo {title} {{Advanced
  LIGO}},}\ }\Doi {10.1088/0264-9381/32/7/074001} {\bibfield  {journal}
  {\bibinfo  {journal} {Classical Quantum Gravity},\ }\textbf {\bibinfo
  {volume} {32}},\ \bibinfo {pages} {074001} (\bibinfo {year} {2015})},\
  \Eprint {http://arxiv.org/abs/1411.4547} {arXiv:1411.4547 [gr-qc]}
  \BibitemShut {NoStop}%
\bibitem [{\citenamefont {Acernese}\ \emph {et~al.}(2015)\citenamefont
  {Acernese} \emph {et~al.}}]{TheVirgo:2014hva}%
  \BibitemOpen
  \bibfield  {author} {\bibinfo {author} {\bibfnamefont {F.}~\bibnamefont
  {Acernese}} \emph {et~al.} (\bibinfo {collaboration} {Virgo Collaboration}),\
  }\bibfield  {title} {\enquote {\bibinfo {title} {{Advanced Virgo: a
  second-generation interferometric gravitational wave detector}},}\ }\Doi
  {10.1088/0264-9381/32/2/024001} {\bibfield  {journal} {\bibinfo  {journal}
  {Classical Quantum Gravity},\ }\textbf {\bibinfo {volume} {32}},\ \bibinfo
  {pages} {024001} (\bibinfo {year} {2015})},\ \Eprint
  {http://arxiv.org/abs/1408.3978} {arXiv:1408.3978 [gr-qc]} \BibitemShut
  {NoStop}%
\bibitem [{\citenamefont {Abbott}\ \emph
  {et~al.}(2016){\natexlab{a}}\citenamefont {Abbott} \emph
  {et~al.}}]{GW150914:TGR}%
  \BibitemOpen
  \bibfield  {author} {\bibinfo {author} {\bibfnamefont {B.~P.}\ \bibnamefont
  {Abbott}} \emph {et~al.} (\bibinfo {collaboration} {LIGO Scientific
  Collaboration and Virgo Collaboration}),\ }\bibfield  {title} {\enquote
  {\bibinfo {title} {{Tests of general relativity with GW150914}},}\ }\Doi
  {10.1103/PhysRevLett.116.221101} {\bibfield  {journal} {\bibinfo  {journal}
  {Phys. Rev. Lett.},\ }\textbf {\bibinfo {volume} {116}},\ \bibinfo {pages}
  {221101} (\bibinfo {year} {2016}{\natexlab{a}})},\ \bibinfo {note}
  {\href{http://dx.doi.org/10.1103/PhysRevLett.121.129902}{{\textbf{121}},
  129902(E) (2018)}},\ \Eprint {http://arxiv.org/abs/1602.03841}
  {arXiv:1602.03841 [gr-qc]} \BibitemShut {NoStop}%
\bibitem [{\citenamefont {Abbott}\ \emph
  {et~al.}(2016){\natexlab{b}}\citenamefont {Abbott} \emph {et~al.}}]{O1:BBH}%
  \BibitemOpen
  \bibfield  {author} {\bibinfo {author} {\bibfnamefont {B.~P.}\ \bibnamefont
  {Abbott}} \emph {et~al.} (\bibinfo {collaboration} {LIGO Scientific
  Collaboration and Virgo Collaboration}),\ }\bibfield  {title} {\enquote
  {\bibinfo {title} {{Binary Black Hole Mergers in the first Advanced LIGO
  Observing Run}},}\ }\Doi {10.1103/PhysRevX.6.041015} {\bibfield  {journal}
  {\bibinfo  {journal} {Phys. Rev. X},\ }\textbf {\bibinfo {volume} {6}},\
  \bibinfo {pages} {041015} (\bibinfo {year} {2016}{\natexlab{b}})},\ \bibinfo
  {note} {\href{http://dx.doi.org/10.1103/PhysRevX.8.039903}{{\bf{8}},
  039903(E) (2018)}},\ \Eprint {http://arxiv.org/abs/1606.04856}
  {arXiv:1606.04856 [gr-qc]} \BibitemShut {NoStop}%
\bibitem [{\citenamefont {Abbott}\ \emph
  {et~al.}(2017){\natexlab{a}}\citenamefont {Abbott} \emph
  {et~al.}}]{GW170104}%
  \BibitemOpen
  \bibfield  {author} {\bibinfo {author} {\bibfnamefont {B.~P.}\ \bibnamefont
  {Abbott}} \emph {et~al.} (\bibinfo {collaboration} {LIGO Scientific
  Collaboration and Virgo Collaboration}),\ }\bibfield  {title} {\enquote
  {\bibinfo {title} {{GW170104: Observation of a 50-Solar-Mass Binary Black
  Hole Coalescence at Redshift 0.2}},}\ }\Doi {10.1103/PhysRevLett.118.221101}
  {\bibfield  {journal} {\bibinfo  {journal} {Phys. Rev. Lett.},\ }\textbf
  {\bibinfo {volume} {118}},\ \bibinfo {pages} {221101} (\bibinfo {year}
  {2017}{\natexlab{a}})},\ \bibinfo {note}
  {\href{http://dx.doi.org/10.1103/PhysRevLett.121.129901}{{\bf{121}},
  129901(E) (2018)}},\ \Eprint {http://arxiv.org/abs/1706.01812}
  {arXiv:1706.01812 [gr-qc]} \BibitemShut {NoStop}%
\bibitem [{\citenamefont {Abbott}\ \emph
  {et~al.}(2017){\natexlab{b}}\citenamefont {Abbott} \emph
  {et~al.}}]{GW170814paper}%
  \BibitemOpen
  \bibfield  {author} {\bibinfo {author} {\bibfnamefont {B.~P.}\ \bibnamefont
  {Abbott}} \emph {et~al.} (\bibinfo {collaboration} {LIGO Scientific
  Collaboration and Virgo Collaboration}),\ }\bibfield  {title} {\enquote
  {\bibinfo {title} {{GW170814: A Three-Detector Observation of Gravitational
  Waves from a Binary Black Hole Coalescence}},}\ }\Doi
  {10.1103/PhysRevLett.119.141101} {\bibfield  {journal} {\bibinfo  {journal}
  {Phys. Rev. Lett.},\ }\textbf {\bibinfo {volume} {119}},\ \bibinfo {pages}
  {141101} (\bibinfo {year} {2017}{\natexlab{b}})},\ \Eprint
  {http://arxiv.org/abs/1709.09660} {arXiv:1709.09660 [gr-qc]} \BibitemShut
  {NoStop}%
\bibitem [{\citenamefont {Abbott}\ \emph
  {et~al.}(2019){\natexlab{a}}\citenamefont {Abbott} \emph {et~al.}}]{bns-tgr}%
  \BibitemOpen
  \bibfield  {author} {\bibinfo {author} {\bibfnamefont {B.~P.}\ \bibnamefont
  {Abbott}} \emph {et~al.} (\bibinfo {collaboration} {LIGO Scientific
  Collaboration and Virgo Collaboration}),\ }\bibfield  {title} {\enquote
  {\bibinfo {title} {{Tests of General Relativity with GW170817}},}\ }\Doi
  {10.1103/PhysRevLett.123.011102} {\bibfield  {journal} {\bibinfo  {journal}
  {Phys. Rev. Lett.},\ }\textbf {\bibinfo {volume} {123}},\ \bibinfo {pages}
  {011102} (\bibinfo {year} {2019}{\natexlab{a}})},\ \Eprint
  {http://arxiv.org/abs/1811.00364} {arXiv:1811.00364 [gr-qc]} \BibitemShut
  {NoStop}%
\bibitem [{\citenamefont {{LIGO Scientific Collaboration and Virgo
  Collaboration}}(2018){\natexlab{a}}}]{GWOSC:GWTC}%
  \BibitemOpen
  \bibfield  {author} {\bibinfo {author} {\bibnamefont {{LIGO Scientific
  Collaboration and Virgo Collaboration}}},\ }\href@noop {} {\enquote {\bibinfo
  {title} {{GWTC-1}},}\ }\bibinfo {howpublished}
  {\href{https://doi.org/10.7935/82H3-HH23}{https://doi.org/10.7935/82H3-HH23}}
  (\bibinfo {year} {2018}{\natexlab{a}})\BibitemShut {NoStop}%
\bibitem [{\citenamefont {{Abbott}}\ \emph {et~al.}(2016)\citenamefont
  {{Abbott}} \emph {et~al.}}]{GW150914_paper}%
  \BibitemOpen
  \bibfield  {author} {\bibinfo {author} {\bibfnamefont {B.~P.}\ \bibnamefont
  {{Abbott}}} \emph {et~al.} (\bibinfo {collaboration} {LIGO Scientific
  Collaboration and Virgo Collaboration}),\ }\bibfield  {title} {\enquote
  {\bibinfo {title} {{Observation of Gravitational Waves from a Binary Black
  Hole Merger}},}\ }\Doi {10.1103/PhysRevLett.116.061102} {\bibfield  {journal}
  {\bibinfo  {journal} {Phys. Rev. Lett.},\ }\textbf {\bibinfo {volume}
  {116}},\ \bibinfo {eid} {061102} (\bibinfo {year} {2016})},\ \Eprint
  {http://arxiv.org/abs/1602.03837} {arXiv:1602.03837 [gr-qc]} \BibitemShut
  {NoStop}%
\bibitem [{\citenamefont {Abbott}\ \emph
  {et~al.}(2016){\natexlab{a}}\citenamefont {Abbott} \emph
  {et~al.}}]{TheLIGOScientific:2016qqj}%
  \BibitemOpen
  \bibfield  {author} {\bibinfo {author} {\bibfnamefont {B.~P.}\ \bibnamefont
  {Abbott}} \emph {et~al.} (\bibinfo {collaboration} {LIGO Scientific
  Collaboration and Virgo Collaboration}),\ }\bibfield  {title} {\enquote
  {\bibinfo {title} {{GW150914: First results from the search for binary black
  hole coalescence with Advanced LIGO}},}\ }\Doi {10.1103/PhysRevD.93.122003}
  {\bibfield  {journal} {\bibinfo  {journal} {Phys. Rev. D},\ }\textbf
  {\bibinfo {volume} {93}},\ \bibinfo {pages} {122003} (\bibinfo {year}
  {2016}{\natexlab{a}})},\ \Eprint {http://arxiv.org/abs/1602.03839}
  {arXiv:1602.03839 [gr-qc]} \BibitemShut {NoStop}%
\bibitem [{\citenamefont {Abbott}\ \emph
  {et~al.}(2016){\natexlab{b}}\citenamefont {Abbott} \emph
  {et~al.}}]{GW151226}%
  \BibitemOpen
  \bibfield  {author} {\bibinfo {author} {\bibfnamefont {B.~P.}\ \bibnamefont
  {Abbott}} \emph {et~al.} (\bibinfo {collaboration} {LIGO Scientific
  Collaboration and Virgo Collaboration}),\ }\bibfield  {title} {\enquote
  {\bibinfo {title} {{GW151226: Observation of gravitational waves from a
  22-solar-mass binary black hole coalescence}},}\ }\Doi
  {10.1103/PhysRevLett.116.241103} {\bibfield  {journal} {\bibinfo  {journal}
  {Phys. Rev. Lett.},\ }\textbf {\bibinfo {volume} {116}},\ \bibinfo {pages}
  {241103} (\bibinfo {year} {2016}{\natexlab{b}})},\ \Eprint
  {http://arxiv.org/abs/1606.04855} {arXiv:1606.04855 [gr-qc]} \BibitemShut
  {NoStop}%
\bibitem [{\citenamefont {Abbott}\ \emph
  {et~al.}(2017){\natexlab{c}}\citenamefont {Abbott} \emph
  {et~al.}}]{GW170608}%
  \BibitemOpen
  \bibfield  {author} {\bibinfo {author} {\bibfnamefont {B.~P.}\ \bibnamefont
  {Abbott}} \emph {et~al.} (\bibinfo {collaboration} {LIGO Scientific
  Collaboration and Virgo Collaboration}),\ }\bibfield  {title} {\enquote
  {\bibinfo {title} {{GW170608: Observation of a 19 solar-mass binary black
  hole coalescence}},}\ }\Doi {10.3847/2041-8213/aa9f0c} {\bibfield  {journal}
  {\bibinfo  {journal} {Astrophys. J. Lett.},\ }\textbf {\bibinfo {volume}
  {851}},\ \bibinfo {pages} {L35} (\bibinfo {year} {2017}{\natexlab{c}})},\
  \Eprint {http://arxiv.org/abs/1711.05578} {arXiv:1711.05578 [astro-ph.HE]}
  \BibitemShut {NoStop}%
\bibitem [{\citenamefont {Abbott}\ \emph
  {et~al.}(2019){\natexlab{b}}\citenamefont {Abbott} \emph
  {et~al.}}]{O2:Catalog}%
  \BibitemOpen
  \bibfield  {author} {\bibinfo {author} {\bibfnamefont {B.~P.}\ \bibnamefont
  {Abbott}} \emph {et~al.} (\bibinfo {collaboration} {LIGO Scientific
  Collaboration and Virgo Collaboration}),\ }\bibfield  {title} {\enquote
  {\bibinfo {title} {{GWTC-1: A Gravitational-Wave Transient Catalog of Compact
  Binary Mergers Observed by LIGO and Virgo during the First and Second
  Observing Runs}},}\ }\Doi {10.1103/PhysRevX.9.031040} {\bibfield  {journal}
  {\bibinfo  {journal} {Phys. Rev. X},\ }\textbf {\bibinfo {volume} {9}},\
  \bibinfo {pages} {031040} (\bibinfo {year} {2019}{\natexlab{b}})},\ \Eprint
  {http://arxiv.org/abs/1811.12907} {arXiv:1811.12907 [astro-ph.HE]}
  \BibitemShut {NoStop}%
\bibitem [{\citenamefont {Abbott}\ \emph
  {et~al.}(2016){\natexlab{c}}\citenamefont {Abbott} \emph
  {et~al.}}]{TheLIGOScientific:2016uux}%
  \BibitemOpen
  \bibfield  {author} {\bibinfo {author} {\bibfnamefont {B.~P.}\ \bibnamefont
  {Abbott}} \emph {et~al.} (\bibinfo {collaboration} {LIGO Scientific
  Collaboration and Virgo Collaboration}),\ }\bibfield  {title} {\enquote
  {\bibinfo {title} {{Observing gravitational-wave transient GW150914 with
  minimal assumptions}},}\ }\Doi {10.1103/PhysRevD.93.122004} {\bibfield
  {journal} {\bibinfo  {journal} {Phys. Rev. D},\ }\textbf {\bibinfo {volume}
  {93}},\ \bibinfo {pages} {122004} (\bibinfo {year} {2016}{\natexlab{c}})},\
  \Eprint {http://arxiv.org/abs/1602.03843} {arXiv:1602.03843 [gr-qc]}
  \BibitemShut {NoStop}%
\bibitem [{\citenamefont {Vallisneri}\ and\ \citenamefont
  {Yunes}(2013)}]{Vallisneri:2013rc}%
  \BibitemOpen
  \bibfield  {author} {\bibinfo {author} {\bibfnamefont {M.}~\bibnamefont
  {Vallisneri}}\ and\ \bibinfo {author} {\bibfnamefont {N.}~\bibnamefont
  {Yunes}},\ }\bibfield  {title} {\enquote {\bibinfo {title} {{Stealth Bias in
  Gravitational-Wave Parameter Estimation}},}\ }\Doi
  {10.1103/PhysRevD.87.102002} {\bibfield  {journal} {\bibinfo  {journal}
  {Phys. Rev. D},\ }\textbf {\bibinfo {volume} {87}},\ \bibinfo {pages}
  {102002} (\bibinfo {year} {2013})},\ \Eprint {http://arxiv.org/abs/1301.2627}
  {arXiv:1301.2627 [gr-qc]} \BibitemShut {NoStop}%
\bibitem [{\citenamefont {Vitale}\ and\ \citenamefont
  {Del~Pozzo}(2014)}]{Vitale:2013bma}%
  \BibitemOpen
  \bibfield  {author} {\bibinfo {author} {\bibfnamefont {S.}~\bibnamefont
  {Vitale}}\ and\ \bibinfo {author} {\bibfnamefont {W.}~\bibnamefont
  {Del~Pozzo}},\ }\bibfield  {title} {\enquote {\bibinfo {title} {{How serious
  can the stealth bias be in gravitational wave parameter estimation?}}}\ }\Doi
  {10.1103/PhysRevD.89.022002} {\bibfield  {journal} {\bibinfo  {journal}
  {Phys. Rev. D},\ }\textbf {\bibinfo {volume} {89}},\ \bibinfo {pages}
  {022002} (\bibinfo {year} {2014})},\ \Eprint {http://arxiv.org/abs/1311.2057}
  {arXiv:1311.2057 [gr-qc]} \BibitemShut {NoStop}%
\bibitem [{\citenamefont {Okounkova}\ \emph {et~al.}(2017)\citenamefont
  {Okounkova}, \citenamefont {Stein}, \citenamefont {Scheel},\ and\
  \citenamefont {Hemberger}}]{Okounkova:2017yby}%
  \BibitemOpen
  \bibfield  {author} {\bibinfo {author} {\bibfnamefont {M.}~\bibnamefont
  {Okounkova}}, \bibinfo {author} {\bibfnamefont {L.~C.}\ \bibnamefont
  {Stein}}, \bibinfo {author} {\bibfnamefont {M.~A.}\ \bibnamefont {Scheel}}, \
  and\ \bibinfo {author} {\bibfnamefont {D.~A.}\ \bibnamefont {Hemberger}},\
  }\bibfield  {title} {\enquote {\bibinfo {title} {{Numerical binary black hole
  mergers in dynamical Chern-Simons gravity: Scalar field}},}\ }\Doi
  {10.1103/PhysRevD.96.044020} {\bibfield  {journal} {\bibinfo  {journal}
  {Phys. Rev. D},\ }\textbf {\bibinfo {volume} {96}},\ \bibinfo {pages}
  {044020} (\bibinfo {year} {2017})},\ \Eprint
  {http://arxiv.org/abs/1705.07924} {arXiv:1705.07924 [gr-qc]} \BibitemShut
  {NoStop}%
\bibitem [{\citenamefont {Witek}\ \emph {et~al.}(2019)\citenamefont {Witek},
  \citenamefont {Gualtieri}, \citenamefont {Pani},\ and\ \citenamefont
  {Sotiriou}}]{Witek:2018dmd}%
  \BibitemOpen
  \bibfield  {author} {\bibinfo {author} {\bibfnamefont {H.}~\bibnamefont
  {Witek}}, \bibinfo {author} {\bibfnamefont {L.}~\bibnamefont {Gualtieri}},
  \bibinfo {author} {\bibfnamefont {P.}~\bibnamefont {Pani}}, \ and\ \bibinfo
  {author} {\bibfnamefont {T.~P.}\ \bibnamefont {Sotiriou}},\ }\bibfield
  {title} {\enquote {\bibinfo {title} {{Black holes and binary mergers in
  scalar Gauss-Bonnet gravity: scalar field dynamics}},}\ }\Doi
  {10.1103/PhysRevD.99.064035} {\bibfield  {journal} {\bibinfo  {journal}
  {Phys. Rev. D},\ }\textbf {\bibinfo {volume} {99}},\ \bibinfo {pages}
  {064035} (\bibinfo {year} {2019})},\ \Eprint
  {http://arxiv.org/abs/1810.05177} {arXiv:1810.05177 [gr-qc]} \BibitemShut
  {NoStop}%
\bibitem [{\citenamefont {Okounkova}\ \emph {et~al.}(2019)\citenamefont
  {Okounkova}, \citenamefont {Stein}, \citenamefont {Scheel},\ and\
  \citenamefont {Teukolsky}}]{Okounkova:2019dfo}%
  \BibitemOpen
  \bibfield  {author} {\bibinfo {author} {\bibfnamefont {M.}~\bibnamefont
  {Okounkova}}, \bibinfo {author} {\bibfnamefont {L.~C.}\ \bibnamefont
  {Stein}}, \bibinfo {author} {\bibfnamefont {M.~A.}\ \bibnamefont {Scheel}}, \
  and\ \bibinfo {author} {\bibfnamefont {S.~A.}\ \bibnamefont {Teukolsky}},\
  }\bibfield  {title} {\enquote {\bibinfo {title} {{Numerical binary black hole
  collisions in dynamical Chern-Simons gravity}},}\ }\href@noop {} { (\bibinfo
  {year} {2019})},\ \Eprint {http://arxiv.org/abs/1906.08789} {arXiv:1906.08789
  [gr-qc]} \BibitemShut {NoStop}%
\bibitem [{\citenamefont {Bohé}\ \emph {et~al.}(2017)\citenamefont {Bohé}
  \emph {et~al.}}]{Bohe:2016gbl}%
  \BibitemOpen
  \bibfield  {author} {\bibinfo {author} {\bibfnamefont {A.}~\bibnamefont
  {Bohé}} \emph {et~al.},\ }\bibfield  {title} {\enquote {\bibinfo {title}
  {{Improved effective-one-body model of spinning, nonprecessing binary black
  holes for the era of gravitational-wave astrophysics with advanced
  detectors}},}\ }\Doi {10.1103/PhysRevD.95.044028} {\bibfield  {journal}
  {\bibinfo  {journal} {Phys. Rev. D},\ }\textbf {\bibinfo {volume} {95}},\
  \bibinfo {pages} {044028} (\bibinfo {year} {2017})},\ \Eprint
  {http://arxiv.org/abs/1611.03703} {arXiv:1611.03703 [gr-qc]} \BibitemShut
  {NoStop}%
\bibitem [{\citenamefont {Khan}\ \emph {et~al.}(2016)\citenamefont {Khan},
  \citenamefont {Husa}, \citenamefont {Hannam}, \citenamefont {Ohme},
  \citenamefont {Pürrer}, \citenamefont {Jiménez~Forteza},\ and\
  \citenamefont {Bohé}}]{Khan:2015jqa}%
  \BibitemOpen
  \bibfield  {author} {\bibinfo {author} {\bibfnamefont {S.}~\bibnamefont
  {Khan}}, \bibinfo {author} {\bibfnamefont {S.}~\bibnamefont {Husa}}, \bibinfo
  {author} {\bibfnamefont {M.}~\bibnamefont {Hannam}}, \bibinfo {author}
  {\bibfnamefont {F.}~\bibnamefont {Ohme}}, \bibinfo {author} {\bibfnamefont
  {M.}~\bibnamefont {Pürrer}}, \bibinfo {author} {\bibfnamefont
  {X.}~\bibnamefont {Jiménez~Forteza}}, \ and\ \bibinfo {author}
  {\bibfnamefont {A.}~\bibnamefont {Bohé}},\ }\bibfield  {title} {\enquote
  {\bibinfo {title} {{Frequency-domain gravitational waves from non-precessing
  black-hole binaries. II. A phenomenological model for the advanced detector
  era}},}\ }\Doi {10.1103/PhysRevD.93.044007} {\bibfield  {journal} {\bibinfo
  {journal} {Phys. Rev. D},\ }\textbf {\bibinfo {volume} {93}},\ \bibinfo
  {pages} {044007} (\bibinfo {year} {2016})},\ \Eprint
  {http://arxiv.org/abs/1508.07253} {arXiv:1508.07253 [gr-qc]} \BibitemShut
  {NoStop}%
\bibitem [{\citenamefont {Blackman}\ \emph
  {et~al.}(2017){\natexlab{a}}\citenamefont {Blackman}, \citenamefont {Field},
  \citenamefont {Scheel}, \citenamefont {Galley}, \citenamefont {Hemberger},
  \citenamefont {Schmidt},\ and\ \citenamefont {Smith}}]{Blackman:2017dfb}%
  \BibitemOpen
  \bibfield  {author} {\bibinfo {author} {\bibfnamefont {J.}~\bibnamefont
  {Blackman}}, \bibinfo {author} {\bibfnamefont {S.~E.}\ \bibnamefont {Field}},
  \bibinfo {author} {\bibfnamefont {M.~A.}\ \bibnamefont {Scheel}}, \bibinfo
  {author} {\bibfnamefont {C.~R.}\ \bibnamefont {Galley}}, \bibinfo {author}
  {\bibfnamefont {D.~A.}\ \bibnamefont {Hemberger}}, \bibinfo {author}
  {\bibfnamefont {P.}~\bibnamefont {Schmidt}}, \ and\ \bibinfo {author}
  {\bibfnamefont {R.}~\bibnamefont {Smith}},\ }\bibfield  {title} {\enquote
  {\bibinfo {title} {{A Surrogate Model of Gravitational Waveforms from
  Numerical Relativity Simulations of Precessing Binary Black Hole Mergers}},}\
  }\Doi {10.1103/PhysRevD.95.104023} {\bibfield  {journal} {\bibinfo  {journal}
  {Phys. Rev. D},\ }\textbf {\bibinfo {volume} {95}},\ \bibinfo {pages}
  {104023} (\bibinfo {year} {2017}{\natexlab{a}})},\ \Eprint
  {http://arxiv.org/abs/1701.00550} {arXiv:1701.00550 [gr-qc]} \BibitemShut
  {NoStop}%
\bibitem [{\citenamefont {Khan}\ \emph {et~al.}(2019)\citenamefont {Khan},
  \citenamefont {Chatziioannou}, \citenamefont {Hannam},\ and\ \citenamefont
  {Ohme}}]{Khan:2018fmp}%
  \BibitemOpen
  \bibfield  {author} {\bibinfo {author} {\bibfnamefont {S.}~\bibnamefont
  {Khan}}, \bibinfo {author} {\bibfnamefont {K.}~\bibnamefont {Chatziioannou}},
  \bibinfo {author} {\bibfnamefont {M.}~\bibnamefont {Hannam}}, \ and\ \bibinfo
  {author} {\bibfnamefont {F.}~\bibnamefont {Ohme}},\ }\bibfield  {title}
  {\enquote {\bibinfo {title} {{Phenomenological model for the
  gravitational-wave signal from precessing binary black holes with two-spin
  effects}},}\ }\Doi {10.1103/PhysRevD.100.024059} {\bibfield  {journal}
  {\bibinfo  {journal} {Phys. Rev. D},\ }\textbf {\bibinfo {volume} {100}},\
  \bibinfo {pages} {024059} (\bibinfo {year} {2019})},\ \Eprint
  {http://arxiv.org/abs/1809.10113} {arXiv:1809.10113 [gr-qc]} \BibitemShut
  {NoStop}%
\bibitem [{\citenamefont {Abbott}\ \emph
  {et~al.}(2017){\natexlab{d}}\citenamefont {Abbott} \emph
  {et~al.}}]{Abbott:2016wiq}%
  \BibitemOpen
  \bibfield  {author} {\bibinfo {author} {\bibfnamefont {B.~P.}\ \bibnamefont
  {Abbott}} \emph {et~al.} (\bibinfo {collaboration} {LIGO Scientific
  Collaboration and Virgo Collaboration}),\ }\bibfield  {title} {\enquote
  {\bibinfo {title} {{Effects of waveform model systematics on the
  interpretation of GW150914}},}\ }\Doi {10.1088/1361-6382/aa6854} {\bibfield
  {journal} {\bibinfo  {journal} {Classical Quantum Gravity},\ }\textbf
  {\bibinfo {volume} {34}},\ \bibinfo {pages} {104002} (\bibinfo {year}
  {2017}{\natexlab{d}})},\ \Eprint {http://arxiv.org/abs/1611.07531}
  {arXiv:1611.07531 [gr-qc]} \BibitemShut {NoStop}%
\bibitem [{\citenamefont {{LIGO Scientific Collaboration and Virgo
  Collaboration}}(2019)}]{GWOSC:O2TGR}%
  \BibitemOpen
  \bibfield  {author} {\bibinfo {author} {\bibnamefont {{LIGO Scientific
  Collaboration and Virgo Collaboration}}},\ }\href@noop {} {\enquote {\bibinfo
  {title} {{Data release for testing GR with GWTC-1}},}\ }\bibinfo
  {howpublished}
  {\href{https://dcc.ligo.org/LIGO-P1900087/public}{https://dcc.ligo.org/LIGO-P1900087/public}}
  (\bibinfo {year} {2019})\BibitemShut {NoStop}%
\bibitem [{\citenamefont {{LIGO Scientific Collaboration, Virgo
  Collaboration}}(2018)}]{GWOSC}%
  \BibitemOpen
  \bibfield  {author} {\bibinfo {author} {\bibnamefont {{LIGO Scientific
  Collaboration, Virgo Collaboration}}},\ }\href@noop {} {\enquote {\bibinfo
  {title} {{Gravitational Wave Open Science Center}},}\ }\bibinfo
  {howpublished}
  {\href{https://www.gw-openscience.org}{https://www.gw-openscience.org}}
  (\bibinfo {year} {2018})\BibitemShut {NoStop}%
\bibitem [{\citenamefont {Karki}\ \emph {et~al.}(2016)\citenamefont {Karki}
  \emph {et~al.}}]{Karki:2016pht}%
  \BibitemOpen
  \bibfield  {author} {\bibinfo {author} {\bibfnamefont {S.}~\bibnamefont
  {Karki}} \emph {et~al.},\ }\bibfield  {title} {\enquote {\bibinfo {title}
  {{The Advanced LIGO Photon Calibrators}},}\ }\Doi {10.1063/1.4967303}
  {\bibfield  {journal} {\bibinfo  {journal} {Rev. Sci. Instrum.},\ }\textbf
  {\bibinfo {volume} {87}},\ \bibinfo {pages} {114503} (\bibinfo {year}
  {2016})},\ \Eprint {http://arxiv.org/abs/1608.05055} {arXiv:1608.05055
  [astro-ph.IM]} \BibitemShut {NoStop}%
\bibitem [{\citenamefont {Cahillane}\ \emph {et~al.}(2017)\citenamefont
  {Cahillane} \emph {et~al.}}]{Cahillane:2017vkb}%
  \BibitemOpen
  \bibfield  {author} {\bibinfo {author} {\bibfnamefont {C.}~\bibnamefont
  {Cahillane}} \emph {et~al.},\ }\bibfield  {title} {\enquote {\bibinfo {title}
  {{Calibration uncertainty for Advanced LIGO’s first and second observing
  runs}},}\ }\Doi {10.1103/PhysRevD.96.102001} {\bibfield  {journal} {\bibinfo
  {journal} {Phys. Rev. D},\ }\textbf {\bibinfo {volume} {96}},\ \bibinfo
  {pages} {102001} (\bibinfo {year} {2017})},\ \Eprint
  {http://arxiv.org/abs/1708.03023} {arXiv:1708.03023 [astro-ph.IM]}
  \BibitemShut {NoStop}%
\bibitem [{\citenamefont {Viets}\ \emph {et~al.}(2018)\citenamefont {Viets}
  \emph {et~al.}}]{Viets:2017yvy}%
  \BibitemOpen
  \bibfield  {author} {\bibinfo {author} {\bibfnamefont {A.}~\bibnamefont
  {Viets}} \emph {et~al.},\ }\bibfield  {title} {\enquote {\bibinfo {title}
  {{Reconstructing the calibrated strain signal in the Advanced LIGO
  detectors}},}\ }\Doi {10.1088/1361-6382/aab658} {\bibfield  {journal}
  {\bibinfo  {journal} {Classical Quantum Gravity},\ }\textbf {\bibinfo
  {volume} {35}},\ \bibinfo {pages} {095015} (\bibinfo {year} {2018})},\
  \Eprint {http://arxiv.org/abs/1710.09973} {arXiv:1710.09973 [astro-ph.IM]}
  \BibitemShut {NoStop}%
\bibitem [{\citenamefont {Estevez}\ \emph {et~al.}(2018)\citenamefont {Estevez}
  \emph {et~al.}}]{VIR-0362A-18}%
  \BibitemOpen
  \bibfield  {author} {\bibinfo {author} {\bibfnamefont {D.}~\bibnamefont
  {Estevez}} \emph {et~al.},\ }\href
  {{https://tds.virgo-gw.eu/?content=3&r=14363}} {\emph {\bibinfo {title}
  {{V1O2Repro2A h(t) reprocessing for Virgo O2 data}}}},\ \bibinfo {type}
  {Tech. Rep.}\ \bibinfo {number} {{VIR-0362A-18}}\ (\bibinfo  {institution}
  {Virgo project},\ \bibinfo {year} {2018})\BibitemShut {NoStop}%
\bibitem [{\citenamefont {Acernese}\ \emph {et~al.}(2018)\citenamefont
  {Acernese} \emph {et~al.}}]{Acernese:2018bfl}%
  \BibitemOpen
  \bibfield  {author} {\bibinfo {author} {\bibfnamefont {F.}~\bibnamefont
  {Acernese}} \emph {et~al.} (\bibinfo {collaboration} {Virgo Collaboration}),\
  }\bibfield  {title} {\enquote {\bibinfo {title} {{Calibration of Advanced
  Virgo and Reconstruction of the Gravitational Wave Signal $h(t)$ during the
  Observing Run O2}},}\ }\Doi {10.1088/1361-6382/aadf1a} {\bibfield  {journal}
  {\bibinfo  {journal} {Classical Quantum Gravity},\ }\textbf {\bibinfo
  {volume} {35}},\ \bibinfo {pages} {205004} (\bibinfo {year} {2018})},\
  \Eprint {http://arxiv.org/abs/1807.03275} {arXiv:1807.03275 [gr-qc]}
  \BibitemShut {NoStop}%
\bibitem [{\citenamefont {Farr}\ \emph {et~al.}(2015)\citenamefont {Farr},
  \citenamefont {Farr},\ and\ \citenamefont
  {Littenberg}}]{SplineCalMarg-T1400682}%
  \BibitemOpen
  \bibfield  {author} {\bibinfo {author} {\bibfnamefont {W.~M.}\ \bibnamefont
  {Farr}}, \bibinfo {author} {\bibfnamefont {B.}~\bibnamefont {Farr}}, \ and\
  \bibinfo {author} {\bibfnamefont {T.}~\bibnamefont {Littenberg}},\ }\href
  {https://dcc.ligo.org/T1400682/public} {\emph {\bibinfo {title} {Modelling
  Calibration Errors In CBC Waveforms}}},\ \bibinfo {type} {Tech. Rep.}\
  \bibinfo {number} {{LIGO}-T1400682}\ (\bibinfo  {institution} {{LIGO}
  Project},\ \bibinfo {year} {2015})\BibitemShut {NoStop}%
\bibitem [{\citenamefont {Veitch}\ \emph {et~al.}(2015)\citenamefont {Veitch}
  \emph {et~al.}}]{Veitch:2014wba}%
  \BibitemOpen
  \bibfield  {author} {\bibinfo {author} {\bibfnamefont {J.}~\bibnamefont
  {Veitch}} \emph {et~al.},\ }\bibfield  {title} {\enquote {\bibinfo {title}
  {{Robust parameter estimation for compact binaries with ground-based
  gravitational-wave observations using the LALInference software library}},}\
  }\Doi {10.1103/PhysRevD.91.042003} {\bibfield  {journal} {\bibinfo  {journal}
  {Phys. Rev. D},\ }\textbf {\bibinfo {volume} {91}},\ \bibinfo {pages}
  {042003} (\bibinfo {year} {2015})},\ \Eprint {http://arxiv.org/abs/1409.7215}
  {arXiv:1409.7215 [gr-qc]} \BibitemShut {NoStop}%
\bibitem [{\citenamefont {Abbott}\ \emph
  {et~al.}(2019){\natexlab{c}}\citenamefont {Abbott} \emph
  {et~al.}}]{Abbott:2018wiz}%
  \BibitemOpen
  \bibfield  {author} {\bibinfo {author} {\bibfnamefont {B.~P.}\ \bibnamefont
  {Abbott}} \emph {et~al.} (\bibinfo {collaboration} {LIGO Scientific
  Collaboration and Virgo Collaboration}),\ }\bibfield  {title} {\enquote
  {\bibinfo {title} {{Properties of the binary neutron star merger
  GW170817}},}\ }\Doi {10.1103/PhysRevX.9.011001} {\bibfield  {journal}
  {\bibinfo  {journal} {Phys. Rev. X},\ }\textbf {\bibinfo {volume} {9}},\
  \bibinfo {pages} {011001} (\bibinfo {year} {2019}{\natexlab{c}})},\ \Eprint
  {http://arxiv.org/abs/1805.11579} {arXiv:1805.11579 [gr-qc]} \BibitemShut
  {NoStop}%
\bibitem [{\citenamefont {{Abbott}}\ \emph {et~al.}(2017)\citenamefont
  {{Abbott}} \emph {et~al.}}]{DetectionPaper}%
  \BibitemOpen
  \bibfield  {author} {\bibinfo {author} {\bibfnamefont {B.~P.}\ \bibnamefont
  {{Abbott}}} \emph {et~al.} (\bibinfo {collaboration} {LIGO Scientific
  Collaboration and Virgo Collaboration}),\ }\bibfield  {title} {\enquote
  {\bibinfo {title} {{GW170817: Observation of Gravitational Waves from a
  Binary Neutron Star Inspiral}},}\ }\Doi {10.1103/PhysRevLett.119.141101}
  {\bibfield  {journal} {\bibinfo  {journal} {\prl},\ }\textbf {\bibinfo
  {volume} {119}},\ \bibinfo {pages} {161101} (\bibinfo {year}
  {2017})}\BibitemShut {NoStop}%
\bibitem [{\citenamefont {Driggers}\ \emph {et~al.}(2019)\citenamefont
  {Driggers} \emph {et~al.}}]{Driggers:2018gii}%
  \BibitemOpen
  \bibfield  {author} {\bibinfo {author} {\bibfnamefont {J.~C.}\ \bibnamefont
  {Driggers}} \emph {et~al.} (\bibinfo {collaboration} {LIGO Scientific
  Collaboration Instrument Science Authors}),\ }\bibfield  {title} {\enquote
  {\bibinfo {title} {{Improving astrophysical parameter estimation via offline
  noise subtraction for Advanced LIGO}},}\ }\Doi {10.1103/PhysRevD.99.042001}
  {\bibfield  {journal} {\bibinfo  {journal} {Phys. Rev. D},\ }\textbf
  {\bibinfo {volume} {99}},\ \bibinfo {pages} {042001} (\bibinfo {year}
  {2019})},\ \Eprint {http://arxiv.org/abs/1806.00532} {arXiv:1806.00532
  [astro-ph.IM]} \BibitemShut {NoStop}%
\bibitem [{\citenamefont {Davis}\ \emph {et~al.}(2019)\citenamefont {Davis},
  \citenamefont {Massinger}, \citenamefont {Lundgren}, \citenamefont
  {Driggers}, \citenamefont {Urban},\ and\ \citenamefont
  {Nuttall}}]{Davis:2018yrz}%
  \BibitemOpen
  \bibfield  {author} {\bibinfo {author} {\bibfnamefont {D.}~\bibnamefont
  {Davis}}, \bibinfo {author} {\bibfnamefont {T.~J.}\ \bibnamefont
  {Massinger}}, \bibinfo {author} {\bibfnamefont {A.~P.}\ \bibnamefont
  {Lundgren}}, \bibinfo {author} {\bibfnamefont {J.~C.}\ \bibnamefont
  {Driggers}}, \bibinfo {author} {\bibfnamefont {A.~L.}\ \bibnamefont {Urban}},
  \ and\ \bibinfo {author} {\bibfnamefont {L.~K.}\ \bibnamefont {Nuttall}},\
  }\bibfield  {title} {\enquote {\bibinfo {title} {{Improving the Sensitivity
  of Advanced LIGO Using Noise Subtraction}},}\ }\Doi
  {10.1088/1361-6382/ab01c5} {\bibfield  {journal} {\bibinfo  {journal}
  {Classical Quantum Gravity},\ }\textbf {\bibinfo {volume} {36}},\ \bibinfo
  {pages} {055011} (\bibinfo {year} {2019})},\ \Eprint
  {http://arxiv.org/abs/1809.05348} {arXiv:1809.05348 [astro-ph.IM]}
  \BibitemShut {NoStop}%
\bibitem [{\citenamefont {Klimenko}\ \emph {et~al.}(2008)\citenamefont
  {Klimenko}, \citenamefont {Yakushin}, \citenamefont {Mercer},\ and\
  \citenamefont {Mitselmakher}}]{Klimenko:2008fu}%
  \BibitemOpen
  \bibfield  {author} {\bibinfo {author} {\bibfnamefont {S.}~\bibnamefont
  {Klimenko}}, \bibinfo {author} {\bibfnamefont {I.}~\bibnamefont {Yakushin}},
  \bibinfo {author} {\bibfnamefont {A.}~\bibnamefont {Mercer}}, \ and\ \bibinfo
  {author} {\bibfnamefont {G.}~\bibnamefont {Mitselmakher}},\ }\bibfield
  {title} {\enquote {\bibinfo {title} {{Coherent method for detection of
  gravitational wave bursts}},}\ }\Doi {10.1088/0264-9381/25/11/114029}
  {\bibfield  {journal} {\bibinfo  {journal} {Classical Quantum Gravity},\
  }\textbf {\bibinfo {volume} {25}},\ \bibinfo {pages} {114029} (\bibinfo
  {year} {2008})},\ \Eprint {http://arxiv.org/abs/0802.3232} {arXiv:0802.3232
  [gr-qc]} \BibitemShut {NoStop}%
\bibitem [{\citenamefont {Klimenko}\ \emph {et~al.}(2016)\citenamefont
  {Klimenko} \emph {et~al.}}]{Klimenko:2015ypf}%
  \BibitemOpen
  \bibfield  {author} {\bibinfo {author} {\bibfnamefont {S.}~\bibnamefont
  {Klimenko}} \emph {et~al.},\ }\bibfield  {title} {\enquote {\bibinfo {title}
  {{Method for detection and reconstruction of gravitational wave transients
  with networks of advanced detectors}},}\ }\Doi {10.1103/PhysRevD.93.042004}
  {\bibfield  {journal} {\bibinfo  {journal} {Phys. Rev. D},\ }\textbf
  {\bibinfo {volume} {93}},\ \bibinfo {pages} {042004} (\bibinfo {year}
  {2016})},\ \Eprint {http://arxiv.org/abs/1511.05999} {arXiv:1511.05999
  [gr-qc]} \BibitemShut {NoStop}%
\bibitem [{\citenamefont {Ghosh}\ \emph {et~al.}(2018)\citenamefont {Ghosh},
  \citenamefont {Johnson-McDaniel}, \citenamefont {Ghosh}, \citenamefont
  {Mishra}, \citenamefont {Ajith}, \citenamefont {Del~Pozzo}, \citenamefont
  {Berry}, \citenamefont {Nielsen},\ and\ \citenamefont
  {London}}]{Ghosh:2017gfp}%
  \BibitemOpen
  \bibfield  {author} {\bibinfo {author} {\bibfnamefont {A.}~\bibnamefont
  {Ghosh}}, \bibinfo {author} {\bibfnamefont {N.~K.}\ \bibnamefont
  {Johnson-McDaniel}}, \bibinfo {author} {\bibfnamefont {A.}~\bibnamefont
  {Ghosh}}, \bibinfo {author} {\bibfnamefont {C.~K.}\ \bibnamefont {Mishra}},
  \bibinfo {author} {\bibfnamefont {P.}~\bibnamefont {Ajith}}, \bibinfo
  {author} {\bibfnamefont {W.}~\bibnamefont {Del~Pozzo}}, \bibinfo {author}
  {\bibfnamefont {C.~P.~L.}\ \bibnamefont {Berry}}, \bibinfo {author}
  {\bibfnamefont {A.~B.}\ \bibnamefont {Nielsen}}, \ and\ \bibinfo {author}
  {\bibfnamefont {L.}~\bibnamefont {London}},\ }\bibfield  {title} {\enquote
  {\bibinfo {title} {{Testing general relativity using gravitational wave
  signals from the inspiral, merger and ringdown of binary black holes}},}\
  }\Doi {10.1088/1361-6382/aa972e} {\bibfield  {journal} {\bibinfo  {journal}
  {Classical Quantum Gravity},\ }\textbf {\bibinfo {volume} {35}},\ \bibinfo
  {pages} {014002} (\bibinfo {year} {2018})},\ \Eprint
  {http://arxiv.org/abs/1704.06784} {arXiv:1704.06784 [gr-qc]} \BibitemShut
  {NoStop}%
\bibitem [{\citenamefont {Agathos}\ \emph {et~al.}(2014)\citenamefont
  {Agathos}, \citenamefont {Del~Pozzo}, \citenamefont {Li}, \citenamefont {Van
  Den~Broeck}, \citenamefont {Veitch},\ and\ \citenamefont
  {Vitale}}]{Agathos:2013upa}%
  \BibitemOpen
  \bibfield  {author} {\bibinfo {author} {\bibfnamefont {M.}~\bibnamefont
  {Agathos}}, \bibinfo {author} {\bibfnamefont {W.}~\bibnamefont {Del~Pozzo}},
  \bibinfo {author} {\bibfnamefont {T.~G.~F.}\ \bibnamefont {Li}}, \bibinfo
  {author} {\bibfnamefont {C.}~\bibnamefont {Van Den~Broeck}}, \bibinfo
  {author} {\bibfnamefont {J.}~\bibnamefont {Veitch}}, \ and\ \bibinfo {author}
  {\bibfnamefont {S.}~\bibnamefont {Vitale}},\ }\bibfield  {title} {\enquote
  {\bibinfo {title} {{TIGER: A data analysis pipeline for testing the
  strong-field dynamics of general relativity with gravitational wave signals
  from coalescing compact binaries}},}\ }\Doi {10.1103/PhysRevD.89.082001}
  {\bibfield  {journal} {\bibinfo  {journal} {Phys. Rev. D},\ }\textbf
  {\bibinfo {volume} {89}},\ \bibinfo {pages} {082001} (\bibinfo {year}
  {2014})},\ \Eprint {http://arxiv.org/abs/1311.0420} {arXiv:1311.0420 [gr-qc]}
  \BibitemShut {NoStop}%
\bibitem [{\citenamefont {Zimmerman}\ \emph {et~al.}(2019)\citenamefont
  {Zimmerman}, \citenamefont {Haster},\ and\ \citenamefont
  {Chatziioannou}}]{Zimmerman:2019wzo}%
  \BibitemOpen
  \bibfield  {author} {\bibinfo {author} {\bibfnamefont {A.}~\bibnamefont
  {Zimmerman}}, \bibinfo {author} {\bibfnamefont {C.-J.}\ \bibnamefont
  {Haster}}, \ and\ \bibinfo {author} {\bibfnamefont {K.}~\bibnamefont
  {Chatziioannou}},\ }\bibfield  {title} {\enquote {\bibinfo {title} {{On
  combining information from multiple gravitational wave sources}},}\ }\Doi
  {10.1103/PhysRevD.99.124044} {\bibfield  {journal} {\bibinfo  {journal}
  {Phys. Rev. D},\ }\textbf {\bibinfo {volume} {99}},\ \bibinfo {pages}
  {124044} (\bibinfo {year} {2019})},\ \Eprint
  {http://arxiv.org/abs/1903.11008} {arXiv:1903.11008 [astro-ph.IM]}
  \BibitemShut {NoStop}%
\bibitem [{\citenamefont {Isi}\ \emph {et~al.}(2019)\citenamefont {Isi},
  \citenamefont {Chatziioannou},\ and\ \citenamefont {Farr}}]{Isi:2019asy}%
  \BibitemOpen
  \bibfield  {author} {\bibinfo {author} {\bibfnamefont {M.}~\bibnamefont
  {Isi}}, \bibinfo {author} {\bibfnamefont {K.}~\bibnamefont {Chatziioannou}},
  \ and\ \bibinfo {author} {\bibfnamefont {W.~M.}\ \bibnamefont {Farr}},\
  }\bibfield  {title} {\enquote {\bibinfo {title} {{Hierarchical test of
  general relativity with gravitational waves}},}\ }\Doi
  {10.1103/physrevlett.123.121101} {\bibfield  {journal} {\bibinfo  {journal}
  {Phys. Rev. Lett.},\ }\textbf {\bibinfo {volume} {123}},\ \bibinfo {pages}
  {121101} (\bibinfo {year} {2019})},\ \Eprint
  {http://arxiv.org/abs/1904.08011} {arXiv:1904.08011 [gr-qc]} \BibitemShut
  {NoStop}%
\bibitem [{\citenamefont {Nitz}\ \emph {et~al.}(2018)\citenamefont {Nitz} \emph
  {et~al.}}]{pycbc-github}%
  \BibitemOpen
  \bibfield  {author} {\bibinfo {author} {\bibfnamefont {A.~H.}\ \bibnamefont
  {Nitz}} \emph {et~al.},\ }\href@noop {} {\enquote {\bibinfo {title} {{PyCBC
  software}},}\ }\bibinfo {howpublished}
  {\url{https://github.com/ligo-cbc/pycbc}} (\bibinfo {year}
  {2018})\BibitemShut {NoStop}%
\bibitem [{\citenamefont {Dal~Canton}\ \emph {et~al.}(2014)\citenamefont
  {Dal~Canton} \emph {et~al.}}]{Canton:2014ena}%
  \BibitemOpen
  \bibfield  {author} {\bibinfo {author} {\bibfnamefont {T.}~\bibnamefont
  {Dal~Canton}} \emph {et~al.},\ }\bibfield  {title} {\enquote {\bibinfo
  {title} {{Implementing a search for aligned-spin neutron star-black hole
  systems with advanced ground based gravitational wave detectors}},}\ }\Doi
  {10.1103/PhysRevD.90.082004} {\bibfield  {journal} {\bibinfo  {journal}
  {Phys. Rev. D},\ }\textbf {\bibinfo {volume} {90}},\ \bibinfo {pages}
  {082004} (\bibinfo {year} {2014})},\ \Eprint {http://arxiv.org/abs/1405.6731}
  {arXiv:1405.6731 [gr-qc]} \BibitemShut {NoStop}%
\bibitem [{\citenamefont {Usman}\ \emph {et~al.}(2016)\citenamefont {Usman}
  \emph {et~al.}}]{Usman:2015kfa}%
  \BibitemOpen
  \bibfield  {author} {\bibinfo {author} {\bibfnamefont {S.~A.}\ \bibnamefont
  {Usman}} \emph {et~al.},\ }\bibfield  {title} {\enquote {\bibinfo {title}
  {{The PyCBC search for gravitational waves from compact binary
  coalescence}},}\ }\Doi {10.1088/0264-9381/33/21/215004} {\bibfield  {journal}
  {\bibinfo  {journal} {Classical Quantum Gravity},\ }\textbf {\bibinfo
  {volume} {33}},\ \bibinfo {pages} {215004} (\bibinfo {year} {2016})},\
  \Eprint {http://arxiv.org/abs/1508.02357} {arXiv:1508.02357 [gr-qc]}
  \BibitemShut {NoStop}%
\bibitem [{\citenamefont {Sachdev}\ \emph {et~al.}(2019)\citenamefont {Sachdev}
  \emph {et~al.}}]{Sachdev:2019vvd}%
  \BibitemOpen
  \bibfield  {author} {\bibinfo {author} {\bibfnamefont {S.}~\bibnamefont
  {Sachdev}} \emph {et~al.},\ }\bibfield  {title} {\enquote {\bibinfo {title}
  {{The GstLAL Search Analysis Methods for Compact Binary Mergers in Advanced
  LIGO's Second and Advanced Virgo's First Observing Runs}},}\ }\href@noop {} {
  (\bibinfo {year} {2019})},\ \Eprint {http://arxiv.org/abs/1901.08580}
  {arXiv:1901.08580 [gr-qc]} \BibitemShut {NoStop}%
\bibitem [{\citenamefont {Messick}\ \emph {et~al.}(2017)\citenamefont {Messick}
  \emph {et~al.}}]{Messick:2016aqy}%
  \BibitemOpen
  \bibfield  {author} {\bibinfo {author} {\bibfnamefont {C.}~\bibnamefont
  {Messick}} \emph {et~al.},\ }\bibfield  {title} {\enquote {\bibinfo {title}
  {{Analysis Framework for the Prompt Discovery of Compact Binary Mergers in
  Gravitational-wave Data}},}\ }\Doi {10.1103/PhysRevD.95.042001} {\bibfield
  {journal} {\bibinfo  {journal} {Phys. Rev. D},\ }\textbf {\bibinfo {volume}
  {95}},\ \bibinfo {pages} {042001} (\bibinfo {year} {2017})},\ \Eprint
  {http://arxiv.org/abs/1604.04324} {arXiv:1604.04324 [astro-ph.IM]}
  \BibitemShut {NoStop}%
\bibitem [{\citenamefont {Peters}\ and\ \citenamefont
  {Mathews}(1963)}]{PhysRev.131.435}%
  \BibitemOpen
  \bibfield  {author} {\bibinfo {author} {\bibfnamefont {P.~C.}\ \bibnamefont
  {Peters}}\ and\ \bibinfo {author} {\bibfnamefont {J.}~\bibnamefont
  {Mathews}},\ }\bibfield  {title} {\enquote {\bibinfo {title} {{Gravitational
  radiation from point masses in a Keplerian orbit}},}\ }\Doi
  {10.1103/PhysRev.131.435} {\bibfield  {journal} {\bibinfo  {journal} {Phys.
  Rev.},\ }\textbf {\bibinfo {volume} {131}},\ \bibinfo {pages} {435} (\bibinfo
  {year} {1963})}\BibitemShut {NoStop}%
\bibitem [{\citenamefont {Peters}(1964)}]{Peters:1964}%
  \BibitemOpen
  \bibfield  {author} {\bibinfo {author} {\bibfnamefont {P.~C.}\ \bibnamefont
  {Peters}},\ }\bibfield  {title} {\enquote {\bibinfo {title} {{Gravitational
  Radiation and the Motion of Two Point Masses}},}\ }\Doi
  {10.1103/PhysRev.136.B1224} {\bibfield  {journal} {\bibinfo  {journal} {Phys.
  Rev.},\ }\textbf {\bibinfo {volume} {136}},\ \bibinfo {pages} {B1224}
  (\bibinfo {year} {1964})}\BibitemShut {NoStop}%
\bibitem [{\citenamefont {Hinderer}\ and\ \citenamefont
  {Babak}(2017)}]{Hinderer:2017jcs}%
  \BibitemOpen
  \bibfield  {author} {\bibinfo {author} {\bibfnamefont {T.}~\bibnamefont
  {Hinderer}}\ and\ \bibinfo {author} {\bibfnamefont {S.}~\bibnamefont
  {Babak}},\ }\bibfield  {title} {\enquote {\bibinfo {title} {{Foundations of
  an effective-one-body model for coalescing binaries on eccentric orbits}},}\
  }\Doi {10.1103/PhysRevD.96.104048} {\bibfield  {journal} {\bibinfo  {journal}
  {Phys. Rev. D},\ }\textbf {\bibinfo {volume} {96}},\ \bibinfo {pages}
  {104048} (\bibinfo {year} {2017})},\ \Eprint
  {http://arxiv.org/abs/1707.08426} {arXiv:1707.08426 [gr-qc]} \BibitemShut
  {NoStop}%
\bibitem [{\citenamefont {Cao}\ and\ \citenamefont {Han}(2017)}]{Cao:2017ndf}%
  \BibitemOpen
  \bibfield  {author} {\bibinfo {author} {\bibfnamefont {Z.}~\bibnamefont
  {Cao}}\ and\ \bibinfo {author} {\bibfnamefont {W.-B.}\ \bibnamefont {Han}},\
  }\bibfield  {title} {\enquote {\bibinfo {title} {{Waveform model for an
  eccentric binary black hole based on the
  effective-one-body-numerical-relativity formalism}},}\ }\Doi
  {10.1103/PhysRevD.96.044028} {\bibfield  {journal} {\bibinfo  {journal}
  {Phys. Rev. D},\ }\textbf {\bibinfo {volume} {96}},\ \bibinfo {pages}
  {044028} (\bibinfo {year} {2017})},\ \Eprint
  {http://arxiv.org/abs/1708.00166} {arXiv:1708.00166 [gr-qc]} \BibitemShut
  {NoStop}%
\bibitem [{\citenamefont {Hinder}\ \emph {et~al.}(2018)\citenamefont {Hinder},
  \citenamefont {Kidder},\ and\ \citenamefont {Pfeiffer}}]{Hinder:2017sxy}%
  \BibitemOpen
  \bibfield  {author} {\bibinfo {author} {\bibfnamefont {I.}~\bibnamefont
  {Hinder}}, \bibinfo {author} {\bibfnamefont {L.~E.}\ \bibnamefont {Kidder}},
  \ and\ \bibinfo {author} {\bibfnamefont {H.~P.}\ \bibnamefont {Pfeiffer}},\
  }\bibfield  {title} {\enquote {\bibinfo {title} {{Eccentric binary black hole
  inspiral-merger-ringdown gravitational waveform model from numerical
  relativity and post-Newtonian theory}},}\ }\Doi {10.1103/PhysRevD.98.044015}
  {\bibfield  {journal} {\bibinfo  {journal} {Phys. Rev. D},\ }\textbf
  {\bibinfo {volume} {98}},\ \bibinfo {pages} {044015} (\bibinfo {year}
  {2018})},\ \Eprint {http://arxiv.org/abs/1709.02007} {arXiv:1709.02007
  [gr-qc]} \BibitemShut {NoStop}%
\bibitem [{\citenamefont {Huerta}\ \emph {et~al.}(2018)\citenamefont {Huerta}
  \emph {et~al.}}]{Huerta:2017kez}%
  \BibitemOpen
  \bibfield  {author} {\bibinfo {author} {\bibfnamefont {E.~A.}\ \bibnamefont
  {Huerta}} \emph {et~al.},\ }\bibfield  {title} {\enquote {\bibinfo {title}
  {{Eccentric, nonspinning, inspiral, Gaussian-process merger approximant for
  the detection and characterization of eccentric binary black hole
  mergers}},}\ }\Doi {10.1103/PhysRevD.97.024031} {\bibfield  {journal}
  {\bibinfo  {journal} {Phys. Rev. D},\ }\textbf {\bibinfo {volume} {97}},\
  \bibinfo {pages} {024031} (\bibinfo {year} {2018})},\ \Eprint
  {http://arxiv.org/abs/1711.06276} {arXiv:1711.06276 [gr-qc]} \BibitemShut
  {NoStop}%
\bibitem [{\citenamefont {Klein}\ \emph {et~al.}(2018)\citenamefont {Klein},
  \citenamefont {Boetzel}, \citenamefont {Gopakumar}, \citenamefont {Jetzer},\
  and\ \citenamefont {de~Vittori}}]{Klein:2018ybm}%
  \BibitemOpen
  \bibfield  {author} {\bibinfo {author} {\bibfnamefont {A.}~\bibnamefont
  {Klein}}, \bibinfo {author} {\bibfnamefont {Y.}~\bibnamefont {Boetzel}},
  \bibinfo {author} {\bibfnamefont {A.}~\bibnamefont {Gopakumar}}, \bibinfo
  {author} {\bibfnamefont {P.}~\bibnamefont {Jetzer}}, \ and\ \bibinfo {author}
  {\bibfnamefont {L.}~\bibnamefont {de~Vittori}},\ }\bibfield  {title}
  {\enquote {\bibinfo {title} {{Fourier domain gravitational waveforms for
  precessing eccentric binaries}},}\ }\Doi {10.1103/PhysRevD.98.104043}
  {\bibfield  {journal} {\bibinfo  {journal} {Phys. Rev. D},\ }\textbf
  {\bibinfo {volume} {98}},\ \bibinfo {pages} {104043} (\bibinfo {year}
  {2018})},\ \Eprint {http://arxiv.org/abs/1801.08542} {arXiv:1801.08542
  [gr-qc]} \BibitemShut {NoStop}%
\bibitem [{\citenamefont {Moore}\ \emph {et~al.}(2018)\citenamefont {Moore},
  \citenamefont {Robson}, \citenamefont {Loutrel},\ and\ \citenamefont
  {Yunes}}]{Moore:2018kvz}%
  \BibitemOpen
  \bibfield  {author} {\bibinfo {author} {\bibfnamefont {B.}~\bibnamefont
  {Moore}}, \bibinfo {author} {\bibfnamefont {T.}~\bibnamefont {Robson}},
  \bibinfo {author} {\bibfnamefont {N.}~\bibnamefont {Loutrel}}, \ and\
  \bibinfo {author} {\bibfnamefont {N.}~\bibnamefont {Yunes}},\ }\bibfield
  {title} {\enquote {\bibinfo {title} {{Towards a Fourier domain waveform for
  non-spinning binaries with arbitrary eccentricity}},}\ }\Doi
  {10.1088/1361-6382/aaea00} {\bibfield  {journal} {\bibinfo  {journal}
  {Classical Quantum Gravity},\ }\textbf {\bibinfo {volume} {35}},\ \bibinfo
  {pages} {235006} (\bibinfo {year} {2018})},\ \Eprint
  {http://arxiv.org/abs/1807.07163} {arXiv:1807.07163 [gr-qc]} \BibitemShut
  {NoStop}%
\bibitem [{\citenamefont {Moore}\ and\ \citenamefont
  {Yunes}(2019)}]{Moore:2019xkm}%
  \BibitemOpen
  \bibfield  {author} {\bibinfo {author} {\bibfnamefont {B.}~\bibnamefont
  {Moore}}\ and\ \bibinfo {author} {\bibfnamefont {N.}~\bibnamefont {Yunes}},\
  }\bibfield  {title} {\enquote {\bibinfo {title} {{A 3PN Fourier Domain
  Waveform for Non-Spinning Binaries with Moderate Eccentricity}},}\ }\Doi
  {10.1088/1361-6382/ab3778} {\bibfield  {journal} {\bibinfo  {journal}
  {Classical Quantum Gravity},\ }\textbf {\bibinfo {volume} {36}},\ \bibinfo
  {pages} {185003} (\bibinfo {year} {2019})},\ \Eprint
  {http://arxiv.org/abs/1903.05203} {arXiv:1903.05203 [gr-qc]} \BibitemShut
  {NoStop}%
\bibitem [{\citenamefont {Tiwari}\ \emph {et~al.}(2019)\citenamefont {Tiwari},
  \citenamefont {Achamveedu}, \citenamefont {Haney},\ and\ \citenamefont
  {Hemantakumar}}]{Tiwari:2019jtz}%
  \BibitemOpen
  \bibfield  {author} {\bibinfo {author} {\bibfnamefont {S.}~\bibnamefont
  {Tiwari}}, \bibinfo {author} {\bibfnamefont {G.}~\bibnamefont {Achamveedu}},
  \bibinfo {author} {\bibfnamefont {M.}~\bibnamefont {Haney}}, \ and\ \bibinfo
  {author} {\bibfnamefont {P.}~\bibnamefont {Hemantakumar}},\ }\bibfield
  {title} {\enquote {\bibinfo {title} {{Ready-to-use Fourier domain templates
  for compact binaries inspiraling along moderately eccentric orbits}},}\ }\Doi
  {10.1103/PhysRevD.99.124008} {\bibfield  {journal} {\bibinfo  {journal}
  {Phys. Rev. D},\ }\textbf {\bibinfo {volume} {99}},\ \bibinfo {pages}
  {124008} (\bibinfo {year} {2019})},\ \Eprint
  {http://arxiv.org/abs/1905.07956} {arXiv:1905.07956 [gr-qc]} \BibitemShut
  {NoStop}%
\bibitem [{\citenamefont {Samsing}(2018)}]{Samsing:2017xmd}%
  \BibitemOpen
  \bibfield  {author} {\bibinfo {author} {\bibfnamefont {J.}~\bibnamefont
  {Samsing}},\ }\bibfield  {title} {\enquote {\bibinfo {title} {{Eccentric
  Black Hole Mergers Forming in Globular Clusters}},}\ }\Doi
  {10.1103/PhysRevD.97.103014} {\bibfield  {journal} {\bibinfo  {journal}
  {Phys. Rev. D},\ }\textbf {\bibinfo {volume} {97}},\ \bibinfo {pages}
  {103014} (\bibinfo {year} {2018})},\ \Eprint
  {http://arxiv.org/abs/1711.07452} {arXiv:1711.07452 [astro-ph.HE]}
  \BibitemShut {NoStop}%
\bibitem [{\citenamefont {Rodriguez}\ \emph
  {et~al.}(2018){\natexlab{a}}\citenamefont {Rodriguez}, \citenamefont
  {Amaro-Seoane}, \citenamefont {Chatterjee},\ and\ \citenamefont
  {Rasio}}]{Rodriguez:2017pec}%
  \BibitemOpen
  \bibfield  {author} {\bibinfo {author} {\bibfnamefont {C.~L.}\ \bibnamefont
  {Rodriguez}}, \bibinfo {author} {\bibfnamefont {P.}~\bibnamefont
  {Amaro-Seoane}}, \bibinfo {author} {\bibfnamefont {S.}~\bibnamefont
  {Chatterjee}}, \ and\ \bibinfo {author} {\bibfnamefont {F.~A.}\ \bibnamefont
  {Rasio}},\ }\bibfield  {title} {\enquote {\bibinfo {title} {{Post-Newtonian
  Dynamics in Dense Star Clusters: Highly-Eccentric, Highly-Spinning, and
  Repeated Binary Black Hole Mergers}},}\ }\Doi
  {10.1103/PhysRevLett.120.151101} {\bibfield  {journal} {\bibinfo  {journal}
  {Phys. Rev. Lett.},\ }\textbf {\bibinfo {volume} {120}},\ \bibinfo {pages}
  {151101} (\bibinfo {year} {2018}{\natexlab{a}})},\ \Eprint
  {http://arxiv.org/abs/1712.04937} {arXiv:1712.04937 [astro-ph.HE]}
  \BibitemShut {NoStop}%
\bibitem [{\citenamefont {Zevin}\ \emph {et~al.}(2019)\citenamefont {Zevin},
  \citenamefont {Samsing}, \citenamefont {Rodriguez}, \citenamefont {Haster},\
  and\ \citenamefont {Ramirez-Ruiz}}]{Zevin:2018kzq}%
  \BibitemOpen
  \bibfield  {author} {\bibinfo {author} {\bibfnamefont {M.}~\bibnamefont
  {Zevin}}, \bibinfo {author} {\bibfnamefont {J.}~\bibnamefont {Samsing}},
  \bibinfo {author} {\bibfnamefont {C.}~\bibnamefont {Rodriguez}}, \bibinfo
  {author} {\bibfnamefont {C.-J.}\ \bibnamefont {Haster}}, \ and\ \bibinfo
  {author} {\bibfnamefont {E.}~\bibnamefont {Ramirez-Ruiz}},\ }\bibfield
  {title} {\enquote {\bibinfo {title} {{Eccentric Black Hole Mergers in Dense
  Star Clusters: The Role of Binary-Binary Encounters}},}\ }\Doi
  {10.3847/1538-4357/aaf6ec} {\bibfield  {journal} {\bibinfo  {journal}
  {Astrophys. J.},\ }\textbf {\bibinfo {volume} {871}},\ \bibinfo {pages} {1}
  (\bibinfo {year} {2019})},\ \Eprint {http://arxiv.org/abs/1810.00901}
  {arXiv:1810.00901 [astro-ph.HE]} \BibitemShut {NoStop}%
\bibitem [{\citenamefont {Rodriguez}\ \emph
  {et~al.}(2018){\natexlab{b}}\citenamefont {Rodriguez}, \citenamefont
  {Amaro-Seoane}, \citenamefont {Chatterjee}, \citenamefont {Kremer},
  \citenamefont {Rasio}, \citenamefont {Samsing}, \citenamefont {Ye},\ and\
  \citenamefont {Zevin}}]{Rodriguez:2018pss}%
  \BibitemOpen
  \bibfield  {author} {\bibinfo {author} {\bibfnamefont {C.~L.}\ \bibnamefont
  {Rodriguez}}, \bibinfo {author} {\bibfnamefont {P.}~\bibnamefont
  {Amaro-Seoane}}, \bibinfo {author} {\bibfnamefont {S.}~\bibnamefont
  {Chatterjee}}, \bibinfo {author} {\bibfnamefont {K.}~\bibnamefont {Kremer}},
  \bibinfo {author} {\bibfnamefont {F.~A.}\ \bibnamefont {Rasio}}, \bibinfo
  {author} {\bibfnamefont {J.}~\bibnamefont {Samsing}}, \bibinfo {author}
  {\bibfnamefont {C.~S.}\ \bibnamefont {Ye}}, \ and\ \bibinfo {author}
  {\bibfnamefont {M.}~\bibnamefont {Zevin}},\ }\bibfield  {title} {\enquote
  {\bibinfo {title} {{Post-Newtonian Dynamics in Dense Star Clusters:
  Formation, Masses, and Merger Rates of Highly-Eccentric Black Hole
  Binaries}},}\ }\Doi {10.1103/PhysRevD.98.123005} {\bibfield  {journal}
  {\bibinfo  {journal} {Phys. Rev. D},\ }\textbf {\bibinfo {volume} {98}},\
  \bibinfo {pages} {123005} (\bibinfo {year} {2018}{\natexlab{b}})},\ \Eprint
  {http://arxiv.org/abs/1811.04926} {arXiv:1811.04926 [astro-ph.HE]}
  \BibitemShut {NoStop}%
\bibitem [{\citenamefont {Husa}\ \emph {et~al.}(2016)\citenamefont {Husa},
  \citenamefont {Khan}, \citenamefont {Hannam}, \citenamefont {P{\"u}rrer},
  \citenamefont {Ohme}, \citenamefont {Forteza},\ and\ \citenamefont
  {Boh{\'e}}}]{Husa:2015iqa}%
  \BibitemOpen
  \bibfield  {author} {\bibinfo {author} {\bibfnamefont {S.}~\bibnamefont
  {Husa}}, \bibinfo {author} {\bibfnamefont {S.}~\bibnamefont {Khan}}, \bibinfo
  {author} {\bibfnamefont {M.}~\bibnamefont {Hannam}}, \bibinfo {author}
  {\bibfnamefont {M.}~\bibnamefont {P{\"u}rrer}}, \bibinfo {author}
  {\bibfnamefont {F.}~\bibnamefont {Ohme}}, \bibinfo {author} {\bibfnamefont
  {X.~J.}\ \bibnamefont {Forteza}}, \ and\ \bibinfo {author} {\bibfnamefont
  {A.}~\bibnamefont {Boh{\'e}}},\ }\bibfield  {title} {\enquote {\bibinfo
  {title} {{Frequency-domain gravitational waves from nonprecessing black-hole
  binaries. I. New numerical waveforms and anatomy of the signal}},}\ }\Doi
  {10.1103/PhysRevD.93.044006} {\bibfield  {journal} {\bibinfo  {journal}
  {Phys. Rev. D},\ }\textbf {\bibinfo {volume} {93}},\ \bibinfo {pages}
  {044006} (\bibinfo {year} {2016})},\ \Eprint
  {http://arxiv.org/abs/1508.07250} {arXiv:1508.07250 [gr-qc]} \BibitemShut
  {NoStop}%
\bibitem [{\citenamefont {Hannam}\ \emph {et~al.}(2014)\citenamefont {Hannam},
  \citenamefont {Schmidt}, \citenamefont {Boh{\'e}}, \citenamefont {Haegel},
  \citenamefont {Husa}, \citenamefont {Ohme}, \citenamefont {Pratten},\ and\
  \citenamefont {P{\"u}rrer}}]{Hannam:2013oca}%
  \BibitemOpen
  \bibfield  {author} {\bibinfo {author} {\bibfnamefont {M.}~\bibnamefont
  {Hannam}}, \bibinfo {author} {\bibfnamefont {P.}~\bibnamefont {Schmidt}},
  \bibinfo {author} {\bibfnamefont {A.}~\bibnamefont {Boh{\'e}}}, \bibinfo
  {author} {\bibfnamefont {L.}~\bibnamefont {Haegel}}, \bibinfo {author}
  {\bibfnamefont {S.}~\bibnamefont {Husa}}, \bibinfo {author} {\bibfnamefont
  {F.}~\bibnamefont {Ohme}}, \bibinfo {author} {\bibfnamefont {G.}~\bibnamefont
  {Pratten}}, \ and\ \bibinfo {author} {\bibfnamefont {M.}~\bibnamefont
  {P{\"u}rrer}},\ }\bibfield  {title} {\enquote {\bibinfo {title} {{Simple
  Model of Complete Precessing Black-Hole-Binary Gravitational Waveforms}},}\
  }\Doi {10.1103/PhysRevLett.113.151101} {\bibfield  {journal} {\bibinfo
  {journal} {Phys. Rev. Lett.},\ }\textbf {\bibinfo {volume} {113}},\ \bibinfo
  {pages} {151101} (\bibinfo {year} {2014})},\ \Eprint
  {http://arxiv.org/abs/1308.3271} {arXiv:1308.3271 [gr-qc]} \BibitemShut
  {NoStop}%
\bibitem [{\citenamefont {{Taracchini}}\ \emph {et~al.}(2014)\citenamefont
  {{Taracchini}} \emph {et~al.}}]{Taracchini:2013}%
  \BibitemOpen
  \bibfield  {author} {\bibinfo {author} {\bibfnamefont {A.}~\bibnamefont
  {{Taracchini}}} \emph {et~al.},\ }\bibfield  {title} {\enquote {\bibinfo
  {title} {{Effective-one-body model for black-hole binaries with generic mass
  ratios and spins}},}\ }\Doi {10.1103/PhysRevD.89.061502} {\bibfield
  {journal} {\bibinfo  {journal} {\prd},\ }\textbf {\bibinfo {volume} {89}},\
  \bibinfo {eid} {061502} (\bibinfo {year} {2014})},\ \Eprint
  {http://arxiv.org/abs/1311.2544} {arXiv:1311.2544 [gr-qc]} \BibitemShut
  {NoStop}%
\bibitem [{\citenamefont {Babak}\ \emph {et~al.}(2017)\citenamefont {Babak},
  \citenamefont {Taracchini},\ and\ \citenamefont {Buonanno}}]{Babak:2016tgq}%
  \BibitemOpen
  \bibfield  {author} {\bibinfo {author} {\bibfnamefont {S.}~\bibnamefont
  {Babak}}, \bibinfo {author} {\bibfnamefont {A.}~\bibnamefont {Taracchini}}, \
  and\ \bibinfo {author} {\bibfnamefont {A.}~\bibnamefont {Buonanno}},\
  }\bibfield  {title} {\enquote {\bibinfo {title} {{Validating the
  effective-one-body model of spinning, precessing binary black holes against
  numerical relativity}},}\ }\Doi {10.1103/PhysRevD.95.024010} {\bibfield
  {journal} {\bibinfo  {journal} {Phys. Rev. D},\ }\textbf {\bibinfo {volume}
  {95}},\ \bibinfo {pages} {024010} (\bibinfo {year} {2017})},\ \Eprint
  {http://arxiv.org/abs/1607.05661} {arXiv:1607.05661 [gr-qc]} \BibitemShut
  {NoStop}%
\bibitem [{\citenamefont {Blackman}\ \emph
  {et~al.}(2017){\natexlab{b}}\citenamefont {Blackman}, \citenamefont {Field},
  \citenamefont {Scheel}, \citenamefont {Galley}, \citenamefont {Ott},
  \citenamefont {Boyle}, \citenamefont {Kidder}, \citenamefont {Pfeiffer},\
  and\ \citenamefont {Szil{\'a}gyi}}]{Blackman:2017pcm}%
  \BibitemOpen
  \bibfield  {author} {\bibinfo {author} {\bibfnamefont {J.}~\bibnamefont
  {Blackman}}, \bibinfo {author} {\bibfnamefont {S.~E.}\ \bibnamefont {Field}},
  \bibinfo {author} {\bibfnamefont {M.~A.}\ \bibnamefont {Scheel}}, \bibinfo
  {author} {\bibfnamefont {C.~R.}\ \bibnamefont {Galley}}, \bibinfo {author}
  {\bibfnamefont {C.~D.}\ \bibnamefont {Ott}}, \bibinfo {author} {\bibfnamefont
  {M.}~\bibnamefont {Boyle}}, \bibinfo {author} {\bibfnamefont {L.~E.}\
  \bibnamefont {Kidder}}, \bibinfo {author} {\bibfnamefont {H.~P.}\
  \bibnamefont {Pfeiffer}}, \ and\ \bibinfo {author} {\bibfnamefont
  {B.}~\bibnamefont {Szil{\'a}gyi}},\ }\bibfield  {title} {\enquote {\bibinfo
  {title} {{Numerical relativity waveform surrogate model for generically
  precessing binary black hole mergers}},}\ }\Doi {10.1103/PhysRevD.96.024058}
  {\bibfield  {journal} {\bibinfo  {journal} {Phys. Rev. D},\ }\textbf
  {\bibinfo {volume} {96}},\ \bibinfo {pages} {024058} (\bibinfo {year}
  {2017}{\natexlab{b}})},\ \Eprint {http://arxiv.org/abs/1705.07089}
  {arXiv:1705.07089 [gr-qc]} \BibitemShut {NoStop}%
\bibitem [{\citenamefont {Varma}\ \emph
  {et~al.}(2019){\natexlab{a}}\citenamefont {Varma}, \citenamefont {Field},
  \citenamefont {Scheel}, \citenamefont {Blackman}, \citenamefont {Gerosa},
  \citenamefont {Stein}, \citenamefont {Kidder},\ and\ \citenamefont
  {Pfeiffer}}]{Varma:2019csw}%
  \BibitemOpen
  \bibfield  {author} {\bibinfo {author} {\bibfnamefont {V.}~\bibnamefont
  {Varma}}, \bibinfo {author} {\bibfnamefont {S.~E.}\ \bibnamefont {Field}},
  \bibinfo {author} {\bibfnamefont {M.~A.}\ \bibnamefont {Scheel}}, \bibinfo
  {author} {\bibfnamefont {J.}~\bibnamefont {Blackman}}, \bibinfo {author}
  {\bibfnamefont {D.}~\bibnamefont {Gerosa}}, \bibinfo {author} {\bibfnamefont
  {L.~C.}\ \bibnamefont {Stein}}, \bibinfo {author} {\bibfnamefont {L.~E.}\
  \bibnamefont {Kidder}}, \ and\ \bibinfo {author} {\bibfnamefont {H.~P.}\
  \bibnamefont {Pfeiffer}},\ }\bibfield  {title} {\enquote {\bibinfo {title}
  {{Surrogate models for precessing binary black hole simulations with unequal
  masses}},}\ }\href@noop {} { (\bibinfo {year} {2019}{\natexlab{a}})},\
  \Eprint {http://arxiv.org/abs/1905.09300} {arXiv:1905.09300 [gr-qc]}
  \BibitemShut {NoStop}%
\bibitem [{\citenamefont {Kamaretsos}\ \emph {et~al.}(2012)\citenamefont
  {Kamaretsos}, \citenamefont {Hannam}, \citenamefont {Husa},\ and\
  \citenamefont {Sathyaprakash}}]{PhysRevD.85.024018}%
  \BibitemOpen
  \bibfield  {author} {\bibinfo {author} {\bibfnamefont {I.}~\bibnamefont
  {Kamaretsos}}, \bibinfo {author} {\bibfnamefont {M.}~\bibnamefont {Hannam}},
  \bibinfo {author} {\bibfnamefont {S.}~\bibnamefont {Husa}}, \ and\ \bibinfo
  {author} {\bibfnamefont {B.~S.}\ \bibnamefont {Sathyaprakash}},\ }\bibfield
  {title} {\enquote {\bibinfo {title} {Black-hole hair loss: Learning about
  binary progenitors from ringdown signals},}\ }\Doi
  {10.1103/PhysRevD.85.024018} {\bibfield  {journal} {\bibinfo  {journal}
  {Phys. Rev. D},\ }\textbf {\bibinfo {volume} {85}},\ \bibinfo {pages}
  {024018} (\bibinfo {year} {2012})},\ \Eprint {http://arxiv.org/abs/1107.0854}
  {arXiv:1107.0854 [gr-qc]} \BibitemShut {NoStop}%
\bibitem [{\citenamefont {London}\ \emph {et~al.}(2014)\citenamefont {London},
  \citenamefont {Shoemaker},\ and\ \citenamefont {Healy}}]{PhysRevD.90.124032}%
  \BibitemOpen
  \bibfield  {author} {\bibinfo {author} {\bibfnamefont {L.}~\bibnamefont
  {London}}, \bibinfo {author} {\bibfnamefont {D.}~\bibnamefont {Shoemaker}}, \
  and\ \bibinfo {author} {\bibfnamefont {J.}~\bibnamefont {Healy}},\ }\bibfield
   {title} {\enquote {\bibinfo {title} {Modeling ringdown: Beyond the
  fundamental quasinormal modes},}\ }\Doi {10.1103/PhysRevD.90.124032}
  {\bibfield  {journal} {\bibinfo  {journal} {Phys. Rev. D},\ }\textbf
  {\bibinfo {volume} {90}},\ \bibinfo {pages} {124032} (\bibinfo {year}
  {2014})},\ \bibinfo {note}
  {\href{http://dx.doi.org/10.1103/PhysRevD.94.069902}{{\bf{94}}, 069902(E)
  (2016)}},\ \Eprint {http://arxiv.org/abs/1404.3197} {arXiv:1404.3197 [gr-qc]}
  \BibitemShut {NoStop}%
\bibitem [{\citenamefont {London}\ \emph {et~al.}(2018)\citenamefont {London},
  \citenamefont {Khan}, \citenamefont {Fauchon-Jones}, \citenamefont {García},
  \citenamefont {Hannam}, \citenamefont {Husa}, \citenamefont
  {Jim{\'e}nez-Forteza}, \citenamefont {Kalaghatgi}, \citenamefont {Ohme},\
  and\ \citenamefont {Pannarale}}]{London:2017bcn}%
  \BibitemOpen
  \bibfield  {author} {\bibinfo {author} {\bibfnamefont {L.}~\bibnamefont
  {London}}, \bibinfo {author} {\bibfnamefont {S.}~\bibnamefont {Khan}},
  \bibinfo {author} {\bibfnamefont {E.}~\bibnamefont {Fauchon-Jones}}, \bibinfo
  {author} {\bibfnamefont {C.}~\bibnamefont {García}}, \bibinfo {author}
  {\bibfnamefont {M.}~\bibnamefont {Hannam}}, \bibinfo {author} {\bibfnamefont
  {S.}~\bibnamefont {Husa}}, \bibinfo {author} {\bibfnamefont {X.}~\bibnamefont
  {Jim{\'e}nez-Forteza}}, \bibinfo {author} {\bibfnamefont {C.}~\bibnamefont
  {Kalaghatgi}}, \bibinfo {author} {\bibfnamefont {F.}~\bibnamefont {Ohme}}, \
  and\ \bibinfo {author} {\bibfnamefont {F.}~\bibnamefont {Pannarale}},\
  }\bibfield  {title} {\enquote {\bibinfo {title} {{First higher-multipole
  model of gravitational waves from spinning and coalescing black-hole
  binaries}},}\ }\Doi {10.1103/PhysRevLett.120.161102} {\bibfield  {journal}
  {\bibinfo  {journal} {Phys. Rev. Lett.},\ }\textbf {\bibinfo {volume}
  {120}},\ \bibinfo {pages} {161102} (\bibinfo {year} {2018})},\ \Eprint
  {http://arxiv.org/abs/1708.00404} {arXiv:1708.00404 [gr-qc]} \BibitemShut
  {NoStop}%
\bibitem [{\citenamefont {Cotesta}\ \emph {et~al.}(2018)\citenamefont
  {Cotesta}, \citenamefont {Buonanno}, \citenamefont {Boh{\'e}}, \citenamefont
  {Taracchini}, \citenamefont {Hinder},\ and\ \citenamefont
  {Ossokine}}]{Cotesta:2018fcv}%
  \BibitemOpen
  \bibfield  {author} {\bibinfo {author} {\bibfnamefont {R.}~\bibnamefont
  {Cotesta}}, \bibinfo {author} {\bibfnamefont {A.}~\bibnamefont {Buonanno}},
  \bibinfo {author} {\bibfnamefont {A.}~\bibnamefont {Boh{\'e}}}, \bibinfo
  {author} {\bibfnamefont {A.}~\bibnamefont {Taracchini}}, \bibinfo {author}
  {\bibfnamefont {I.}~\bibnamefont {Hinder}}, \ and\ \bibinfo {author}
  {\bibfnamefont {S.}~\bibnamefont {Ossokine}},\ }\bibfield  {title} {\enquote
  {\bibinfo {title} {{Enriching the Symphony of Gravitational Waves from Binary
  Black Holes by Tuning Higher Harmonics}},}\ }\Doi
  {10.1103/PhysRevD.98.084028} {\bibfield  {journal} {\bibinfo  {journal}
  {Phys. Rev. D},\ }\textbf {\bibinfo {volume} {98}},\ \bibinfo {pages}
  {084028} (\bibinfo {year} {2018})},\ \Eprint
  {http://arxiv.org/abs/1803.10701} {arXiv:1803.10701 [gr-qc]} \BibitemShut
  {NoStop}%
\bibitem [{\citenamefont {Varma}\ \emph
  {et~al.}(2019){\natexlab{b}}\citenamefont {Varma}, \citenamefont {Field},
  \citenamefont {Scheel}, \citenamefont {Blackman}, \citenamefont {Kidder},\
  and\ \citenamefont {Pfeiffer}}]{Varma:2018mmi}%
  \BibitemOpen
  \bibfield  {author} {\bibinfo {author} {\bibfnamefont {V.}~\bibnamefont
  {Varma}}, \bibinfo {author} {\bibfnamefont {S.~E.}\ \bibnamefont {Field}},
  \bibinfo {author} {\bibfnamefont {M.~A.}\ \bibnamefont {Scheel}}, \bibinfo
  {author} {\bibfnamefont {J.}~\bibnamefont {Blackman}}, \bibinfo {author}
  {\bibfnamefont {L.~E.}\ \bibnamefont {Kidder}}, \ and\ \bibinfo {author}
  {\bibfnamefont {H.~P.}\ \bibnamefont {Pfeiffer}},\ }\bibfield  {title}
  {\enquote {\bibinfo {title} {{Surrogate model of hybridized numerical
  relativity binary black hole waveforms}},}\ }\Doi
  {10.1103/PhysRevD.99.064045} {\bibfield  {journal} {\bibinfo  {journal}
  {Phys. Rev. D},\ }\textbf {\bibinfo {volume} {99}},\ \bibinfo {pages}
  {064045} (\bibinfo {year} {2019}{\natexlab{b}})},\ \Eprint
  {http://arxiv.org/abs/1812.07865} {arXiv:1812.07865 [gr-qc]} \BibitemShut
  {NoStop}%
\bibitem [{\citenamefont {Varma}\ and\ \citenamefont
  {Ajith}(2017)}]{Varma:2016dnf}%
  \BibitemOpen
  \bibfield  {author} {\bibinfo {author} {\bibfnamefont {V.}~\bibnamefont
  {Varma}}\ and\ \bibinfo {author} {\bibfnamefont {P.}~\bibnamefont {Ajith}},\
  }\bibfield  {title} {\enquote {\bibinfo {title} {{Effects of nonquadrupole
  modes in the detection and parameter estimation of black hole binaries with
  nonprecessing spins}},}\ }\Doi {10.1103/PhysRevD.96.124024} {\bibfield
  {journal} {\bibinfo  {journal} {Phys. Rev. D},\ }\textbf {\bibinfo {volume}
  {96}},\ \bibinfo {pages} {124024} (\bibinfo {year} {2017})},\ \Eprint
  {http://arxiv.org/abs/1612.05608} {arXiv:1612.05608 [gr-qc]} \BibitemShut
  {NoStop}%
\bibitem [{\citenamefont {Pang}\ \emph {et~al.}(2018)\citenamefont {Pang},
  \citenamefont {Calder{\'o}n~Bustillo}, \citenamefont {Wang},\ and\
  \citenamefont {Li}}]{Pang:2018hjb}%
  \BibitemOpen
  \bibfield  {author} {\bibinfo {author} {\bibfnamefont {P.~T.~H.}\
  \bibnamefont {Pang}}, \bibinfo {author} {\bibfnamefont {J.}~\bibnamefont
  {Calder{\'o}n~Bustillo}}, \bibinfo {author} {\bibfnamefont {Y.}~\bibnamefont
  {Wang}}, \ and\ \bibinfo {author} {\bibfnamefont {T.~G.~F.}\ \bibnamefont
  {Li}},\ }\bibfield  {title} {\enquote {\bibinfo {title} {{Potential
  observations of false deviations from general relativity in gravitational
  wave signals from binary black holes}},}\ }\Doi {10.1103/PhysRevD.98.024019}
  {\bibfield  {journal} {\bibinfo  {journal} {Phys. Rev. D},\ }\textbf
  {\bibinfo {volume} {98}},\ \bibinfo {pages} {024019} (\bibinfo {year}
  {2018})},\ \Eprint {http://arxiv.org/abs/1802.03306} {arXiv:1802.03306
  [gr-qc]} \BibitemShut {NoStop}%
\bibitem [{\citenamefont {{LIGO Scientific Collaboration and Virgo
  Collaboration}}(2018){\natexlab{b}}}]{lalsuite}%
  \BibitemOpen
  \bibfield  {author} {\bibinfo {author} {\bibnamefont {{LIGO Scientific
  Collaboration and Virgo Collaboration}}},\ }\Doi {10.7935/GT1W-FZ16}
  {\enquote {\bibinfo {title} {{LALSuite software}},}\ } (\bibinfo {year}
  {2018}{\natexlab{b}})\BibitemShut {NoStop}%
\bibitem [{\citenamefont {Cornish}\ and\ \citenamefont
  {Littenberg}(2015)}]{Cornish:2014kda}%
  \BibitemOpen
  \bibfield  {author} {\bibinfo {author} {\bibfnamefont {N.~J.}\ \bibnamefont
  {Cornish}}\ and\ \bibinfo {author} {\bibfnamefont {T.~B.}\ \bibnamefont
  {Littenberg}},\ }\bibfield  {title} {\enquote {\bibinfo {title} {{BayesWave:
  Bayesian Inference for Gravitational Wave Bursts and Instrument Glitches}},}\
  }\Doi {10.1088/0264-9381/32/13/135012} {\bibfield  {journal} {\bibinfo
  {journal} {Classical Quantum Gravity},\ }\textbf {\bibinfo {volume} {32}},\
  \bibinfo {pages} {135012} (\bibinfo {year} {2015})},\ \Eprint
  {http://arxiv.org/abs/1410.3835} {arXiv:1410.3835 [gr-qc]} \BibitemShut
  {NoStop}%
\bibitem [{\citenamefont {Littenberg}\ and\ \citenamefont
  {Cornish}(2015)}]{Littenberg:2014oda}%
  \BibitemOpen
  \bibfield  {author} {\bibinfo {author} {\bibfnamefont {T.~B.}\ \bibnamefont
  {Littenberg}}\ and\ \bibinfo {author} {\bibfnamefont {N.~J.}\ \bibnamefont
  {Cornish}},\ }\bibfield  {title} {\enquote {\bibinfo {title} {{Bayesian
  inference for spectral estimation of gravitational wave detector noise}},}\
  }\Doi {10.1103/PhysRevD.91.084034} {\bibfield  {journal} {\bibinfo  {journal}
  {Phys. Rev. D},\ }\textbf {\bibinfo {volume} {91}},\ \bibinfo {pages}
  {084034} (\bibinfo {year} {2015})},\ \Eprint {http://arxiv.org/abs/1410.3852}
  {arXiv:1410.3852 [gr-qc]} \BibitemShut {NoStop}%
\bibitem [{\citenamefont {Apostolatos}(1995)}]{Apostolatos:1995}%
  \BibitemOpen
  \bibfield  {author} {\bibinfo {author} {\bibfnamefont {T.~A.}\ \bibnamefont
  {Apostolatos}},\ }\bibfield  {title} {\enquote {\bibinfo {title} {Search
  templates for gravitational waves from precessing, inspiraling binaries},}\
  }\Doi {10.1103/PhysRevD.52.605} {\bibfield  {journal} {\bibinfo  {journal}
  {\prd},\ }\textbf {\bibinfo {volume} {52}},\ \bibinfo {pages} {605} (\bibinfo
  {year} {1995})}\BibitemShut {NoStop}%
\bibitem [{\citenamefont {Fisher}(1948)}]{Fisher1948}%
  \BibitemOpen
  \bibfield  {author} {\bibinfo {author} {\bibfnamefont {R.~A.}\ \bibnamefont
  {Fisher}},\ }\bibfield  {title} {\enquote {\bibinfo {title} {Questions and
  answers},}\ }\Doi {10.2307/2681650} {\bibfield  {journal} {\bibinfo
  {journal} {Am. Statistician},\ }\textbf {\bibinfo {volume} {2}},\ \bibinfo
  {pages} {30} (\bibinfo {year} {1948})}\BibitemShut {NoStop}%
\bibitem [{\citenamefont {Ghosh}\ \emph {et~al.}(2016)\citenamefont {Ghosh}
  \emph {et~al.}}]{Ghosh:2016qgn}%
  \BibitemOpen
  \bibfield  {author} {\bibinfo {author} {\bibfnamefont {A.}~\bibnamefont
  {Ghosh}} \emph {et~al.},\ }\bibfield  {title} {\enquote {\bibinfo {title}
  {{Testing general relativity using golden black-hole binaries}},}\ }\Doi
  {10.1103/PhysRevD.94.021101} {\bibfield  {journal} {\bibinfo  {journal}
  {Phys. Rev. D},\ }\textbf {\bibinfo {volume} {94}},\ \bibinfo {pages}
  {021101(R)} (\bibinfo {year} {2016})},\ \Eprint
  {http://arxiv.org/abs/1602.02453} {arXiv:1602.02453 [gr-qc]} \BibitemShut
  {NoStop}%
\bibitem [{\citenamefont {Bardeen}\ \emph {et~al.}(1972)\citenamefont
  {Bardeen}, \citenamefont {Press},\ and\ \citenamefont
  {Teukolsky}}]{Bardeen:1972fi}%
  \BibitemOpen
  \bibfield  {author} {\bibinfo {author} {\bibfnamefont {J.~M.}\ \bibnamefont
  {Bardeen}}, \bibinfo {author} {\bibfnamefont {W.~H.}\ \bibnamefont {Press}},
  \ and\ \bibinfo {author} {\bibfnamefont {S.~A.}\ \bibnamefont {Teukolsky}},\
  }\bibfield  {title} {\enquote {\bibinfo {title} {{Rotating black holes:
  Locally nonrotating frames, energy extraction, and scalar synchrotron
  radiation}},}\ }\Doi {10.1086/151796} {\bibfield  {journal} {\bibinfo
  {journal} {Astrophys. J.},\ }\textbf {\bibinfo {volume} {178}},\ \bibinfo
  {pages} {347} (\bibinfo {year} {1972})}\BibitemShut {NoStop}%
\bibitem [{\citenamefont {Healy}\ and\ \citenamefont
  {Lousto}(2017)}]{Healy:2016lce}%
  \BibitemOpen
  \bibfield  {author} {\bibinfo {author} {\bibfnamefont {J.}~\bibnamefont
  {Healy}}\ and\ \bibinfo {author} {\bibfnamefont {C.~O.}\ \bibnamefont
  {Lousto}},\ }\bibfield  {title} {\enquote {\bibinfo {title} {{Remnant of
  binary black-hole mergers: New simulations and peak luminosity studies}},}\
  }\Doi {10.1103/PhysRevD.95.024037} {\bibfield  {journal} {\bibinfo  {journal}
  {Phys. Rev. D},\ }\textbf {\bibinfo {volume} {95}},\ \bibinfo {pages}
  {024037} (\bibinfo {year} {2017})},\ \Eprint
  {http://arxiv.org/abs/1610.09713} {arXiv:1610.09713 [gr-qc]} \BibitemShut
  {NoStop}%
\bibitem [{\citenamefont {Hofmann}\ \emph {et~al.}(2016)\citenamefont
  {Hofmann}, \citenamefont {Barausse},\ and\ \citenamefont
  {Rezzolla}}]{2041-8205-825-2-L19}%
  \BibitemOpen
  \bibfield  {author} {\bibinfo {author} {\bibfnamefont {F.}~\bibnamefont
  {Hofmann}}, \bibinfo {author} {\bibfnamefont {E.}~\bibnamefont {Barausse}}, \
  and\ \bibinfo {author} {\bibfnamefont {L.}~\bibnamefont {Rezzolla}},\
  }\bibfield  {title} {\enquote {\bibinfo {title} {The final spin from binary
  black holes in quasi-circular orbits},}\ }\Doi {10.3847/2041-8205/825/2/L19}
  {\bibfield  {journal} {\bibinfo  {journal} {Astrophys. J. Lett.},\ }\textbf
  {\bibinfo {volume} {825}},\ \bibinfo {pages} {L19} (\bibinfo {year}
  {2016})},\ \Eprint {http://arxiv.org/abs/1605.01938} {arXiv:1605.01938
  [gr-qc]} \BibitemShut {NoStop}%
\bibitem [{\citenamefont {Jim\'enez-Forteza}\ \emph {et~al.}(2017)\citenamefont
  {Jim\'enez-Forteza}, \citenamefont {Keitel}, \citenamefont {Husa},
  \citenamefont {Hannam}, \citenamefont {Khan},\ and\ \citenamefont
  {P\"urrer}}]{PhysRevD.95.064024}%
  \BibitemOpen
  \bibfield  {author} {\bibinfo {author} {\bibfnamefont {X.}~\bibnamefont
  {Jim\'enez-Forteza}}, \bibinfo {author} {\bibfnamefont {D.}~\bibnamefont
  {Keitel}}, \bibinfo {author} {\bibfnamefont {S.}~\bibnamefont {Husa}},
  \bibinfo {author} {\bibfnamefont {M.}~\bibnamefont {Hannam}}, \bibinfo
  {author} {\bibfnamefont {S.}~\bibnamefont {Khan}}, \ and\ \bibinfo {author}
  {\bibfnamefont {M.}~\bibnamefont {P\"urrer}},\ }\bibfield  {title} {\enquote
  {\bibinfo {title} {Hierarchical data-driven approach to fitting numerical
  relativity data for nonprecessing binary black holes with an application to
  final spin and radiated energy},}\ }\Doi {10.1103/PhysRevD.95.064024}
  {\bibfield  {journal} {\bibinfo  {journal} {Phys. Rev. D},\ }\textbf
  {\bibinfo {volume} {95}},\ \bibinfo {pages} {064024} (\bibinfo {year}
  {2017})}\BibitemShut {NoStop}%
\bibitem [{\citenamefont {{Johnson-McDaniel}}\ \emph
  {et~al.}(2016)\citenamefont {{Johnson-McDaniel}} \emph
  {et~al.}}]{spinfit-T1600168}%
  \BibitemOpen
  \bibfield  {author} {\bibinfo {author} {\bibfnamefont {N.~K.}\ \bibnamefont
  {{Johnson-McDaniel}}} \emph {et~al.},\ }\href
  {https://dcc.ligo.org/T1600168/public} {\emph {\bibinfo {title} {Determining
  the final spin of a binary black hole system including in-plane spins: Method
  and checks of accuracy}}},\ \bibinfo {type} {Tech. Rep.}\ \bibinfo {number}
  {{LIGO}-T1600168}\ (\bibinfo  {institution} {{LIGO} Project},\ \bibinfo
  {year} {2016})\ \bibinfo {note}
  {\url{https://dcc.ligo.org/LIGO-T1600168/public/main}}\BibitemShut {NoStop}%
\bibitem [{\citenamefont {Blanchet}\ \emph {et~al.}(1995)\citenamefont
  {Blanchet}, \citenamefont {Damour}, \citenamefont {Iyer}, \citenamefont
  {Will},\ and\ \citenamefont {Wiseman}}]{Blanchet:1995ez}%
  \BibitemOpen
  \bibfield  {author} {\bibinfo {author} {\bibfnamefont {L.}~\bibnamefont
  {Blanchet}}, \bibinfo {author} {\bibfnamefont {T.}~\bibnamefont {Damour}},
  \bibinfo {author} {\bibfnamefont {B.~R.}\ \bibnamefont {Iyer}}, \bibinfo
  {author} {\bibfnamefont {C.~M.}\ \bibnamefont {Will}}, \ and\ \bibinfo
  {author} {\bibfnamefont {A.~G.}\ \bibnamefont {Wiseman}},\ }\bibfield
  {title} {\enquote {\bibinfo {title} {Gravitational-radiation damping of
  compact binary systems to second post-{N}ewtonian order},}\ }\Doi
  {10.1103/PhysRevLett.74.3515} {\bibfield  {journal} {\bibinfo  {journal}
  {\prl},\ }\textbf {\bibinfo {volume} {74}},\ \bibinfo {pages} {3515}
  (\bibinfo {year} {1995})},\ \Eprint {http://arxiv.org/abs/gr-qc/9501027}
  {arXiv:gr-qc/9501027} \BibitemShut {NoStop}%
\bibitem [{\citenamefont {Blanchet}\ \emph {et~al.}(2004)\citenamefont
  {Blanchet}, \citenamefont {Damour}, \citenamefont {Esposito-Far\`ese},\ and\
  \citenamefont {Iyer}}]{Blanchet:2004ek}%
  \BibitemOpen
  \bibfield  {author} {\bibinfo {author} {\bibfnamefont {L.}~\bibnamefont
  {Blanchet}}, \bibinfo {author} {\bibfnamefont {T.}~\bibnamefont {Damour}},
  \bibinfo {author} {\bibfnamefont {G.}~\bibnamefont {Esposito-Far\`ese}}, \
  and\ \bibinfo {author} {\bibfnamefont {B.~R.}\ \bibnamefont {Iyer}},\
  }\bibfield  {title} {\enquote {\bibinfo {title} {Gravitational radiation from
  inspiralling compact binaries completed at the third post-{N}ewtonian
  order},}\ }\Doi {10.1103/PhysRevLett.93.091101} {\bibfield  {journal}
  {\bibinfo  {journal} {\prl},\ }\textbf {\bibinfo {volume} {93}},\ \bibinfo
  {pages} {091101} (\bibinfo {year} {2004})},\ \Eprint
  {http://arxiv.org/abs/gr-qc/0406012} {arXiv:gr-qc/0406012} \BibitemShut
  {NoStop}%
\bibitem [{\citenamefont {Blanchet}\ \emph {et~al.}(2005)\citenamefont
  {Blanchet}, \citenamefont {Damour}, \citenamefont {Esposito-Far{\`e}se},\
  and\ \citenamefont {Iyer}}]{Blanchet:2005tk}%
  \BibitemOpen
  \bibfield  {author} {\bibinfo {author} {\bibfnamefont {L.}~\bibnamefont
  {Blanchet}}, \bibinfo {author} {\bibfnamefont {T.}~\bibnamefont {Damour}},
  \bibinfo {author} {\bibfnamefont {G.}~\bibnamefont {Esposito-Far{\`e}se}}, \
  and\ \bibinfo {author} {\bibfnamefont {B.~R.}\ \bibnamefont {Iyer}},\
  }\bibfield  {title} {\enquote {\bibinfo {title} {{Dimensional regularization
  of the third post-Newtonian gravitational wave generation from two point
  masses}},}\ }\Doi {10.1103/PhysRevD.71.124004} {\bibfield  {journal}
  {\bibinfo  {journal} {Phys. Rev. D},\ }\textbf {\bibinfo {volume} {71}},\
  \bibinfo {pages} {124004} (\bibinfo {year} {2005})},\ \Eprint
  {http://arxiv.org/abs/gr-qc/0503044} {arXiv:gr-qc/0503044} \BibitemShut
  {NoStop}%
\bibitem [{\citenamefont {Blanchet}(2014)}]{Blanchet:2013haa}%
  \BibitemOpen
  \bibfield  {author} {\bibinfo {author} {\bibfnamefont {L.}~\bibnamefont
  {Blanchet}},\ }\bibfield  {title} {\enquote {\bibinfo {title} {{Gravitational
  Radiation from Post-Newtonian Sources and Inspiralling Compact Binaries}},}\
  }\Doi {10.12942/lrr-2014-2} {\bibfield  {journal} {\bibinfo  {journal}
  {Living Rev. Relativity},\ }\textbf {\bibinfo {volume} {17}},\ \bibinfo
  {pages} {2} (\bibinfo {year} {2014})},\ \Eprint
  {http://arxiv.org/abs/1310.1528} {arXiv:1310.1528 [gr-qc]} \BibitemShut
  {NoStop}%
\bibitem [{\citenamefont {Arun}\ \emph
  {et~al.}(2006){\natexlab{a}}\citenamefont {Arun}, \citenamefont {Iyer},
  \citenamefont {Qusailah},\ and\ \citenamefont {Sathyaprakash}}]{Arun:2006hn}%
  \BibitemOpen
  \bibfield  {author} {\bibinfo {author} {\bibfnamefont {K.~G.}\ \bibnamefont
  {Arun}}, \bibinfo {author} {\bibfnamefont {B.~R.}\ \bibnamefont {Iyer}},
  \bibinfo {author} {\bibfnamefont {M.~S.~S.}\ \bibnamefont {Qusailah}}, \ and\
  \bibinfo {author} {\bibfnamefont {B.~S.}\ \bibnamefont {Sathyaprakash}},\
  }\bibfield  {title} {\enquote {\bibinfo {title} {{Probing the non-linear
  structure of general relativity with black hole binaries}},}\ }\Doi
  {10.1103/PhysRevD.74.024006} {\bibfield  {journal} {\bibinfo  {journal}
  {Phys. Rev. D},\ }\textbf {\bibinfo {volume} {74}},\ \bibinfo {pages}
  {024006} (\bibinfo {year} {2006}{\natexlab{a}})},\ \Eprint
  {http://arxiv.org/abs/gr-qc/0604067} {arXiv:gr-qc/0604067} \BibitemShut
  {NoStop}%
\bibitem [{\citenamefont {Arun}\ \emph
  {et~al.}(2006){\natexlab{b}}\citenamefont {Arun}, \citenamefont {Iyer},
  \citenamefont {Qusailah},\ and\ \citenamefont {Sathyaprakash}}]{Arun:2006yw}%
  \BibitemOpen
  \bibfield  {author} {\bibinfo {author} {\bibfnamefont {K.~G.}\ \bibnamefont
  {Arun}}, \bibinfo {author} {\bibfnamefont {B.~R.}\ \bibnamefont {Iyer}},
  \bibinfo {author} {\bibfnamefont {M.~S.~S.}\ \bibnamefont {Qusailah}}, \ and\
  \bibinfo {author} {\bibfnamefont {B.~S.}\ \bibnamefont {Sathyaprakash}},\
  }\bibfield  {title} {\enquote {\bibinfo {title} {{Testing post-Newtonian
  theory with gravitational wave observations}},}\ }\Doi
  {10.1088/0264-9381/23/9/L01} {\bibfield  {journal} {\bibinfo  {journal}
  {Classical Quantum Gravity},\ }\textbf {\bibinfo {volume} {23}},\ \bibinfo
  {pages} {L37} (\bibinfo {year} {2006}{\natexlab{b}})},\ \Eprint
  {http://arxiv.org/abs/gr-qc/0604018} {arXiv:gr-qc/0604018} \BibitemShut
  {NoStop}%
\bibitem [{\citenamefont {Mishra}\ \emph {et~al.}(2010)\citenamefont {Mishra},
  \citenamefont {Arun}, \citenamefont {Iyer},\ and\ \citenamefont
  {Sathyaprakash}}]{Mishra:2010tp}%
  \BibitemOpen
  \bibfield  {author} {\bibinfo {author} {\bibfnamefont {C.~K.}\ \bibnamefont
  {Mishra}}, \bibinfo {author} {\bibfnamefont {K.~G.}\ \bibnamefont {Arun}},
  \bibinfo {author} {\bibfnamefont {B.~R.}\ \bibnamefont {Iyer}}, \ and\
  \bibinfo {author} {\bibfnamefont {B.~S.}\ \bibnamefont {Sathyaprakash}},\
  }\bibfield  {title} {\enquote {\bibinfo {title} {{Parametrized tests of
  post-Newtonian theory using Advanced LIGO and Einstein Telescope}},}\ }\Doi
  {10.1103/PhysRevD.82.064010} {\bibfield  {journal} {\bibinfo  {journal}
  {Phys. Rev. D},\ }\textbf {\bibinfo {volume} {82}},\ \bibinfo {pages}
  {064010} (\bibinfo {year} {2010})},\ \Eprint {http://arxiv.org/abs/1005.0304}
  {arXiv:1005.0304 [gr-qc]} \BibitemShut {NoStop}%
\bibitem [{\citenamefont {Yunes}\ and\ \citenamefont
  {Pretorius}(2009)}]{Yunes:2009ke}%
  \BibitemOpen
  \bibfield  {author} {\bibinfo {author} {\bibfnamefont {N.}~\bibnamefont
  {Yunes}}\ and\ \bibinfo {author} {\bibfnamefont {F.}~\bibnamefont
  {Pretorius}},\ }\bibfield  {title} {\enquote {\bibinfo {title} {{Fundamental
  Theoretical Bias in Gravitational Wave Astrophysics and the Parameterized
  Post-Einsteinian Framework}},}\ }\Doi {10.1103/PhysRevD.80.122003} {\bibfield
   {journal} {\bibinfo  {journal} {Phys. Rev. D},\ }\textbf {\bibinfo {volume}
  {80}},\ \bibinfo {pages} {122003} (\bibinfo {year} {2009})},\ \Eprint
  {http://arxiv.org/abs/0909.3328} {arXiv:0909.3328 [gr-qc]} \BibitemShut
  {NoStop}%
\bibitem [{\citenamefont {{Li}}\ \emph
  {et~al.}(2012){\natexlab{a}}\citenamefont {{Li}}, \citenamefont {{Del
  Pozzo}}, \citenamefont {{Vitale}}, \citenamefont {{Van Den Broeck}},
  \citenamefont {{Agathos}}, \citenamefont {{Veitch}}, \citenamefont
  {{Grover}}, \citenamefont {{Sidery}}, \citenamefont {{Sturani}},\ and\
  \citenamefont {{Vecchio}}}]{LiEtAl:2012a}%
  \BibitemOpen
  \bibfield  {author} {\bibinfo {author} {\bibfnamefont {T.~G.~F.}\
  \bibnamefont {{Li}}}, \bibinfo {author} {\bibfnamefont {W.}~\bibnamefont
  {{Del Pozzo}}}, \bibinfo {author} {\bibfnamefont {S.}~\bibnamefont
  {{Vitale}}}, \bibinfo {author} {\bibfnamefont {C.}~\bibnamefont {{Van Den
  Broeck}}}, \bibinfo {author} {\bibfnamefont {M.}~\bibnamefont {{Agathos}}},
  \bibinfo {author} {\bibfnamefont {J.}~\bibnamefont {{Veitch}}}, \bibinfo
  {author} {\bibfnamefont {K.}~\bibnamefont {{Grover}}}, \bibinfo {author}
  {\bibfnamefont {T.}~\bibnamefont {{Sidery}}}, \bibinfo {author}
  {\bibfnamefont {R.}~\bibnamefont {{Sturani}}}, \ and\ \bibinfo {author}
  {\bibfnamefont {A.}~\bibnamefont {{Vecchio}}},\ }\bibfield  {title} {\enquote
  {\bibinfo {title} {{Towards a generic test of the strong field dynamics of
  general relativity using compact binary coalescence}},}\ }\Doi
  {10.1103/PhysRevD.85.082003} {\bibfield  {journal} {\bibinfo  {journal}
  {\prd},\ }\textbf {\bibinfo {volume} {85}},\ \bibinfo {eid} {082003}
  (\bibinfo {year} {2012}{\natexlab{a}})},\ \Eprint
  {http://arxiv.org/abs/1110.0530} {arXiv:1110.0530 [gr-qc]} \BibitemShut
  {NoStop}%
\bibitem [{\citenamefont {{Li}}\ \emph
  {et~al.}(2012){\natexlab{b}}\citenamefont {{Li}}, \citenamefont {{Del
  Pozzo}}, \citenamefont {{Vitale}}, \citenamefont {{Van Den Broeck}},
  \citenamefont {{Agathos}}, \citenamefont {{Veitch}}, \citenamefont
  {{Grover}}, \citenamefont {{Sidery}}, \citenamefont {{Sturani}},\ and\
  \citenamefont {{Vecchio}}}]{LiEtAl:2012b}%
  \BibitemOpen
  \bibfield  {author} {\bibinfo {author} {\bibfnamefont {T.~G.~F.}\
  \bibnamefont {{Li}}}, \bibinfo {author} {\bibfnamefont {W.}~\bibnamefont
  {{Del Pozzo}}}, \bibinfo {author} {\bibfnamefont {S.}~\bibnamefont
  {{Vitale}}}, \bibinfo {author} {\bibfnamefont {C.}~\bibnamefont {{Van Den
  Broeck}}}, \bibinfo {author} {\bibfnamefont {M.}~\bibnamefont {{Agathos}}},
  \bibinfo {author} {\bibfnamefont {J.}~\bibnamefont {{Veitch}}}, \bibinfo
  {author} {\bibfnamefont {K.}~\bibnamefont {{Grover}}}, \bibinfo {author}
  {\bibfnamefont {T.}~\bibnamefont {{Sidery}}}, \bibinfo {author}
  {\bibfnamefont {R.}~\bibnamefont {{Sturani}}}, \ and\ \bibinfo {author}
  {\bibfnamefont {A.}~\bibnamefont {{Vecchio}}},\ }\bibfield  {title} {\enquote
  {\bibinfo {title} {{Towards a generic test of the strong field dynamics of
  general relativity using compact binary coalescence: Further
  investigations}},}\ }\Doi {10.1088/1742-6596/363/1/012028} {\bibfield
  {journal} {\bibinfo  {journal} {J. Phys. Conf. Ser.},\ }\textbf {\bibinfo
  {volume} {363}},\ \bibinfo {pages} {012028} (\bibinfo {year}
  {2012}{\natexlab{b}})},\ \Eprint {http://arxiv.org/abs/1111.5274}
  {arXiv:1111.5274 [gr-qc]} \BibitemShut {NoStop}%
\bibitem [{\citenamefont {Cornish}\ \emph {et~al.}(2011)\citenamefont
  {Cornish}, \citenamefont {Sampson}, \citenamefont {Yunes},\ and\
  \citenamefont {Pretorius}}]{Cornish:2011ys}%
  \BibitemOpen
  \bibfield  {author} {\bibinfo {author} {\bibfnamefont {N.}~\bibnamefont
  {Cornish}}, \bibinfo {author} {\bibfnamefont {L.}~\bibnamefont {Sampson}},
  \bibinfo {author} {\bibfnamefont {N.}~\bibnamefont {Yunes}}, \ and\ \bibinfo
  {author} {\bibfnamefont {F.}~\bibnamefont {Pretorius}},\ }\bibfield  {title}
  {\enquote {\bibinfo {title} {{Gravitational Wave Tests of General Relativity
  with the Parameterized Post-Einsteinian Framework}},}\ }\Doi
  {10.1103/PhysRevD.84.062003} {\bibfield  {journal} {\bibinfo  {journal}
  {\prd},\ }\textbf {\bibinfo {volume} {84}},\ \bibinfo {pages} {062003}
  (\bibinfo {year} {2011})},\ \Eprint {http://arxiv.org/abs/1105.2088}
  {arXiv:1105.2088 [gr-qc]} \BibitemShut {NoStop}%
\bibitem [{\citenamefont {Sampson}\ \emph {et~al.}(2014)\citenamefont
  {Sampson}, \citenamefont {Cornish},\ and\ \citenamefont
  {Yunes}}]{Sampson:2013jpa}%
  \BibitemOpen
  \bibfield  {author} {\bibinfo {author} {\bibfnamefont {L.}~\bibnamefont
  {Sampson}}, \bibinfo {author} {\bibfnamefont {N.}~\bibnamefont {Cornish}}, \
  and\ \bibinfo {author} {\bibfnamefont {N.}~\bibnamefont {Yunes}},\ }\bibfield
   {title} {\enquote {\bibinfo {title} {{Mismodeling in gravitational-wave
  astronomy: The trouble with templates}},}\ }\Doi {10.1103/PhysRevD.89.064037}
  {\bibfield  {journal} {\bibinfo  {journal} {Phys. Rev. D},\ }\textbf
  {\bibinfo {volume} {89}},\ \bibinfo {pages} {064037} (\bibinfo {year}
  {2014})},\ \Eprint {http://arxiv.org/abs/1311.4898} {arXiv:1311.4898 [gr-qc]}
  \BibitemShut {NoStop}%
\bibitem [{\citenamefont {Meidam}\ \emph {et~al.}(2018)\citenamefont {Meidam}
  \emph {et~al.}}]{Meidam:2017dgf}%
  \BibitemOpen
  \bibfield  {author} {\bibinfo {author} {\bibfnamefont {J.}~\bibnamefont
  {Meidam}} \emph {et~al.},\ }\bibfield  {title} {\enquote {\bibinfo {title}
  {{Parametrized tests of the strong-field dynamics of general relativity using
  gravitational wave signals from coalescing binary black holes: Fast
  likelihood calculations and sensitivity of the method}},}\ }\Doi
  {10.1103/PhysRevD.97.044033} {\bibfield  {journal} {\bibinfo  {journal}
  {Phys. Rev. D},\ }\textbf {\bibinfo {volume} {97}},\ \bibinfo {pages}
  {044033} (\bibinfo {year} {2018})},\ \Eprint
  {http://arxiv.org/abs/1712.08772} {arXiv:1712.08772 [gr-qc]} \BibitemShut
  {NoStop}%
\bibitem [{\citenamefont {Yunes}\ \emph {et~al.}(2016)\citenamefont {Yunes},
  \citenamefont {Yagi},\ and\ \citenamefont {Pretorius}}]{Yunes:2016jcc}%
  \BibitemOpen
  \bibfield  {author} {\bibinfo {author} {\bibfnamefont {N.}~\bibnamefont
  {Yunes}}, \bibinfo {author} {\bibfnamefont {K.}~\bibnamefont {Yagi}}, \ and\
  \bibinfo {author} {\bibfnamefont {F.}~\bibnamefont {Pretorius}},\ }\bibfield
  {title} {\enquote {\bibinfo {title} {{Theoretical Physics Implications of the
  Binary Black-Hole Mergers GW150914 and GW151226}},}\ }\Doi
  {10.1103/PhysRevD.94.084002} {\bibfield  {journal} {\bibinfo  {journal}
  {Phys. Rev. D},\ }\textbf {\bibinfo {volume} {94}},\ \bibinfo {pages}
  {084002} (\bibinfo {year} {2016})},\ \Eprint
  {http://arxiv.org/abs/1603.08955} {arXiv:1603.08955 [gr-qc]} \BibitemShut
  {NoStop}%
\bibitem [{\citenamefont {Yagi}\ \emph {et~al.}(2016)\citenamefont {Yagi},
  \citenamefont {Stein},\ and\ \citenamefont {Yunes}}]{Yagi:2015oca}%
  \BibitemOpen
  \bibfield  {author} {\bibinfo {author} {\bibfnamefont {K.}~\bibnamefont
  {Yagi}}, \bibinfo {author} {\bibfnamefont {L.~C.}\ \bibnamefont {Stein}}, \
  and\ \bibinfo {author} {\bibfnamefont {N.}~\bibnamefont {Yunes}},\ }\bibfield
   {title} {\enquote {\bibinfo {title} {{Challenging the Presence of Scalar
  Charge and Dipolar Radiation in Binary Pulsars}},}\ }\Doi
  {10.1103/PhysRevD.93.024010} {\bibfield  {journal} {\bibinfo  {journal}
  {Phys. Rev. D},\ }\textbf {\bibinfo {volume} {93}},\ \bibinfo {pages}
  {024010} (\bibinfo {year} {2016})},\ \Eprint
  {http://arxiv.org/abs/1510.02152} {arXiv:1510.02152 [gr-qc]} \BibitemShut
  {NoStop}%
\bibitem [{\citenamefont {Yagi}\ and\ \citenamefont
  {Stein}(2016)}]{Yagi:2016jml}%
  \BibitemOpen
  \bibfield  {author} {\bibinfo {author} {\bibfnamefont {K.}~\bibnamefont
  {Yagi}}\ and\ \bibinfo {author} {\bibfnamefont {L.~C.}\ \bibnamefont
  {Stein}},\ }\bibfield  {title} {\enquote {\bibinfo {title} {{Black Hole Based
  Tests of General Relativity}},}\ }\Doi {10.1088/0264-9381/33/5/054001}
  {\bibfield  {journal} {\bibinfo  {journal} {Classical Quantum Gravity},\
  }\textbf {\bibinfo {volume} {33}},\ \bibinfo {pages} {054001} (\bibinfo
  {year} {2016})},\ \Eprint {http://arxiv.org/abs/1602.02413} {arXiv:1602.02413
  [gr-qc]} \BibitemShut {NoStop}%
\bibitem [{\citenamefont {Barausse}\ \emph {et~al.}(2016)\citenamefont
  {Barausse}, \citenamefont {Yunes},\ and\ \citenamefont
  {Chamberlain}}]{Barausse:2016eii}%
  \BibitemOpen
  \bibfield  {author} {\bibinfo {author} {\bibfnamefont {E.}~\bibnamefont
  {Barausse}}, \bibinfo {author} {\bibfnamefont {N.}~\bibnamefont {Yunes}}, \
  and\ \bibinfo {author} {\bibfnamefont {K.}~\bibnamefont {Chamberlain}},\
  }\bibfield  {title} {\enquote {\bibinfo {title} {{Theory-Agnostic Constraints
  on Black-Hole Dipole Radiation with Multiband Gravitational-Wave
  Astrophysics}},}\ }\Doi {10.1103/PhysRevLett.116.241104} {\bibfield
  {journal} {\bibinfo  {journal} {Phys. Rev. Lett.},\ }\textbf {\bibinfo
  {volume} {116}},\ \bibinfo {pages} {241104} (\bibinfo {year} {2016})},\
  \Eprint {http://arxiv.org/abs/1603.04075} {arXiv:1603.04075 [gr-qc]}
  \BibitemShut {NoStop}%
\bibitem [{\citenamefont {Arun}(2012)}]{Arun:2012hf}%
  \BibitemOpen
  \bibfield  {author} {\bibinfo {author} {\bibfnamefont {K.~G.}\ \bibnamefont
  {Arun}},\ }\bibfield  {title} {\enquote {\bibinfo {title} {{Generic bounds on
  dipolar gravitational radiation from inspiralling compact binaries}},}\ }\Doi
  {10.1088/0264-9381/29/7/075011} {\bibfield  {journal} {\bibinfo  {journal}
  {Classical Quantum Gravity},\ }\textbf {\bibinfo {volume} {29}},\ \bibinfo
  {pages} {075011} (\bibinfo {year} {2012})},\ \Eprint
  {http://arxiv.org/abs/1202.5911} {arXiv:1202.5911 [gr-qc]} \BibitemShut
  {NoStop}%
\bibitem [{\citenamefont {Sampson}\ \emph {et~al.}(2013)\citenamefont
  {Sampson}, \citenamefont {Cornish},\ and\ \citenamefont
  {Yunes}}]{Sampson:2013lpa}%
  \BibitemOpen
  \bibfield  {author} {\bibinfo {author} {\bibfnamefont {L.}~\bibnamefont
  {Sampson}}, \bibinfo {author} {\bibfnamefont {N.}~\bibnamefont {Cornish}}, \
  and\ \bibinfo {author} {\bibfnamefont {N.}~\bibnamefont {Yunes}},\ }\bibfield
   {title} {\enquote {\bibinfo {title} {{Gravitational Wave Tests of Strong
  Field General Relativity with Binary Inspirals: Realistic Injections and
  Optimal Model Selection}},}\ }\Doi {10.1103/PhysRevD.87.102001} {\bibfield
  {journal} {\bibinfo  {journal} {Phys. Rev. D},\ }\textbf {\bibinfo {volume}
  {87}},\ \bibinfo {pages} {102001} (\bibinfo {year} {2013})},\ \Eprint
  {http://arxiv.org/abs/1303.1185} {arXiv:1303.1185 [gr-qc]} \BibitemShut
  {NoStop}%
\bibitem [{\citenamefont {Kramer}\ \emph {et~al.}(2006)\citenamefont {Kramer}
  \emph {et~al.}}]{Kramer:2006nb}%
  \BibitemOpen
  \bibfield  {author} {\bibinfo {author} {\bibfnamefont {M.}~\bibnamefont
  {Kramer}} \emph {et~al.},\ }\bibfield  {title} {\enquote {\bibinfo {title}
  {{Tests of general relativity from timing the double pulsar}},}\ }\Doi
  {10.1126/science.1132305} {\bibfield  {journal} {\bibinfo  {journal}
  {Science},\ }\textbf {\bibinfo {volume} {314}},\ \bibinfo {pages} {97}
  (\bibinfo {year} {2006})},\ \Eprint {http://arxiv.org/abs/astro-ph/0609417}
  {arXiv:astro-ph/0609417 [astro-ph]} \BibitemShut {NoStop}%
\bibitem [{\citenamefont {{Yunes}}\ and\ \citenamefont
  {{Hughes}}(2010)}]{YunesHughes2010}%
  \BibitemOpen
  \bibfield  {author} {\bibinfo {author} {\bibfnamefont {N.}~\bibnamefont
  {{Yunes}}}\ and\ \bibinfo {author} {\bibfnamefont {S.~A.}\ \bibnamefont
  {{Hughes}}},\ }\bibfield  {title} {\enquote {\bibinfo {title} {{Binary pulsar
  constraints on the parametrized post-Einsteinian framework}},}\ }\Doi
  {10.1103/PhysRevD.82.082002} {\bibfield  {journal} {\bibinfo  {journal}
  {\prd},\ }\textbf {\bibinfo {volume} {82}},\ \bibinfo {eid} {082002}
  (\bibinfo {year} {2010})},\ \Eprint {http://arxiv.org/abs/1007.1995}
  {arXiv:1007.1995 [gr-qc]} \BibitemShut {NoStop}%
\bibitem [{\citenamefont {Mirshekari}\ \emph {et~al.}(2012)\citenamefont
  {Mirshekari}, \citenamefont {Yunes},\ and\ \citenamefont
  {Will}}]{Mirshekari:2011yq}%
  \BibitemOpen
  \bibfield  {author} {\bibinfo {author} {\bibfnamefont {S.}~\bibnamefont
  {Mirshekari}}, \bibinfo {author} {\bibfnamefont {N.}~\bibnamefont {Yunes}}, \
  and\ \bibinfo {author} {\bibfnamefont {C.~M.}\ \bibnamefont {Will}},\
  }\bibfield  {title} {\enquote {\bibinfo {title} {{Constraining Generic
  Lorentz Violation and the Speed of the Graviton with Gravitational Waves}},}\
  }\Doi {10.1103/PhysRevD.85.024041} {\bibfield  {journal} {\bibinfo  {journal}
  {Phys. Rev. D},\ }\textbf {\bibinfo {volume} {85}},\ \bibinfo {pages}
  {024041} (\bibinfo {year} {2012})},\ \Eprint {http://arxiv.org/abs/1110.2720}
  {arXiv:1110.2720 [gr-qc]} \BibitemShut {NoStop}%
\bibitem [{\citenamefont {Abbott}\ \emph {et~al.}(2017)\citenamefont {Abbott}
  \emph {et~al.}}]{Monitor:2017mdv}%
  \BibitemOpen
  \bibfield  {author} {\bibinfo {author} {\bibfnamefont {B.~P.}\ \bibnamefont
  {Abbott}} \emph {et~al.} (\bibinfo {collaboration} {LIGO Scientific
  Collaboration, Virgo Collaboration, Fermi-GBM Collaboration, and INTEGRAL
  Collaboration}),\ }\bibfield  {title} {\enquote {\bibinfo {title}
  {{Gravitational Waves and Gamma-rays from a Binary Neutron Star Merger:
  GW170817 and GRB 170817A}},}\ }\Doi {10.3847/2041-8213/aa920c} {\bibfield
  {journal} {\bibinfo  {journal} {Astrophys. J. Lett.},\ }\textbf {\bibinfo
  {volume} {848}},\ \bibinfo {pages} {L13} (\bibinfo {year} {2017})},\ \Eprint
  {http://arxiv.org/abs/1710.05834} {arXiv:1710.05834 [astro-ph.HE]}
  \BibitemShut {NoStop}%
\bibitem [{\citenamefont {Will}(1998)}]{Will:1997bb}%
  \BibitemOpen
  \bibfield  {author} {\bibinfo {author} {\bibfnamefont {C.~M.}\ \bibnamefont
  {Will}},\ }\bibfield  {title} {\enquote {\bibinfo {title} {{Bounding the mass
  of the graviton using gravitational wave observations of inspiralling compact
  binaries}},}\ }\Doi {10.1103/PhysRevD.57.2061} {\bibfield  {journal}
  {\bibinfo  {journal} {Phys. Rev. D},\ }\textbf {\bibinfo {volume} {57}},\
  \bibinfo {pages} {2061} (\bibinfo {year} {1998})},\ \Eprint
  {http://arxiv.org/abs/gr-qc/9709011} {arXiv:gr-qc/9709011 [gr-qc]}
  \BibitemShut {NoStop}%
\bibitem [{\citenamefont {Calcagni}(2010)}]{Calcagni:2009kc}%
  \BibitemOpen
  \bibfield  {author} {\bibinfo {author} {\bibfnamefont {G.}~\bibnamefont
  {Calcagni}},\ }\bibfield  {title} {\enquote {\bibinfo {title} {{Fractal
  universe and quantum gravity}},}\ }\Doi {10.1103/PhysRevLett.104.251301}
  {\bibfield  {journal} {\bibinfo  {journal} {Phys. Rev. Lett.},\ }\textbf
  {\bibinfo {volume} {104}},\ \bibinfo {pages} {251301} (\bibinfo {year}
  {2010})},\ \Eprint {http://arxiv.org/abs/0912.3142} {arXiv:0912.3142
  [hep-th]} \BibitemShut {NoStop}%
\bibitem [{\citenamefont {Amelino-Camelia}(2002)}]{AmelinoCamelia:2002wr}%
  \BibitemOpen
  \bibfield  {author} {\bibinfo {author} {\bibfnamefont {G.}~\bibnamefont
  {Amelino-Camelia}},\ }\bibfield  {title} {\enquote {\bibinfo {title} {{Doubly
  special relativity}},}\ }\Doi {10.1038/418034a} {\bibfield  {journal}
  {\bibinfo  {journal} {Nature (London)},\ }\textbf {\bibinfo {volume} {418}},\
  \bibinfo {pages} {34} (\bibinfo {year} {2002})},\ \Eprint
  {http://arxiv.org/abs/gr-qc/0207049} {arXiv:gr-qc/0207049 [gr-qc]}
  \BibitemShut {NoStop}%
\bibitem [{\citenamefont {Ho{\v{r}}ava}(2009)}]{Horava:2009uw}%
  \BibitemOpen
  \bibfield  {author} {\bibinfo {author} {\bibfnamefont {P.}~\bibnamefont
  {Ho{\v{r}}ava}},\ }\bibfield  {title} {\enquote {\bibinfo {title} {{Quantum
  Gravity at a Lifshitz Point}},}\ }\Doi {10.1103/PhysRevD.79.084008}
  {\bibfield  {journal} {\bibinfo  {journal} {Phys. Rev. D},\ }\textbf
  {\bibinfo {volume} {79}},\ \bibinfo {pages} {084008} (\bibinfo {year}
  {2009})},\ \Eprint {http://arxiv.org/abs/0901.3775} {arXiv:0901.3775
  [hep-th]} \BibitemShut {NoStop}%
\bibitem [{\citenamefont {Sefiedgar}\ \emph {et~al.}(2011)\citenamefont
  {Sefiedgar}, \citenamefont {Nozari},\ and\ \citenamefont
  {Sepangi}}]{Sefiedgar:2010we}%
  \BibitemOpen
  \bibfield  {author} {\bibinfo {author} {\bibfnamefont {A.~S.}\ \bibnamefont
  {Sefiedgar}}, \bibinfo {author} {\bibfnamefont {K.}~\bibnamefont {Nozari}}, \
  and\ \bibinfo {author} {\bibfnamefont {H.~R.}\ \bibnamefont {Sepangi}},\
  }\bibfield  {title} {\enquote {\bibinfo {title} {{Modified dispersion
  relations in extra dimensions}},}\ }\Doi {10.1016/j.physletb.2010.11.067}
  {\bibfield  {journal} {\bibinfo  {journal} {Phys. Lett.},\ }\textbf {\bibinfo
  {volume} {B696}},\ \bibinfo {pages} {119} (\bibinfo {year} {2011})},\ \Eprint
  {http://arxiv.org/abs/1012.1406} {arXiv:1012.1406 [gr-qc]} \BibitemShut
  {NoStop}%
\bibitem [{\citenamefont {Kosteleck{\'y}}\ and\ \citenamefont
  {Mewes}(2016)}]{Kostelecky:2016kfm}%
  \BibitemOpen
  \bibfield  {author} {\bibinfo {author} {\bibfnamefont {V.~A.}\ \bibnamefont
  {Kosteleck{\'y}}}\ and\ \bibinfo {author} {\bibfnamefont {M.}~\bibnamefont
  {Mewes}},\ }\bibfield  {title} {\enquote {\bibinfo {title} {{Testing local
  Lorentz invariance with gravitational waves}},}\ }\Doi
  {10.1016/j.physletb.2016.04.040} {\bibfield  {journal} {\bibinfo  {journal}
  {Phys. Lett.},\ }\textbf {\bibinfo {volume} {B757}},\ \bibinfo {pages} {510}
  (\bibinfo {year} {2016})},\ \Eprint {http://arxiv.org/abs/1602.04782}
  {arXiv:1602.04782 [gr-qc]} \BibitemShut {NoStop}%
\bibitem [{\citenamefont {Ade}\ \emph {et~al.}(2016)\citenamefont {Ade} \emph
  {et~al.}}]{Ade:2015xua}%
  \BibitemOpen
  \bibfield  {author} {\bibinfo {author} {\bibfnamefont {P.~A.~R.}\
  \bibnamefont {Ade}} \emph {et~al.} (\bibinfo {collaboration} {Planck
  Collaboration}),\ }\bibfield  {title} {\enquote {\bibinfo {title} {{Planck
  2015 results. XIII. Cosmological parameters}},}\ }\Doi
  {10.1051/0004-6361/201525830} {\bibfield  {journal} {\bibinfo  {journal}
  {Astron. Astrophys.},\ }\textbf {\bibinfo {volume} {594}},\ \bibinfo {pages}
  {A13} (\bibinfo {year} {2016})},\ \Eprint {http://arxiv.org/abs/1502.01589}
  {arXiv:1502.01589 [astro-ph.CO]} \BibitemShut {NoStop}%
\bibitem [{\citenamefont {Aghanim}\ \emph {et~al.}(2018)\citenamefont {Aghanim}
  \emph {et~al.}}]{Aghanim:2018eyx}%
  \BibitemOpen
  \bibfield  {author} {\bibinfo {author} {\bibfnamefont {N.}~\bibnamefont
  {Aghanim}} \emph {et~al.} (\bibinfo {collaboration} {Planck Collaboration}),\
  }\bibfield  {title} {\enquote {\bibinfo {title} {{Planck 2018 results. VI.
  Cosmological parameters}},}\ }\href@noop {} { (\bibinfo {year} {2018})},\
  \Eprint {http://arxiv.org/abs/1807.06209} {arXiv:1807.06209 [astro-ph.CO]}
  \BibitemShut {NoStop}%
\bibitem [{\citenamefont {Amelino-Camelia}\ \emph {et~al.}(2006)\citenamefont
  {Amelino-Camelia}, \citenamefont {Arzano}, \citenamefont {Ling},\ and\
  \citenamefont {Mandanici}}]{AmelinoCamelia:2005ik}%
  \BibitemOpen
  \bibfield  {author} {\bibinfo {author} {\bibfnamefont {G.}~\bibnamefont
  {Amelino-Camelia}}, \bibinfo {author} {\bibfnamefont {M.}~\bibnamefont
  {Arzano}}, \bibinfo {author} {\bibfnamefont {Y.}~\bibnamefont {Ling}}, \ and\
  \bibinfo {author} {\bibfnamefont {G.}~\bibnamefont {Mandanici}},\ }\bibfield
  {title} {\enquote {\bibinfo {title} {{Black-hole thermodynamics with modified
  dispersion relations and generalized uncertainty principles}},}\ }\Doi
  {10.1088/0264-9381/23/7/022} {\bibfield  {journal} {\bibinfo  {journal}
  {Classical Quantum Gravity},\ }\textbf {\bibinfo {volume} {23}},\ \bibinfo
  {pages} {2585} (\bibinfo {year} {2006})},\ \Eprint
  {http://arxiv.org/abs/gr-qc/0506110} {arXiv:gr-qc/0506110 [gr-qc]}
  \BibitemShut {NoStop}%
\bibitem [{\citenamefont {Calcagni}(2017)}]{Calcagni:2016zqv}%
  \BibitemOpen
  \bibfield  {author} {\bibinfo {author} {\bibfnamefont {G.}~\bibnamefont
  {Calcagni}},\ }\bibfield  {title} {\enquote {\bibinfo {title} {{Lorentz
  violations in multifractal spacetimes}},}\ }\Doi
  {10.1140/epjc/s10052-017-4841-6} {\bibfield  {journal} {\bibinfo  {journal}
  {Eur. Phys. J.},\ }\textbf {\bibinfo {volume} {C77}},\ \bibinfo {pages} {291}
  (\bibinfo {year} {2017})},\ \Eprint {http://arxiv.org/abs/1603.03046}
  {arXiv:1603.03046 [gr-qc]} \BibitemShut {NoStop}%
\bibitem [{\citenamefont {Bernus}\ \emph {et~al.}(2019)\citenamefont {Bernus},
  \citenamefont {Minazzoli}, \citenamefont {Fienga}, \citenamefont {Gastineau},
  \citenamefont {Laskar},\ and\ \citenamefont {Deram}}]{Bernus:2019rgl}%
  \BibitemOpen
  \bibfield  {author} {\bibinfo {author} {\bibfnamefont {L.}~\bibnamefont
  {Bernus}}, \bibinfo {author} {\bibfnamefont {O.}~\bibnamefont {Minazzoli}},
  \bibinfo {author} {\bibfnamefont {A.}~\bibnamefont {Fienga}}, \bibinfo
  {author} {\bibfnamefont {M.}~\bibnamefont {Gastineau}}, \bibinfo {author}
  {\bibfnamefont {J.}~\bibnamefont {Laskar}}, \ and\ \bibinfo {author}
  {\bibfnamefont {P.}~\bibnamefont {Deram}},\ }\bibfield  {title} {\enquote
  {\bibinfo {title} {{Constraining the mass of the graviton with the planetary
  ephemeris INPOP}},}\ }\href@noop {} { (\bibinfo {year} {2019})},\ \Eprint
  {http://arxiv.org/abs/1901.04307} {arXiv:1901.04307 [gr-qc]} \BibitemShut
  {NoStop}%
\bibitem [{\citenamefont {Will}(2018)}]{Will:2018gku}%
  \BibitemOpen
  \bibfield  {author} {\bibinfo {author} {\bibfnamefont {C.~M.}\ \bibnamefont
  {Will}},\ }\bibfield  {title} {\enquote {\bibinfo {title} {{Solar system vs.
  gravitational-wave bounds on the graviton mass}},}\ }\Doi
  {10.1088/1361-6382/aad13c} {\bibfield  {journal} {\bibinfo  {journal}
  {Classical Quantum Gravity},\ }\textbf {\bibinfo {volume} {35}},\ \bibinfo
  {pages} {17LT01} (\bibinfo {year} {2018})},\ \Eprint
  {http://arxiv.org/abs/1805.10523} {arXiv:1805.10523 [gr-qc]} \BibitemShut
  {NoStop}%
\bibitem [{\citenamefont {de~Rham}\ \emph {et~al.}(2017)\citenamefont
  {de~Rham}, \citenamefont {Deskins}, \citenamefont {Tolley},\ and\
  \citenamefont {Zhou}}]{deRham:2016nuf}%
  \BibitemOpen
  \bibfield  {author} {\bibinfo {author} {\bibfnamefont {C.}~\bibnamefont
  {de~Rham}}, \bibinfo {author} {\bibfnamefont {J.~T.}\ \bibnamefont
  {Deskins}}, \bibinfo {author} {\bibfnamefont {A.~J.}\ \bibnamefont {Tolley}},
  \ and\ \bibinfo {author} {\bibfnamefont {S.-Y.}\ \bibnamefont {Zhou}},\
  }\bibfield  {title} {\enquote {\bibinfo {title} {{Graviton Mass Bounds}},}\
  }\Doi {10.1103/RevModPhys.89.025004} {\bibfield  {journal} {\bibinfo
  {journal} {Rev. Mod. Phys.},\ }\textbf {\bibinfo {volume} {89}},\ \bibinfo
  {pages} {025004} (\bibinfo {year} {2017})},\ \Eprint
  {http://arxiv.org/abs/1606.08462} {arXiv:1606.08462 [astro-ph.CO]}
  \BibitemShut {NoStop}%
\bibitem [{\citenamefont {Rubin}(1981)}]{rubin1981}%
  \BibitemOpen
  \bibfield  {author} {\bibinfo {author} {\bibfnamefont {D.~B.}\ \bibnamefont
  {Rubin}},\ }\bibfield  {title} {\enquote {\bibinfo {title} {{The Bayesian
  Bootstrap}},}\ }\Doi {10.1214/aos/1176345338} {\bibfield  {journal} {\bibinfo
   {journal} {Ann. Statist.},\ }\textbf {\bibinfo {volume} {9}},\ \bibinfo
  {pages} {130} (\bibinfo {year} {1981})}\BibitemShut {NoStop}%
\bibitem [{\citenamefont {Eardley}\ \emph {et~al.}(1973)\citenamefont
  {Eardley}, \citenamefont {Lee},\ and\ \citenamefont
  {Lightman}}]{Eardley:1974nw}%
  \BibitemOpen
  \bibfield  {author} {\bibinfo {author} {\bibfnamefont {D.~M.}\ \bibnamefont
  {Eardley}}, \bibinfo {author} {\bibfnamefont {D.~L.}\ \bibnamefont {Lee}}, \
  and\ \bibinfo {author} {\bibfnamefont {A.~P.}\ \bibnamefont {Lightman}},\
  }\bibfield  {title} {\enquote {\bibinfo {title} {{Gravitational-wave
  observations as a tool for testing relativistic gravity}},}\ }\Doi
  {10.1103/PhysRevD.8.3308} {\bibfield  {journal} {\bibinfo  {journal} {Phys.
  Rev. D},\ }\textbf {\bibinfo {volume} {8}},\ \bibinfo {pages} {3308}
  (\bibinfo {year} {1973})}\BibitemShut {NoStop}%
\bibitem [{\citenamefont {Chatziioannou}\ \emph {et~al.}(2012)\citenamefont
  {Chatziioannou}, \citenamefont {Yunes},\ and\ \citenamefont
  {Cornish}}]{Chatziioannou:2012rf}%
  \BibitemOpen
  \bibfield  {author} {\bibinfo {author} {\bibfnamefont {K.}~\bibnamefont
  {Chatziioannou}}, \bibinfo {author} {\bibfnamefont {N.}~\bibnamefont
  {Yunes}}, \ and\ \bibinfo {author} {\bibfnamefont {N.}~\bibnamefont
  {Cornish}},\ }\bibfield  {title} {\enquote {\bibinfo {title}
  {{Model-Independent Test of General Relativity: An Extended post-Einsteinian
  Framework with Complete Polarization Content}},}\ }\Doi
  {10.1103/PhysRevD.86.022004} {\bibfield  {journal} {\bibinfo  {journal}
  {Phys. Rev. D},\ }\textbf {\bibinfo {volume} {86}},\ \bibinfo {pages}
  {022004} (\bibinfo {year} {2012})},\ \bibinfo {note}
  {\href{http://dx.doi.org/10.1103/PhysRevD.95.129901}{{\textbf{95}}, 129901(E)
  (2017)}},\ \Eprint {http://arxiv.org/abs/1204.2585} {arXiv:1204.2585 [gr-qc]}
  \BibitemShut {NoStop}%
\bibitem [{\citenamefont {Isi}\ \emph {et~al.}(2017)\citenamefont {Isi},
  \citenamefont {Pitkin},\ and\ \citenamefont {Weinstein}}]{Isi:2017equ}%
  \BibitemOpen
  \bibfield  {author} {\bibinfo {author} {\bibfnamefont {M.}~\bibnamefont
  {Isi}}, \bibinfo {author} {\bibfnamefont {M.}~\bibnamefont {Pitkin}}, \ and\
  \bibinfo {author} {\bibfnamefont {A.~J.}\ \bibnamefont {Weinstein}},\
  }\bibfield  {title} {\enquote {\bibinfo {title} {{Probing Dynamical Gravity
  with the Polarization of Continuous Gravitational Waves}},}\ }\Doi
  {10.1103/PhysRevD.96.042001} {\bibfield  {journal} {\bibinfo  {journal}
  {Phys. Rev. D},\ }\textbf {\bibinfo {volume} {96}},\ \bibinfo {pages}
  {042001} (\bibinfo {year} {2017})},\ \Eprint
  {http://arxiv.org/abs/1703.07530} {arXiv:1703.07530 [gr-qc]} \BibitemShut
  {NoStop}%
\bibitem [{\citenamefont {Callister}\ \emph {et~al.}(2017)\citenamefont
  {Callister}, \citenamefont {Biscoveanu}, \citenamefont {Christensen},
  \citenamefont {Isi}, \citenamefont {Matas}, \citenamefont {Minazzoli},
  \citenamefont {Regimbau}, \citenamefont {Sakellariadou}, \citenamefont
  {Tasson},\ and\ \citenamefont {Thrane}}]{Callister:2017ocg}%
  \BibitemOpen
  \bibfield  {author} {\bibinfo {author} {\bibfnamefont {T.}~\bibnamefont
  {Callister}}, \bibinfo {author} {\bibfnamefont {A.~S.}\ \bibnamefont
  {Biscoveanu}}, \bibinfo {author} {\bibfnamefont {N.}~\bibnamefont
  {Christensen}}, \bibinfo {author} {\bibfnamefont {M.}~\bibnamefont {Isi}},
  \bibinfo {author} {\bibfnamefont {A.}~\bibnamefont {Matas}}, \bibinfo
  {author} {\bibfnamefont {O.}~\bibnamefont {Minazzoli}}, \bibinfo {author}
  {\bibfnamefont {T.}~\bibnamefont {Regimbau}}, \bibinfo {author}
  {\bibfnamefont {M.}~\bibnamefont {Sakellariadou}}, \bibinfo {author}
  {\bibfnamefont {J.}~\bibnamefont {Tasson}}, \ and\ \bibinfo {author}
  {\bibfnamefont {E.}~\bibnamefont {Thrane}},\ }\bibfield  {title} {\enquote
  {\bibinfo {title} {{Polarization-based Tests of Gravity with the Stochastic
  Gravitational-Wave Background}},}\ }\Doi {10.1103/PhysRevX.7.041058}
  {\bibfield  {journal} {\bibinfo  {journal} {Phys. Rev. X},\ }\textbf
  {\bibinfo {volume} {7}},\ \bibinfo {pages} {041058} (\bibinfo {year}
  {2017})},\ \Eprint {http://arxiv.org/abs/1704.08373} {arXiv:1704.08373
  [gr-qc]} \BibitemShut {NoStop}%
\bibitem [{\citenamefont {Isi}\ and\ \citenamefont
  {Weinstein}(2017)}]{Isi:2017fbj}%
  \BibitemOpen
  \bibfield  {author} {\bibinfo {author} {\bibfnamefont {M.}~\bibnamefont
  {Isi}}\ and\ \bibinfo {author} {\bibfnamefont {A.~J.}\ \bibnamefont
  {Weinstein}},\ }\bibfield  {title} {\enquote {\bibinfo {title} {{Probing
  gravitational wave polarizations with signals from compact binary
  coalescences}},}\ }\href {https://dcc.ligo.org/P1700276/public} {\bibfield
  {journal} {\bibinfo  {journal} {{Tech. Note, LIGO-P1700276}}} (\bibinfo
  {year} {2017})},\ \Eprint {http://arxiv.org/abs/1710.03794} {arXiv:1710.03794
  [gr-qc]} \BibitemShut {NoStop}%
\bibitem [{\citenamefont {B{\l}aut}(2012)}]{Blaut:2012zz}%
  \BibitemOpen
  \bibfield  {author} {\bibinfo {author} {\bibfnamefont {A.}~\bibnamefont
  {B{\l}aut}},\ }\bibfield  {title} {\enquote {\bibinfo {title} {{Angular and
  frequency response of the gravitational wave interferometers in the metric
  theories of gravity}},}\ }\Doi {10.1103/PhysRevD.85.043005} {\bibfield
  {journal} {\bibinfo  {journal} {Phys. Rev. D},\ }\textbf {\bibinfo {volume}
  {85}},\ \bibinfo {pages} {043005} (\bibinfo {year} {2012})}\BibitemShut
  {NoStop}%
\bibitem [{\citenamefont {Abbott}\ \emph {et~al.}(2018)\citenamefont {Abbott}
  \emph {et~al.}}]{Aasi:2013wya}%
  \BibitemOpen
  \bibfield  {author} {\bibinfo {author} {\bibfnamefont {B.~P.}\ \bibnamefont
  {Abbott}} \emph {et~al.} (\bibinfo {collaboration} {KAGRA Collaboration, LIGO
  Scientific Collaboration, and Virgo Collaboration}),\ }\bibfield  {title}
  {\enquote {\bibinfo {title} {{Prospects for Observing and Localizing
  Gravitational-Wave Transients with Advanced LIGO, Advanced Virgo and
  KAGRA}},}\ }\Doi {10.1007/s41114-018-0012-9} {\bibfield  {journal} {\bibinfo
  {journal} {Living Rev. Relativity},\ }\textbf {\bibinfo {volume} {21}},\
  \bibinfo {pages} {3} (\bibinfo {year} {2018})},\ \Eprint
  {http://arxiv.org/abs/1304.0670} {arXiv:1304.0670 [gr-qc]} \BibitemShut
  {NoStop}%
\bibitem [{\citenamefont {Abbott}\ \emph
  {et~al.}(2016){\natexlab{d}}\citenamefont {Abbott} \emph
  {et~al.}}]{Abbott:2016apu}%
  \BibitemOpen
  \bibfield  {author} {\bibinfo {author} {\bibfnamefont {B.~P.}\ \bibnamefont
  {Abbott}} \emph {et~al.} (\bibinfo {collaboration} {LIGO Scientific
  Collaboration and Virgo Collaboration}),\ }\bibfield  {title} {\enquote
  {\bibinfo {title} {{Directly comparing GW150914 with numerical solutions of
  Einstein’s equations for binary black hole coalescence}},}\ }\Doi
  {10.1103/PhysRevD.94.064035} {\bibfield  {journal} {\bibinfo  {journal}
  {Phys. Rev. D},\ }\textbf {\bibinfo {volume} {94}},\ \bibinfo {pages}
  {064035} (\bibinfo {year} {2016}{\natexlab{d}})},\ \Eprint
  {http://arxiv.org/abs/1606.01262} {arXiv:1606.01262 [gr-qc]} \BibitemShut
  {NoStop}%
\bibitem [{\citenamefont {Lange}\ \emph {et~al.}(2017)\citenamefont {Lange}
  \emph {et~al.}}]{Lange:2017wki}%
  \BibitemOpen
  \bibfield  {author} {\bibinfo {author} {\bibfnamefont {J.}~\bibnamefont
  {Lange}} \emph {et~al.},\ }\bibfield  {title} {\enquote {\bibinfo {title}
  {{Parameter estimation method that directly compares gravitational wave
  observations to numerical relativity}},}\ }\Doi {10.1103/PhysRevD.96.104041}
  {\bibfield  {journal} {\bibinfo  {journal} {Phys. Rev. D},\ }\textbf
  {\bibinfo {volume} {96}},\ \bibinfo {pages} {104041} (\bibinfo {year}
  {2017})},\ \Eprint {http://arxiv.org/abs/1705.09833} {arXiv:1705.09833
  [gr-qc]} \BibitemShut {NoStop}%
\bibitem [{\citenamefont {Ashton}\ \emph {et~al.}(2019)\citenamefont {Ashton}
  \emph {et~al.}}]{Ashton:2018jfp}%
  \BibitemOpen
  \bibfield  {author} {\bibinfo {author} {\bibfnamefont {G.}~\bibnamefont
  {Ashton}} \emph {et~al.},\ }\bibfield  {title} {\enquote {\bibinfo {title}
  {{Bilby: A user-friendly Bayesian inference library for gravitational-wave
  astronomy}},}\ }\Doi {10.3847/1538-4365/ab06fc} {\bibfield  {journal}
  {\bibinfo  {journal} {Astrophys. J. Suppl. Ser.},\ }\textbf {\bibinfo
  {volume} {241}},\ \bibinfo {pages} {27} (\bibinfo {year} {2019})},\ \Eprint
  {http://arxiv.org/abs/1811.02042} {arXiv:1811.02042 [astro-ph.IM]}
  \BibitemShut {NoStop}%
\end{thebibliography}%

\clearpage

\iftoggle{endauthorlist}{
 %
 %
 \let\author\myauthor
 \let\affiliation\myaffiliation
 \let\maketitle\mymaketitle
 \title{Authors}
 \pacs{}

 \newpage
 \maketitle
}

\end{document}